\PassOptionsToPackage{desactivate}{linenoaa}
\documentclass{aa}
\usepackage{graphicx}
\usepackage{natbib}
\usepackage{amsmath}

\usepackage[english]{babel}
\usepackage[utf8]{inputenc}
\usepackage{amssymb}
\usepackage{xcolor}
\makeatletter

\newcommand{\Rmnum}[1]{\expandafter\@slowromancap\romannumeral #1@}
\makeatother
\usepackage{txfonts}
\usepackage{verbatim} 
\usepackage[colorlinks=true,linkcolor=bluea,allcolors=blue]{hyperref}
\usepackage{float}
\usepackage{upgreek}
\usepackage{multirow}
\newcommand{\tablesize}{\fontsize{8.5pt}{10pt}\selectfont}
\usepackage{rotating}
\usepackage{booktabs}

\titlerunning{}
\authorrunning{Speranza et al.}

\begin{document}

\title{Beyond the dust: \\ JWST reveals  extinction and ice content in local (U)LIRGs }



\author{ G. Speranza\inst{1}
\and M. Pereira-Santaella\inst{1}
\and I. Garc\'ia-Bernete\inst{2}
\and E. Gonz\'alez-Alfonso\inst{3}
\and S. Arribas\inst{4}
\and I. Lamperti\inst{5, 6}
\and M. Perna\inst{4}
\and A. Alonso-Herrero \inst{2}
\and L. Hermosa Mu\~noz \inst{7}
\and D. Rigopoulou\inst{8, 9}
\and N. Thatte\inst{8}}

\institute{
Instituto de F\'isica Fundamental, CSIC,
Calle Serrano 123, 28006 Madrid, Spain\\
\email{giovanna.speranza@iff.csic.es}
\and Centro de Astrobiolog\'ia (CAB), CSIC-INTA,
Camino Bajo del Castillo s/n,
28692 Villanueva de la Ca\~nada, Spain
\and Universidad de Alcal\'a, Departamento de F\'isica y Matem\'aticas, Campus Universitario, 28871 Alcal\'a de Henares, Spain
\and Centro de Astrobiolog\'ia (CAB), CSIC-INTA, Cra. de Ajalvir km. 4, 28850, Torrej\'on de Ardoz, Madrid, Spain
\and Università di Firenze, Dipartimento di Fisica e Astronomia, Via G. Sansone 1, 50019, Sesto F.no, Firenze, Italy 
\and INAF - Osservatorio Astrofisico di Arcetri, Largo E. Fermi 5, I-50125, Firenze, Italy
\and Departamento de F{\'i}sica, Universidad de Oviedo, Campus de Llamaquique, C/ Calvo Sotelo s/n, 33007 Oviedo, Spain
\and Department of Physics, University of Oxford, Keble Road, Oxford OX1 3RH, UK 
\and School of Sciences, European University Cyprus, Diogenes street, Engomi, 1516 Nicosia, Cyprus}

   \date{}
  \abstract { 

We investigate the spatially resolved extinction levels and ice column densities in the deeply obscured environments of four nearby interacting (Ultra)luminous infrared galaxies ((U)LIRGs;  L$_{\rm IR}>\, 10^{11.6}$ L$_\odot$ and D$_{\rm L}\sim$ 39 -- 107 Mpc). Accurate extinction corrections are essential to constrain the dominant power source and to unveil the most heavily obscured structures. We used the three high-resolution (R$\sim$2700) gratings of the JWST/NIRSpec integral field spectrograph, with spectral coverage of ($\sim$ 0.97 -- 5.14) $\mu$m. This wide range encompasses numerous ($>25$) H recombination and H$_2$ ro-vibrational emission lines, as well as the prominent H$_2$O ($\sim$3.02 $\mu$m) and CO$_2$ ($\sim$4.26 $\mu$m) ice absorption features. We present three methods to derive phase-specific extinction  affecting the ionized gas (A$_{\rm K}^{\rm HII}$), the hot molecular gas (A$_{\rm K}^{\rm H_2}$), and the H$_2$O and CO$_2$ ice column densities (N$_{\rm H_2O}$ and N$_{\rm CO_2}$). Using H and H$_2$ lines detected within the H$_2$O absorption feature, we also derive phase-specific water ice column densities (N$_{\rm H_2O}^{\rm HII}$ and N$_{\rm H_2O}^{\rm H_2}$). We compare results from these methods with classical extinction diagnostics and provide the corresponding correlations. We find that nuclear regions (0.2" radius) exhibit extinction and ice column densities typically about factor of two higher than the extended regions, with A$_{\rm K}^{\rm HII} = (1.25 \pm$ 0.02) and A$_{\rm K}^{\rm H_2} = (0.75 \pm 0.10)$ in the nuclei, compared to A$_{\rm K}^{\rm HII} = (0.61 \pm 0.04)$ and A$_{\rm K}^{\rm H_2} = (0.50 \pm$ 0.07) in the extended regions. The ionized gas is more extincted than the hot molecular counterpart, on average with A$_{\rm K}^{\rm HII} \approx$ 1.5 A$_{\rm K}^{\rm H_2}$, whereas the H$_2$O ice column densities are comparable between the two phases, with median values of $\sim$10$^{18.11}$ cm$^{-2}$ in the nuclei and $\sim$10$^{17.89}$ cm$^{-2}$ in the extended regions. The latter value is consistent, within the uncertainties, with that derived from the continuum, but not in the case of some nuclear regions, where N$_{\rm H_2O}$ increases up to $\sim$10$^{18.81}$ cm$^{-2}$ due to the contribution of compact obscured nuclei. These results suggest that the ionized gas originates from more deeply embedded regions than the hot H$_2$ emission, while the shell of dust grains, where water ice accumulates, envelopes both gas-phases, with an additional ice-rich component surrounding the most obscured nuclei. Finally, from the continuum ice absorption features we derive a median relative abundance of N$_{\rm CO_2}$/N$_{\rm H_2O} \,\sim \,0.10$, a factor $\sim$2 -- 3 lower than typical Galactic values.

}


\maketitle

\section{Introduction}
\label{intro}

Luminous and (Ultra)luminous infrared (IR) galaxies ((U)LIRGs; L$_{\rm IR} > 10^{11}$ L$_{\odot}$) are among the most luminous and heavily obscured objects in the local Universe. 
 They represent a key phase in galaxy evolution (\citealt{Hopkins08}), typically associated with galaxy mergers at different stages. These interactions drive large concentrations of gas and dust toward the central regions, fueling both star formation (SF) and active galactic nucleus (AGN) activity (e.g., \citealt{Stierwalt13, Alonso13}), while simultaneously obscuring the host galaxies and their nuclei.

(U)LIRGs host extremely compact (d$<$200 pc; \citealt{Barcos17, Pereira21}) and dust-embedded nuclei (A$_{\rm V} \gg$ 1000; \citealt{Alfonso15, Falstad21, Bernete22, Bernete25}), where a large fraction of the AGN and SF emission is absorbed by dust and re-emitted at infrared wavelengths. As a consequence, determining whether SF or AGN activity dominates the IR luminosity is challenging. 
Accurate extinction corrections are essential to constrain the dominant power source and to unveil the most heavily obscured structures.
Extinction is commonly estimated from ionized gas tracers, typically using hydrogen recombination line ratios such as the Balmer decrement in the optical (e.g., H$\alpha$/H$\beta$; \citealt{Perna21}). These methods rely on comparing observed line ratios with their intrinsic theoretical values to infer the level of attenuation. Such diagnostics have been widely applied in (U)LIRGs (e.g., \citealt{Takata06, Ceci25}) using different line combinations as  H$\alpha$/H$\beta$ or Pa$\alpha$/Br$\gamma$ (\citealt{Alonso06, Marin09, sanchez26}), or the Br$\gamma$/Br$\delta$ (\citealt{Piqueras13}), and reveal patchy extinction distributions that typically peak in the nuclear regions.

However, these diagnostics probe a limited wavelength range set by the specific line pair used, which may bias the inferred extinction at shorter or longer wavelengths. In addition, they rely on only two emission lines to constrain the entire extinction curve, inevitably providing a limited accuracy of the extinction. 
Indeed, extinction curves can vary significantly in slope, becoming steeper at shorter wavelengths (e.g., \citealt{Cardelli89, Weingartner01}), and two lines may be insufficient to trace such variations, especially for obscured systems as (U)LIRGs. Furthermore, H recombination lines  trace only the ionized gas phase and are not suitable for correcting other components of the interstellar medium (ISM).
Near-infrared spectra, however, contain numerous  H$_2$ ro-vibrational emission lines, which might originate from    regions physically different  than those of ionized lines (e.g., \citealt{Piqueras12,Pereira24}). We propose new approaches capable of probing extinction across the full  near-IR range and within different gas phases in deeply embedded systems.

In the local Universe, the 
  unprecedented sensitivity{, spectral coverage,} and spectral resolution of the JWST Near-Infrared Spectrograph (NIRSpec; \citealt{Jakobsen22, Boker22}) enables the simultaneous detection of several H recombination and H$_2$ ro-vibrational emission lines over a wide wavelength range of  $\sim$0.97--5.14 $\mu$m. 
   These capabilities allow, for the first time in extragalactic sources, the derivation of  extinction curves for the ionized and hot molecular gas phases, a study previously limited to Galactic objects
    (e.g., \citealt{Rosenthal00}). 
This spectral range also includes prominent absorption features from interstellar ices, which provide valuable diagnostics of obscuration, as they form on dust grain surfaces  (see \citealt{Boogert15} for a review). Strong absorption bands from H$_2$O ($\sim$3.02 $\mu$m) and CO$_2$ ($\sim$4.26 $\mu$m) ices have been detected in (U)LIRGs using JWST/NIRSpec (e.g.; \citealt{Buiten24, Bernete26}). 
The water ice absorption feature at $\sim$3 $\mu$m can significantly affect the continuum and adjacent spectral features, including the 3.3 $\mu$m polycyclic aromatic hydrocarbon (PAH) emission  (e.g., \citealt{Rigopoulou24}), 
and serves as a diagnostic of nuclear obscuration when compared to PAH bands and continuum emission at longer wavelengths (e.g., \citealt{Bernete22, Bernete25}). However, its impact on adjacent ionized and hot molecular emission lines in galaxies remains unexplored.
Similarly, the CO$_2$-to-H$_2$O ice abundance ratio is still poorly constrained beyond the Milky Way. While extensively studied in Galactic environments (e.g., \citealt{Gibb04, Pontoppidan08, Smith25}),
 other than for the Large and Small Magellanic Clouds (e.g., \citealt{Oliveira09, demarchi25}), this abundance has been measured in only seven star forming galaxies using AKARI NIR spectroscopy (\citealt{Yamagishi15}).

In this work, we propose three independent methods aimed at deriving phase-specific extinction curves for both the ionized and hot molecular gas in four (U)LIRGs, using JWST/NIRSpec observations. These methods include corrections for water ice absorption affecting emission lines near the 3 $\mu$m feature, and enable constraints on H$_2$O and CO$_2$ ice column densities and their relative abundances.  These methods complement measurements of the extinction affecting the warm molecular gas derived from mid-IR H$_2$ pure rotational emission lines (e.g., \citealt{Donnan24,Rosenthal00, Pereira14}).

This paper is organized as follows: Section~\ref{sec:sample} describes the  selected targets and observations.
Section~\ref{sec:reduction} details the data reduction and the extraction of nuclear spectra. In Section~\ref{sec:analysis}, we present the three methods used to derive phase-specific extinction and ice properties, and report the results for both nuclear and extended emission, and we describe the results. In Section~\ref{sec:discussion}
we discuss the implications of our findings, which we summarize in Section~\ref{sec:conclusions}. Throughout this work we assume the following cosmology: H$_0$ = 70.0 km s$^{-1}$ Mpc$^{-1}$, $\Omega_{\rm M}$ = 0.3, and $\Lambda$ = 0.7.

\section{Observations and sample properties} \label{sec:sample}

\begin{figure*}
\centering
\includegraphics[width=0.9\textwidth]{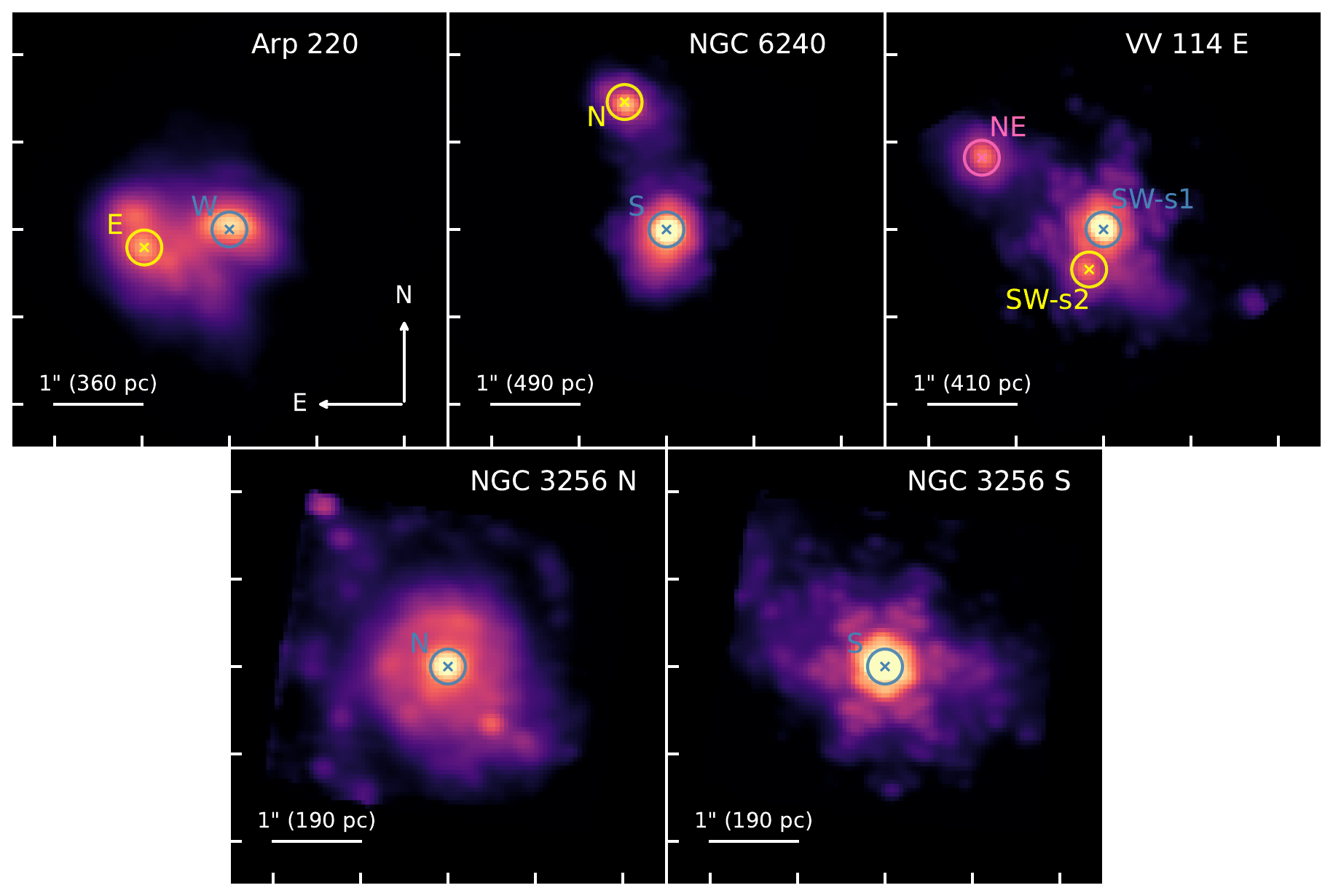}
\caption {NIRSpec continuum images at an observed wavelength of $\sim$4.8 $\mu$m (dominated by the hot dust emission) for all objects in Table~\ref{tab:sample}. North is up and east to the left. Colored circles (0.2" radius) indicate the nuclear extraction apertures, each centered on the cross. The images are 5" $\times$ 5".}
\label{fig:sample}
\end{figure*}

\begin{table}
\caption{Sample of local (U)LIRGs.}
\label{tab:sample}
\centering
\setlength{\tabcolsep}{2.5pt}
\renewcommand{\arraystretch}{0.9}
\begin{tabular}{l l c c l c}
\hline\hline
Object & { Nuc.} $^{(a)}$ & D$_L$$^{(b)}$ & log L$_{\rm IR}$/L$_{\rm \odot}$$^{(b)}$ & d$\rm_N$$^{(c)}$ & Project ID  \\ 
       &                & [Mpc]        &                            &  [pc]                                         &  \\
\hline

\multirow[t]{2}{*}{Arp~220}  
 & E  & \multirow[t]{2}{*}{78}   & \multirow[t]{2}{*}{12.19} & \multirow[t]{2}{*}{370}  &   \multirow[t]{2}{*}{$\#$1267} \\
 & W  &                         &                           &                                                 &    \\

 \multirow[t]{2}{*}{NGC~6240} 
 & N  & \multirow[t]{2}{*}{107}  & \multirow[t]{2}{*}{11.93} & \multirow[t]{2}{*}{740} & \multirow[t]{2}{*}{$\#$1265}\\
 & S  &                         &                           &                                           &        \\

 \multirow[t]{3}{*}{VV~114 E} 
 & NE & \multirow[t]{3}{*}{85.5} & \multirow[t]{3}{*}{11.71} & \multirow[t]{2}{*}{630}  & \multirow[t]{2}{*}{$\#$1328} \\
 & SW--s1 &                         &                           &                                                    & \\
 & SW--s2$^{(\ast)}$     &                         &                           &                                                   &  \\

\multirow[t]{2}{*}{NGC~3256} 
 & N  & \multirow[t]{2}{*}{38.9} & \multirow[t]{2}{*}{11.64} & \multirow[t]{2}{*}{800}  & \multirow[t]{2}{*}{$\#$1328} \\
 & S  &                         &                           &                                              &       \\

\hline
\end{tabular}

\vspace{1mm}
\begin{flushleft}
\footnotesize
\textbf{Notes.}
$^{(a)}$ { Nuclear} regions  studied in this work (other than the extended emission).
$^{(b)}$ Luminosity distance and 8--1000 $\mu$m IR luminosity from \citet{Armus09} and \citet{Pereira21}.
$^{(c)}$ Projected nuclear separation.
$^{(\ast)}$ SW-s2 is located $\sim$100 pc to the south-west of SW--s1. 
\end{flushleft}
\end{table}

The purpose of this work is to study { the near-infrared} extinction in  interacting obscured systems. To do so, we selected local (U)LIRGs observed with the three high-spectral-resolution NIRSpec gratings (G140H 0.97–1.82 $\mu$m, G235H 1.66–3.05 $\mu$m, and G395H 2.87–5.14 $\mu$m), to detect a significant number of H$_2$ and H I recombination emission lines ($\geq$15 and $>$10, respectively),  including emission lines near $\sim$3 $\mu$m that are sensitive to the presence of water ice. 
The final sample consists of four local objects, whose data were retrieved from the JWST archive from different observational programs, including $\#$1265 (PI: A. Alonso Herrero and S. Arribas), $\#$1267 (PI: D. Dicken and S. Arribas), and $\#$1328 (PI: L. Armus and A. Evans).

The main properties of these targets are listed in Table~\ref{tab:sample},  and their morphologies, from the continuum images at $\sim$4.8 $\mu$m, are shown in Fig.~\ref{fig:sample} (the images at $\sim$2 $\mu$m are shown in Fig.~\ref{fig:sample_4.8}). 
The IR luminosities span the range  10$^{11.64}$--10$^{12.19}$ L$_{\odot}$ and their luminosity  distances range between 38.9 and 107 Mpc. All the targets are mergers at different evolutionary stages, each hosting at least two nuclei.
The ULIRG Arp~220 is a late--stage merger that hosts an eastern and western nucleus separated by $\sim$370 pc in projection (e.g., \citealt{Scoville97}). The remaining targets are LIRGs. They include the two advanced mergers NGC~3256 and NGC~6240, both hosting a northern and a southern nucleus, with  projected separation of $\sim$800 pc and $\sim$740 pc, respectively (\citealt{Sakamoto14, Komossa03}), and the early-stage merger VV~114.
{ The latter} consists of two galaxies approaching along the east-west axis (known as VV~114 E and VV~114 W), with a nuclear separation of $\sim$ { 8} kpc (e.g., \citealt{Knop94, Linden23}). Here we focus on VV~114 E, { which dominates the near- and mid-IR emission of the merger}, whose brightest sources are the NE and SW-s1 components, separated by $\sim$630 pc (\citealt{Evans22}).

All these targets are undergoing  star formation and/or AGN episodes due to the large reservoir of gas resulting from the interaction. In the case of Arp~220, multi-phase outflows have been observed in both nuclei (e.g., \citealt{Lamperti22, Perna20}), but it is still debated whether SF or AGN activity dominates the IR luminosity (e.g., \citealt{Perna24}), due to the high extinction affecting the two nuclei (e.g., \citealt{gonzalez04, gonzalez12, Ulivi25}). In NGC~3256, the northern nucleus is powered by a bright starburst (\citealt{Lira08}), while the southern component  is powered by a low-luminosity
(L$_{2-10}$ keV $\sim$ 10$^{40}$ erg s$^{-1}$) obscured AGN (\citealt{Ohyama15}), from which multi-phase bipolar outflows are launched perpendicularly to the galaxy disk (\citealt{Sakamoto14, Emonts14, Pereira24}). 
Both nuclei in NGC~6240 are powered by AGN (\citealt{Komossa03, Armus06}), with multi-phase { starburst- and} AGN-driven outflows detected on different spatial scales (from few parsecs up to more than 10 kpc) and in different directions (mainly on the north-east and east-west; e.g., \citealt{Muller18, Cicone18, Ceci25, Hermosa25}). 
VV~114 E shows a complex morphology. In the near-infrared, SW-s1 dominates the emission, while the NE core dominates at longer wavelengths.  \citet{Donnan23} proposed that NE is a compact obscured nucleus,  
while \citealt{Buiten24} suggested that SW-s1 hosts an AGN, as previously proposed by \citet{Evans22}.

We extracted spectra for nine regions within these systems as reported in Table~\ref{tab:sample} and Fig.~\ref{fig:sample},  corresponding to the nuclei. The SW-s2 core in VV~114 is located $\sim$100 pc to the south-west of the SW--s1 nucleus. { It has been proposed as an intermediate mass black hole (BH) in an early stage of accretion and feedback} (e.g., \citealt{Alfonso24}).
All these nuclei have been classified as deeply obscured according  to the { PAH-EW method, except  Arp~220 E, which is slightly less obscured, and}  NGC~3256 N, which lies in a diagnostic region associated with star-formation, experiencing lower extinction levels (e.g., \citealt{Bernete22, Bernete25}).
Furthermore, \citet{Pereira24b} detected  strong H$_3^+$ absorption features at $>$3.5–3.9 $\mu$m  toward the  hot dust continuum of Arp~220 E and W, NGC~3256 S, and VV~114 NE, SW-s1, and SW-s2, suggesting the presence of a compact ($r<20$ pc) hot dust continuum source.

\section{JWST/NIRSpec data reduction}
\label{sec:reduction}

For the data reduction, we used the JWST calibration pipeline (version 1.12.4; \citealt{Bushouse23}) with context 1254. We followed the standard reduction steps, together by a number of extra steps to mitigate the effects of bad pixels and cosmic rays. Further details on the data reduction are provided in \citet{Bernete24} and \citet{Pereira24}. 
{ In each cube, the final spaxel size is 0.05" $\times$ 0.05". }
The spectra were extracted using circular apertures of 0.40" diameter (80–200 pc depending on the distance), and an aperture correction was applied following the method described in \citet{Bernete22}.

\section{Analysis and results}
\label{sec:analysis}

We used three complementary methods to assess the level of extinction on the line and continuum emissions. The first method uses H recombination lines to measure the extinction affecting the ionized gas (see Section~\ref{sec:rec}), the second relies on sets of ro-vibrational H$_2$ lines to map the extinction of the  hot molecular gas (see Section~\ref{sec:H2}), finally, the third method probes the obscuration { toward} the continuum using absorption features of H$_2$O (3.02 $\mu$m) and CO$_2$ ($\sim$4.26 $\mu$m) ices as diagnostics (see Section~\ref{sec:ices}).  
For simplicity, in this section we present only the results of Arp 220 E,  the same figures for the rest of targets are provided in Appendixes~\ref{appensixA}, ~\ref{appensixB}, and ~\ref{appensixC}.

\begin{sidewaystable*}
\tablesize
\centering
\caption{Extinction and ice columns densities derived using three different methods.}
\label{tab:nucext}
\begin{tabular}{l l 
                ccccc @{\hspace{0.8cm}}
                cccc @{\hspace{0.8cm}}
                ccc}
\hline\hline
Object & Reg. &
\multicolumn{5}{c}{Method 1: H  I rec. lines} &
\multicolumn{4}{c}{Method 2: H$_2$ lines} &
\multicolumn{3}{c}{Method 3: CO$_2$ and H$_2$O  ice absorptions} \\
\cmidrule(r{2.5em}){3-7}
\cmidrule(r{2.5em}){8-11}
\cmidrule(l{0.4em}){12-14}
& & A$_{\rm K}^{\rm HII}$ & A$_{\rm H_2O}^{\rm HII}$ & Q$_{0}$ &N$_{\rm H_2O}^{\rm HII}$ & N$_{\rm H_2O}^{\rm HII}$/N$_{\rm H}^{\rm HII}$&
   A$_{\rm K}^{\rm H_2}$ &  A$_{\rm H_2O}^{\rm H_2}$ & N$_{\rm H_2O}^{\rm H_2}$ & N$_{\rm H_2O}^{\rm H_2}$/N$_{\rm H}^{\rm H_2}$ &
   N$_{\rm CO_2}$ & N$_{\rm H_2O}$ & N$_{\rm CO_2}$/N$_{\rm H_2O}$\\ & & [mag] & [mag] & [10$^{53}$ s$^{-1}$] &[10$^{17}$ cm$^{-2}$] & [10$^{-5}] $& [mag] & [mag]
    &  [10$^{17}$ cm$^{-2}$]& [10$^{-5}$]
   & [10$^{17}$ cm$^{-2}$] & [10$^{17}$ cm$^{-2}$] &\\
\hline
Arp~220 & E    & 1.68$\pm$0.07 & 1.63$\pm$0.20 & 1.58$\pm$0.16 &22.3$\pm$2.7 & 7.9$\pm$1.0 & 1.17$\pm$0.24 & 1.63$\pm$0.45 &22.4$\pm$6.3 & 11.4$\pm$4.0 &2.18$\pm$0.04 & 30.5$\pm$2.7 & 0.07$\pm$0.01\\
  & W    & 1.49$\pm$0.06 & 0.82$\pm$0.17 & 4.42$\pm$0.37 & 11.3$\pm$2.3 & 4.52$\pm$0.93 & 1.13$\pm$0.24 & 1.45$\pm$0.37 &19.9$\pm$5.0 &  10.6$\pm$3.5& 0.93$\pm$0.01 & 14.2$\pm$1.7 & 0.06$\pm$0.01\\
  & Extended   & 1.10$\pm$0.11 & 0.82$\pm$0.27& 13.05$\pm$0.90 &11.5$\pm$3.7 & 5.3$\pm$1.7 & 0.53$\pm$0.15 & 0.70$\pm$0.16 &9.9$\pm$2.2 & 11.9$\pm$2.6 &1.14$\pm$0.08 & 12.7$\pm$2.1 & 0.09$\pm$0.03\\
NGC~6240 & N  & -- & -- & -- & -- & --&0.35$\pm$0.15 & 0.66$\pm$0.21 &9.1$\pm$2.9 & 15.7$\pm$8.5 & 1.18$\pm$0.02 & 9.1$\pm$1.0 & 0.13$\pm$0.02\\
 & S  & -- &  --& -- &--& -- &0.29$\pm$0.23 &  0.60$\pm$0.26&8.2$\pm$3.6 & 17$\pm$15  & 0.76$\pm$0.01& 7.80$\pm$0.50 & 0.10$\pm$0.01 \\
  & Extended & -- & -- & -- & -- & -- &0.39$\pm$0.09 & 0.47$\pm$0.12 &6.7$\pm$1.7 &  8.2$\pm$2.1  & 0.77$\pm$0.04 & 7.1$\pm$1.5 & 0.11$\pm$0.02 \\
VV~114 E & NE     & 1.03$\pm$0.05 & 0.63$\pm$0.17& 1.85$\pm$0.15 & 8.7$\pm$2.4 & 5.1$\pm$1.4 &  0.61$\pm$0.50 & 0.69$\pm$0.43 &9.5$\pm$6.0 & 9.3$\pm$9.7 & 0.38$\pm$0.01 &  7.30$\pm$0.81  &0.05$\pm$0.01\\
 & SW--s1     & 0.94$\pm$0.04 & 0.47$\pm$0.14 & 3.99$\pm$0.26 & 6.5$\pm$1.9 & 4.1$\pm$1.2 & -- &-- &-- & --& 1.13$\pm$0.01& 8.94$\pm$0.23  & 0.13$\pm$0.01\\
 & SW--s2     & 1.10$\pm$0.02 & 0.86$\pm$0.07 & 2.90$\pm$0.97 & 11.8$\pm$1.0 & 6.40$\pm$0.58 & 1.01$\pm$0.14 & 0.55$\pm$0.24 & 7.6$\pm$3.3 & 4.5$\pm$2.1 & 1.60$\pm$0.03 & 15.8$\pm$2.0  & 0.10$\pm$0.01 \\
 & Extended     & 0.60$\pm$0.05 & 0.49$\pm$0.13 & 25.3$\pm$4.2 & 6.9$\pm$1.8 & 5.7$\pm$1.6 &0.49$\pm$0.15 & 0.45$\pm$0.13 & 6.3$\pm$1.7 & 8.5$\pm$2.2 & 0.77$\pm$0.04 & 7.2$\pm$1.7 & 0.11$\pm$0.03 \\
NGC~3256 & N  & 0.64$\pm$0.03 &0.45$\pm$0.08 &   1.85$\pm$0.08& 6.2$\pm$1.1 & 5.8$\pm$1.1 & -- & -- & -- & --& 0.30$\pm$0.01 &  4.87$\pm$0.55 & 0.06$\pm$0.01\\
& Extended & 0.44$\pm$0.04 & 0.39$\pm$0.11 & 11.4$\pm$1.9 & 5.3$\pm$1.5 & 6.0$\pm$1.7 & -- & -- & -- & -- & 0.65$\pm$0.04 & 7.3$\pm$1.4 & 0.09$\pm$0.02 \\
 & S  & 1.86$\pm$0.09 & 1.59$\pm$0.30  & 1.36$\pm$0.20 & 21.8$\pm$4.1 & 7.0$\pm$1.4 &  0.73$\pm$0.21 & 1.30$\pm$0.44 & 17.8$\pm$6.0 &  14.7$\pm$6.5 & 9.7$\pm$1.0 & 65.3$\pm$2.5 & 0.15$\pm$0.02 \\
 & Extended &  0.65$\pm$0.06 & 0.62$\pm$0.16 & 6.66$\pm$0.41 & 8.8$\pm$2.1 &  5.3$\pm$1.4 &0.58$\pm$0.17  & 0.55$\pm$0.15 & 7.6$\pm$2.0 & 8.1$\pm$2.1 & 1.46$\pm$0.27 & 11.7$\pm$2.8 & 0.13$\pm$0.03 \\
\hline
All objects & nuc.  & 1.25$\pm$0.02 & 0.92$\pm$0.07& 2.56$\pm$0.16 & 12.6$\pm$0.90  & 5.84$\pm$0.42 & 0.75$\pm$0.10 & 0.98$\pm$0.13 & 13.5$\pm$1.8 & 11.9$\pm$3.1 & 2.02$\pm$0.11 & 18.20$\pm$0.53 & 0.11$\pm$0.01\\
 & Extended  & 0.61$\pm$0.04 &  0.53$\pm$0.09 & 14.1$\pm$1.8 & 7.6$\pm$2.0 & 5.6$\pm$1.0  & 0.50$\pm$0.07 & 0.55$\pm$0.08 & 7.8$\pm$2.0  & 9.0$\pm$2.0 & 1.01$\pm$0.03 & 10.0$\pm$2.0 & 0.10$\pm$0.03\\
\hline
\end{tabular}

\footnotesize
\textbf{Notes.} Method~1: based on H recombination lines.
Method~2: based on ro-vibrational H$_2$ emission lines.
Method~3: based on the absorption features in the continuum of the H$_2$O ($\sim$3.02 $\mu$m) and CO$_2$ ($\sim$4.26 $\mu$m). For all targets and methods, we report the results associated with the nuclei and the  global median values obtained from the analysis of the extended regions, { excluding the nuclear contribution} ({ "Extended"}).  { The Q$_{\rm 0}$ values for the “Extended” region correspond to the integrated values over all spaxels, excluding the nuclear spaxels.} The last two rows represent the median values { for the nuclear regions listed in the second column and the extended components of} all objects. The hydrogen column densities are derived as N$_{\rm H} = 1.66 \, \times \, 10^{22} $A$_{\rm K}$ (\citealt{Bohlin78}).
\end{sidewaystable*}

\subsection{Method 1: extinction from the H recombination lines} \label{sec:rec}

\begin{figure*}
\centering
\includegraphics[width=0.425\textwidth]{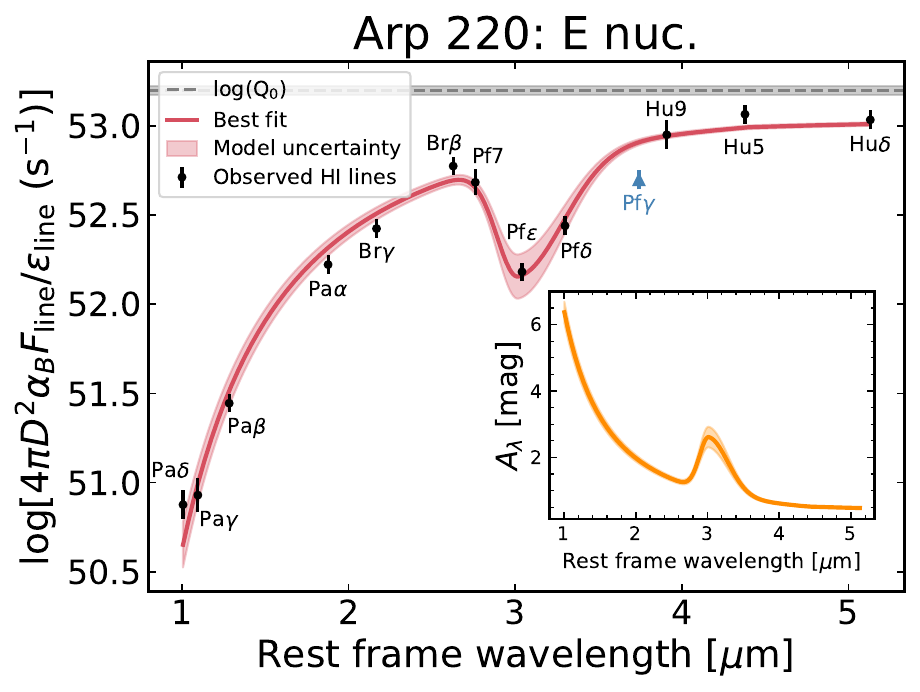}
\includegraphics[width=0.57\textwidth]{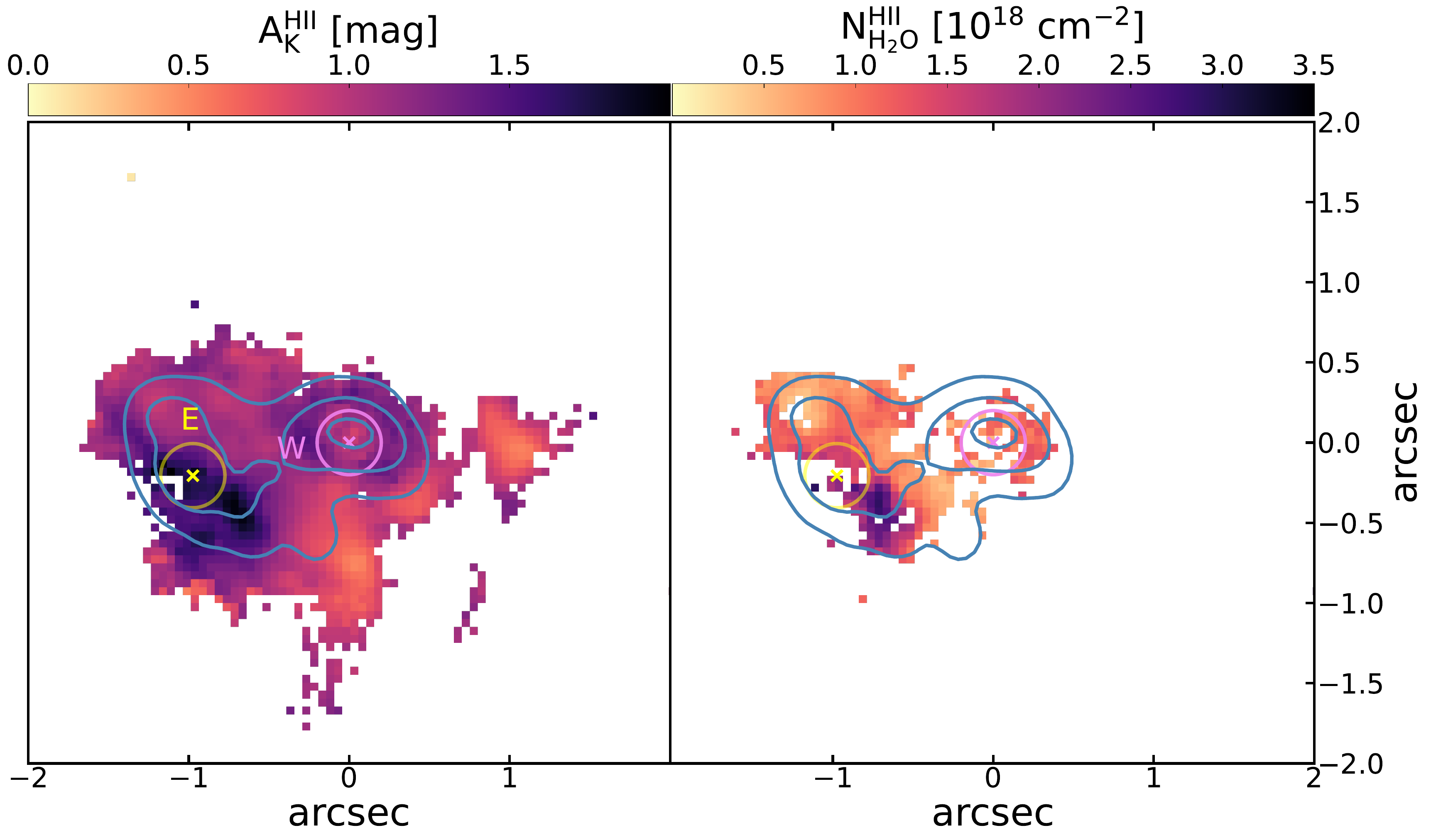}
\caption {Showcase of Method~1 applied to Arp~220 to derive extinction from H recombination lines. Left panel: the first term of Equation~\ref{eq:rec2} plotted as a function of wavelength for the nuclear spectrum of Arp~220 E (black points{,  with the corresponding H transition labeled)}. The best fit to the data and its uncertainty are shown by the red line and shaded area, with the dashed horizontal line and gray shaded area representing Q$_{\rm 0}$ and its uncertainty.  The blue triangle indicates the measurement from the H 8--5 transition, which is excluded from the model optimization { see Section~\ref{sec:rec}}. The inset panel shows the shape of the extinction curve derived from the nuclear A$_{\rm K}^{\rm HII}$ and A$_{\rm H_2O}^{\rm HII}$ parameters { (with fixed $\sigma_{\rm R} = 0.27 \,\mu$m, $\sigma_{\rm B} = 0.13 \,\mu$m,  and $\beta = 1.7$)}. Central panel: spatial distribution of A$_{\rm K}^{\rm HII}$ derived spaxel-by-spaxel from the best-fitting model applied to the rebinned cube of Arp~220. The  yellow and  violet crosses indicate the eastern and western nuclei, respectively,  with circles marking the extraction aperture (0.20$"$ in radius) of the nuclear spectra. Right panel: same as the central, but for N$_{\rm H_2O}^{\rm HII}$ derived from the A$_{\rm H_2O}^{\rm HII}$ parameter. { In both the central and right panels, north is up and east to the left and the blue lines represent the contours of the 5$\%$, 15$\%$ and 68$\%$ of the 4.8 $\mu$m continuum image (Fig.~\ref{fig:sample}).}}
\label{fig:rec}
\end{figure*}

To derive the extinction from the hydrogen recombination lines, we compare the observed line intensities with the theoretical values. Under the Case B recombination assumption  (\citealt{Storey95}), where a dust screen lies  in front of an optically-thin line-emitting  region, the observed flux is given by:
\begin{equation} \label{formula:rec}
    \rm F_{\rm obs} = \epsilon_{\rm line} \frac{\rm Q_0}{\rm 4\pi D_L^2 \alpha_B}\, 10^{-0.4\, A_\lambda}
\end{equation}
where $\rm \epsilon_{\rm line}$ is the line emissivity  defined such that the power radiated per volume in a line is n$\rm _e$n$\rm _H$$\rm \epsilon_{\rm line}$, where n$\rm _e$ and n$\rm _H$ are the electron and H densities, respectively, Q$_{\rm 0}$ denotes the number of hydrogen-ionizing photons (h$\nu$ > 13.6 eV) emitted per second, D$\rm _L$ is the luminosity distance, $\alpha_{\rm B}$ is the total Case B radiative recombination coefficient, and A$_{\lambda}$
represents the extinction at the transition wavelength.
Equation~\ref{formula:rec} can be written in the following form:
\begin{equation} \label{eq:rec2}
\rm log\left(\frac{\rm 4\pi D^2 \alpha_B F_{\rm obs}}{\epsilon_{\rm line}}\right) = \rm log\,(\rm Q_0) -0.4\, A_\lambda
\end{equation}
The first term of Equation~\ref{eq:rec2} is  calculated for each observed hydrogen recombination line.  The $\epsilon_{\rm line}$ values are computed using the PyNeb tool (\citealt{Luridiana15}), and we assume $\alpha_{\rm B} = 2.588\,\times \, 10^{-13}$ cm$^{3}$ s$^{-1}$ for T = 10000 K and n$_ {\rm e}$ = 1000 cm$^{-3}$ (\citealt{Hummer87}), whereas the second term is fitted to the measurements using: 
\begin{small}
\begin{equation} \label{eq:func1}
\rm {A_{\lambda} = A_{\rm K}^{\rm HII} \left(\frac{\lambda_K}{\lambda}\right)^{\beta} + A_{H_2O}^{\rm HII} \times \, 
\begin{cases} 
exp \Biggl( - \frac{(\lambda - \lambda_{ice})^2}{\sigma _{B}}\Biggr)  & \text{for} \,\lambda < \lambda_{ice} \\
exp \Biggl( - \frac{(\lambda - \lambda_{ice})^2}{\sigma _{R}}\Biggr) & \text{for} \,\lambda > \lambda_{ice}
\end{cases}}
\end{equation}
\end{small}
 A$_{\rm K}^{\rm HII}$ denotes the extinction in the K band, with $\lambda_{\rm K}$ = 2.2 $\mu m$, 
while A$_{\rm H_2O}^{\rm HII}$ corresponds to the peak of the asymmetric Gaussian used to model the water ice absorption at $\lambda_{\rm ice} = 3.02 \,\mu$m. The Gaussian has a narrower width on the blue side of the peak ($\sigma_{\rm B} = 0.13\, \mu$m for $\lambda < \lambda_{ice}$) and a broader width on the red side ($\sigma_{\rm R} = 0.27\, \mu$m, for $\lambda > \lambda_{ice}$).  
The sigma of the { red} broader side was set to the average value across the sample to reproduce the measurement from the observed  { Pf$\epsilon$} ($\sim$3.04 $\mu$m) and the { Pf$\delta$} ($\sim$3.30 $\mu$m) transitions  in the ice absorption.
 For all targets, we excluded the  Pf$\gamma$ transition ($\sim$3.74 $\mu$m;  blue triangle in the left panel of Fig.~\ref{fig:rec}) from the analysis because it falls 
within an excess absorption in the redshifted wing of the  H$_2$O  ice absorption feature.
 The origin of this wing will be studied in detail in a forthcoming paper (see also Section~\ref{sec:ices} for details).
We fixed $\beta$ = 1.7, consistent with typical Galactic (e.g.~\citealt{Whittet88, Bertoldi99, Rosenthal00}) and extragalactic values (e.g., \citealt{Donnan24}). We also verified that allowing $\beta$ to vary as a free parameter yields extinction values consistent   with  those obtained assuming $\beta$ = 1.7, for both A$_{\rm K}^{\rm HII}$ and A$_{\rm H_2O}^{\rm HII}$.  In particular, we measured an average value of $\beta = (1.6 \pm 0.3)$ for the nuclear spectra, consistent within the uncertainties with our adopted value.

{ To measure the line fluxes (F$_{\rm obs}$) in} Equation~\ref{eq:rec2}, we performed a linear fit to the continuum adjacent to each H line and subtracted it to obtain baseline-free emission lines. A third-order polynomial fit was used to fit the continuum adjacent to the  Pf$\delta$ transition ($\lambda$ = 3.2970 $\mu$m), which lies on the Polycyclic Aromatic Hydrocarbons (PAH) feature at 3.3 $\mu$m. For this transition, a linear regression does not adequately reproduce the underlying continuum, which is dominated by the PAH emission. After removing the continuum contribution, a single Gaussian was fitted to all H emission lines. { A single component is sufficient to reproduce the line profiles, in  cases where a broad component could be included, its flux is negligible compared to that of the narrow Gaussian.}

The { left-hand member} of Equation~\ref{eq:rec2} was computed for lines with a signal-to-noise ratio greater than 5 (e.g., the black points in Fig.~\ref{fig:rec}). Finally, we determined the best-fitting model by minimizing the second term of Equation~\ref{eq:rec2} with respect to the measurements (see the red curve in Fig.~\ref{fig:rec}). The fit yields  Q$_{\rm 0}$, A$_{\rm K}^{\rm HII}$, and A$_{\rm H_2O}^{\rm HII}$ as output parameters. { For robustness, we consider these parameters reliable only when they are derived from at least ten H emission lines}. A$_{\rm H_2O}^{\rm HII}$ is converted to { H$_2$O ice} column densities (N$_{\rm H_2O}^{\rm HII}$) following Equation~4 of \citet{Rocha22}, by integrating { over the wave number} the optical depth (given by $\tau$ = A$_{\rm H_2O}^{\rm HII}$/1.0857) and dividing by the band strength. We adopt the band strength of  the OH stretching mode of pure H$_2$O  ice at 3.01 $\mu$m provided by \citet{Bouilloud15}, namely 2.2 $\times$ 10$^{-16}$ cm mol$^{-1}$, { measured at 10 K}.

Such analysis is performed  on the { nuclear} spectra, { extracted with a 0.2" radius aperture centered on the nuclei, }  (see Table~\ref{tab:sample}), as well as on their extended regions. To study the extended emission, we used rebinned data cubes obtained by summing 2×2 adjacent spaxels  for each spaxel position. This rebinning increases the signal-to-noise ratio without significantly degrading spatial resolution { and reduces the effect of wiggles in nuclear spaxels (e.g., \citealt{Perna23})}.  

The left panel of Fig.~\ref{fig:rec} shows the best-fitting model for the nuclear spectrum of Arp 220 E, while the central and right panels show the spaxel-by-spaxel distribution of A$_{\rm K}^{\rm HII}$ and N$_{\rm H_2O}^{\rm HII}$ derived from the fitting procedure applied to the rebinned cube of Arp 220.
From both maps, we find that higher extinction and column values affect the eastern nucleus (yellow cross), while the western nucleus is characterized by lower extinction and column but is surrounded by a ring of higher obscuration visible in the A$_{\rm K}^{\rm HII}$ map. { The  N$_{\rm H_2O}^{\rm HII}$ map, whose morphology depends on the detection of the Pf$_{\epsilon}$ and Pf$\delta$ emission lines, does not show values in the nuclear regions of Arp~220 E and W because these lines are not significantly detected owing to obscuration. This is also the case for VV~114 SW-s1 and NGC~3256 N.}
{ In the case of NGC~6240, we do not measure  the extinction following the method described above because the minimum requirement of ten detected H recombination lines is not met in either the nuclear or extended regions.} 

The nuclear values of A$_{\rm K}^{\rm HII}$ and N$_{\rm H_2O}^{\rm HII}$ for the { rest of the} sample are reported in  Table~\ref{tab:nucext}, along with the median value measured from all spaxels. 
H recombination lines are more extincted in the nuclear regions { (e.g., Arp~220 E, VV~114 NE and SW-s2, and NGC~3256 S) or in the region immediately surrounding the nucleus (e.g., Arp~220 W, VV~114 SW-s1, and NGC~3256 N) } than in the more extended region, the water ice column density also increases toward the nuclei { (see maps in Appendix~\ref{appensixA})}. The median value of A$_{\rm K}^{\rm HII}$ is 1.25 $\pm$ 0.02 in the nuclei,  while it decreases to 0.62 $\pm$ 0.05 in the extended regions. 
Similarly, the median water ice column density is higher in the nuclei, with N$_{\rm H_2O}^{\rm HII}$ = (12.6 $\pm \, 0.90) \, \times$ 10$^{17}$ cm$^{-2}$, than in the extended emission, where N$_{\rm H_2O}^{\rm HII}$ = (7.57 $\pm \, 2.00) \, \times$ 10$^{17}$ cm$^{-2}$. However, the  relative abundance of water ice with respect to hydrogen { (derived as N$_{\rm H}^{\rm HII}$ = 1.87 $\times$ 10$^{21}$/8.7 A$_{\rm K}^{\rm HII}$, \citealt{Bohlin78, Cardelli89})} remains  approximately constant between the nuclei and the extended regions,  with N$_{\rm H_2O}^{\rm HII}$/N$_{\rm H}^{\rm HII}$ = (5.84 $\pm$ 0.42) $\times\, 10^{-5}$ and N$_{\rm H_2O}^{\rm HII}$/N$_{\rm H}^{\rm HII}$ = (5.65 $\pm$ 1.51) $\times\, 10^{-5}$, respectively.
All nuclear models and maps for the whole sample are shown in Appendix~\ref{appensixA}.
\subsubsection{Comparison of star formation rates}

\begin{table}
\caption{Comparison of Star formation rates.}
\label{tab:sfr}
\centering
\setlength{\tabcolsep}{2.5pt}
\renewcommand{\arraystretch}{0.9}
\begin{tabular}{l c c c}
\hline\hline
Object &  SFR$_{\rm Q_0}$$^{(a)}$ & SFR$_{\rm IR}$$^{(b)}$ &  SFR$_{\rm IR}$/SFR$_{\rm Q_0}$$^{(c)}$   \\ 
       &         [M$_{\odot}$ yr$^{-1}$]      &     [M$_{\odot}$ yr$^{-1}$]    &                                                                        \\
\hline
\multirow[t]{2}{*}{Arp~220}  
 &  \multirow[t]{2}{*}{9.5}   & \multirow[t]{2}{*}{230.8} & \multirow[t]{2}{*}{24.2}  \\
 &                         &                           &                                                     \\

 \multirow[t]{2}{*}{NGC~6240} 
 &  \multirow[t]{2}{*}{--}  & \multirow[t]{2}{*}{126.8} & \multirow[t]{2}{*}{--} \\
 &                         &                           &                                                  \\

 \multirow[t]{3}{*}{VV~114 E} 
 &  \multirow[t]{3}{*}{18.4} & \multirow[t]{3}{*}{76.4} & \multirow[t]{2}{*}{4.1}   \\
 &                          &                           &                                                     \\

\multirow[t]{2}{*}{NGC~3256} 
 &  \multirow[t]{2}{*}{6.5} & \multirow[t]{2}{*}{65.0} & \multirow[t]{2}{*}{9.9}   \\

\hline
\end{tabular}

\vspace{1mm}
\begin{flushleft}
\footnotesize
\textbf{Notes.}
$^{(a)}$ SFR derived by Q$_{0}$ using Eq.~1 from \citet{Murphy11}. For all targets the values are derived from the Q$_{0}$ measured in the extended regions, for NGC~3256 we used the median value obtain from the extended region of the N and S cubes.
$^{(b)}$ SFR derived by IR luminosities using Eq.~4 from \citet{Murphy11}.
$^{(c)}$ SFR ratios.
\end{flushleft}
\end{table}

The method presented above  provides Q$_0$, the hydrogen-ionizing photon rate (photons s$^{-1}$), which can be used to estimate the star formation rate as SFR$_{\rm Q_0}$ (M$_{\odot}$yr$^{-1}$) = 7.29 $\times$ 10$^{-54}$ Q$_0$ (s$^{-1}$) (Eq. (1) from \citealt{Murphy11}). Using the values listed in the fifth column of Table~\ref{tab:nucext} for the extended regions (i.e., integrating Q$_0$ across the FoV but excluding the nuclear spaxels), we derive median star formation rates of $\sim$11.5 M$_{\odot}$ yr$^{-1}$ (for NGC~3256 we used the median value between the N and S cubes).  The individual values for the extended regions are listed in Table~\ref{tab:sfr}, where we also report the star formation rates derived from the IR luminosity using the relation SFR$_{\rm IR}$ (M$_{\odot}$yr$^{-1}$) = 1.49 $\times$ 10$^{-10}$ L$_{\rm IR}$ (L$_{\odot}$) (Eq. (4) from \citealt{Murphy11}). For comparison, we also list the ratio SFR$_{\rm IR}$/SFR$_{\rm Q_0}$. As expected, the highest ratio is measured for Arp~220, which is characterized by the highest extinction in its extended region.

\subsection{Method 2: extinction from the H$_2$ emission lines}  \label{sec:H2}

\begin{figure*} 
\centering
\includegraphics[width=0.405\textwidth]{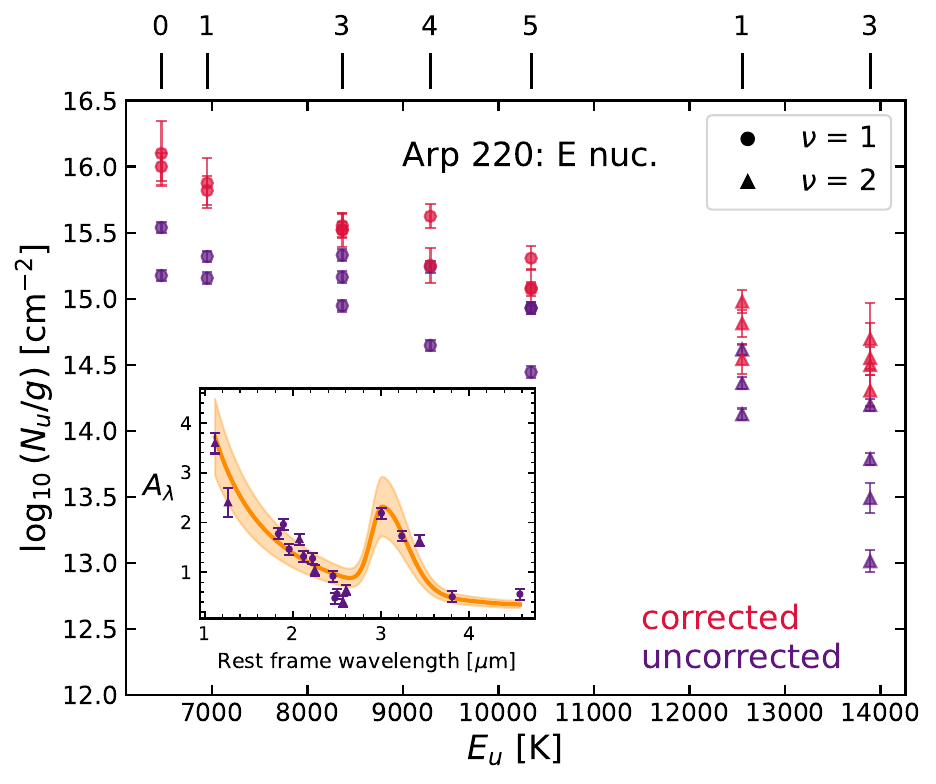}
\includegraphics[width=0.59\textwidth]{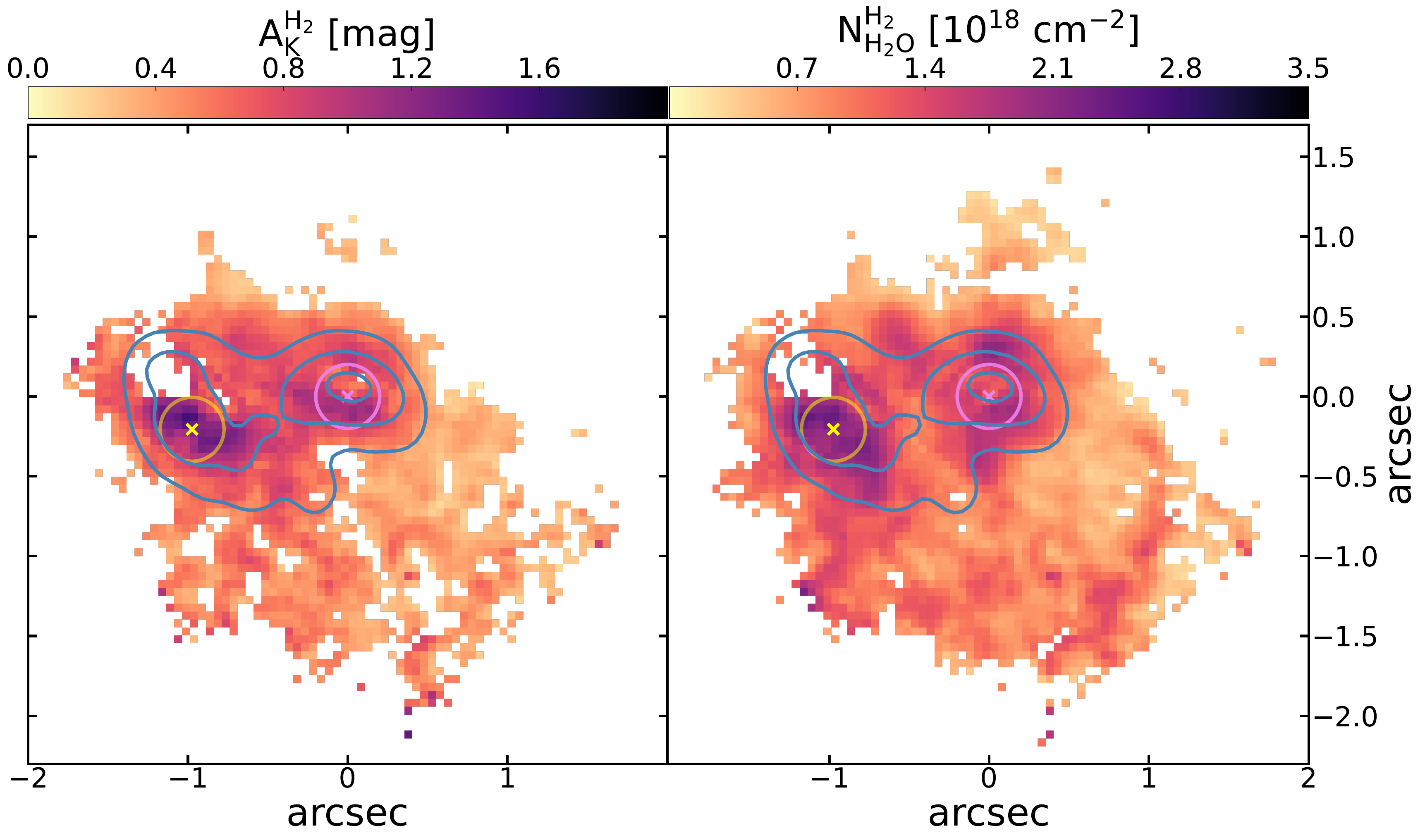}
\caption {Showcase of Method~2 applied to Arp~220 to derive extinction from H$_2$ emission lines. Left panel:  excitation diagram of the eastern nucleus of Arp 220 with N$_u$/g as a function of the upper energy level (E$_{\rm u}$). Measurements for uncorrected H$_2$ lines are shown in purple, while the corresponding extinction-corrected values are shown in red. The rotational J values are indicated at the top, and circles or triangles correspond to $\nu$ = 1 and 2, respectively.
 The inner panel shows how the sets of H$_2$ lines constrain the shape of the extinction curve.  
 Central panel: spatial distribution of A$_{\rm K}^{\rm H_2}$ derived spaxel-by-spaxel from the best-fitting model applied to the rebinned cube of Arp~220. The yellow and violet crosses indicate the eastern and western nuclei, respectively, with circles marking
the extraction aperture (0.20” in radius) of the nuclear spectra. Right panel: same as the central panel, but for N$_{\rm H_2O}^{\rm H_2}$ derived from the A$_{\rm H_2O}^{\rm H_2}$ parameter. { In both the central and right panels, north is up and east to the left and the blue lines represent the contours of the 5$\%$, 15$\%$ and 68$\%$ of the 4.8 $\mu$m continuum image (Fig.~\ref{fig:sample}).} }
\label{fig:h2}
\end{figure*}

To derive the extinction affecting the hot molecular gas from the H$_2$  ro-vibrational emission lines { ($\nu$ = 1--0, 2--1)}, we built excitation diagrams and minimized the dispersion { of the level column densities (N$_{\rm \nu, J}$)} among sets of H$_2$ lines  from the S, Q, and O branches of the vibrational bands sharing the same upper energy level (i.e.  S(J-2), Q(J), and O(J+2), where J is the rotational quantum number { of the lower energy level}). To this end, we computed observed column densities of each line as
\begin{equation}
    \rm {N_{\nu,\,J}^{(i)}} = 4\pi \lambda_{(i)}\, \frac{F_{obs}^{(i)}}{A_{ul(i)}\, h \, c \,\Omega} = N_{\nu, J}^{int} \times 10^{-0.4 A_{\lambda_{(i)}}(A_K^{H_2}, A_{H_2O}^{H_2})}
\end{equation}
where F$_{\rm obs}^{\rm {(i)}}$ is the integrated observed flux of line $i$, A$_{\rm ul(i)}$ is the
Einstein A-coefficient of the transition, { $\lambda$$_{\rm (i)}$ is its rest-frame wavelength},  h is the Planck’s constant, { c is the light-speed,} and $\rm \Omega$ is the solid angle  of the aperture used to measure the observed flux. To measure F$_{\rm obs}$, we applied the same methodology described in Section~\ref{sec:rec} to subtract the continuum and fit a single Gaussian to all H$_2$ lines detected with signal-to-noise $\geq$ 3. { A$_{\lambda_{\rm (i)}}(\rm A_{K}^{H_2},\, A_{H_2O}^{H_2} )$ represent the extinction curve (equivalent to Eq.~\ref{eq:func1}) and} N$_{\rm \nu, J}^{\rm int}$ is  the intrinsic extinction corrected column density common to the set of H$_2$ lines sharing the same { upper} $\nu$ and J,  where $\nu$ is the vibrational quantum number.
Therefore, for  each transition (i),
\begin{equation}\label{eq:fith2}
    \rm log\,{N_{\nu,\,J}^{(i)}} = - \, 0.4 \,  A_{\lambda_{(i)}}(A_K^{H_2}, A_{H_2O}^{H_2}) + log \,N_{\rm \nu, J}^{int}
\end{equation}
  We fitted the observed column densities  to Equation~\ref{eq:fith2}. We use { the superscripts to} emphasize that the derived parameters, A$_{\rm K}^{\rm H_2}$ and A$_{\rm H_2O}^{\rm H_2}$, are obtained from H$_2$ emission lines. We convert A$_{\rm H_2O}^{\rm H_2}$ to the { H$_2$O ice} column density N$_{\rm H_2O}^{\rm H_2}$ following the same methodology described in Section~\ref{sec:rec}. { This method is applied when a minimum number of 15 H$_2$ ro-vibrational lines are detected with S/N $\geq$ 3.}

The left panel of Fig.~\ref{fig:h2} shows the excitation diagram derived from the nuclear spectrum of Arp 220 E, including both uncorrected and corrected measurements (purple and red points, respectively). The dispersion along the N$_{\rm u}$/g axis,  where g = (2J + 1) for para-H$_2$ transitions (even J) and g = 3$\times$ (2J + 1) for ortho-H$_2$ transitions (odd J), decreases significantly after the extinction correction. As in Method 1, this analysis is also applied to the extended regions of the rebinned cubes. 
The central and right panels of Fig.~\ref{fig:h2} show the spatial distribution of A$_{\rm K}^{\rm H_2}$ and N$_{\rm H_2O}^{\rm H_2}$ obtained by applying this method spaxel-by-spaxel to the rebinned cube of Arp~220. Consistent with the results from the ionized gas (see Section~\ref{sec:rec}), the obscuration increases in the proximity of the nuclei,
with the W nucleus surrounded by a ring-like structure of higher values, similar to that observed from the H recombination lines. This structure is present in both the A$_{\rm K}^{\rm H_2}$ and N$_{\rm H_2O}^{\rm H_2}$ maps. 
 The global median values in the nuclear regions { (i.e., 0.2" in radius)} of A$_{\rm K}^{\rm H_2}$ = (0.75 $\pm$ 0.10) are higher than that in the extended regions, where A$_{\rm K}^{\rm H_2}$ = (0.51 $\pm$ 0.14). A similar behavior is found for the water ice column densities, which globally decrease from N$_{\rm H_2O}^{\rm H_2}$ = (13.5 $\pm \, 1.8) \, \times$ 10$^{17}$ cm$^{-2}$ in the nuclei to N$_{\rm H_2O}^{\rm H_2}$ = (7.8 $\pm \, 2.0) \, \times$ 10$^{17}$ cm$^{-2}$ in the extended regions. { Similar to what found when applying Method~1 (see Section~\ref{sec:rec})}, the relative abundance of water ice with respect to hydrogen remains  approximately constant between the nuclei and the extended regions,  with N$_{\rm H_2O}^{\rm H_2}$/N$_{\rm H}^{\rm H_2}$ = (11.9 $\pm$ 3.1) $\times\, 10^{-5}$ and N$_{\rm H_2O}^{\rm H_2}$/N$_{\rm H}^{\rm H_2}$ = (9.0 $\pm$ 2.0) $\times\, 10^{-5}$, respectively. 
The maps and nuclear models for all  targets are shown in Appendix~\ref{appensixB}.

The median extinctions measured from the { ro-vibrational} H$_2$ lines are lower than those derived from the H recombination lines, particularly in the nuclear regions (of a factor >1.6). However, the median values of N$_{\rm H_2O}^{\rm H_2}$ are comparable to those derived from H recombination lines. This results in relative N$_{\rm H_2O}^{\rm H_2}$/N$_{\rm H}^{\rm H_2}$ that are up to a factor of two higher than N$_{\rm H_2O}^{\rm HII}$/N$_{\rm H}^{\rm HII}$. 
We do not report A$_{\rm  K}^{\rm H_2}$ and A$_{\rm  H_2O}^{\rm H_2}$ values for NGC~3256 N because the hot molecular emission is too faint in this target, preventing the detection of a sufficient number of H$_2$ lines to constrain the extinction using the method presented in this Section.

In summary, the extinction values derived from the ro-vibrational H$_2$ lines are lower than those derived from  the ionized gas, while the corresponding ice column densities are comparable { (see Section~\ref{sec:rec_vs_h2} for more details)}. Both tracers show the same overall trend,
with higher extinction and column densities in the nuclei than in the extended regions.

\subsection{Method 3: extinction from the H$_2$O and CO$_2$  ice absorptions} \label{sec:ices}

\begin{figure*} 
\centering
\includegraphics[width=0.415\textwidth]{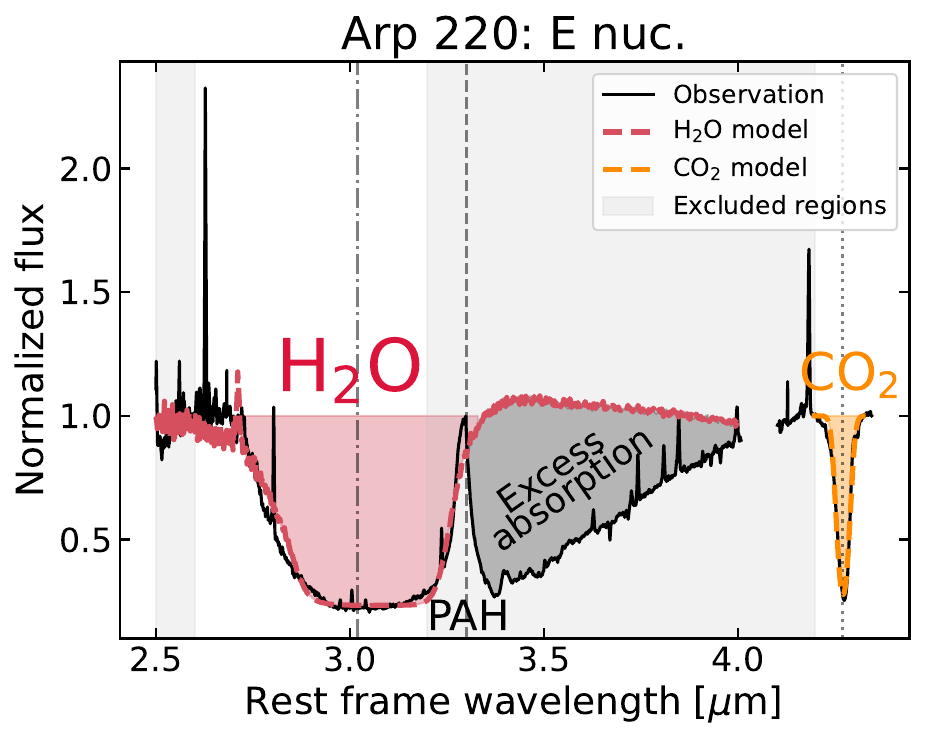}
\includegraphics[width=0.58\textwidth]{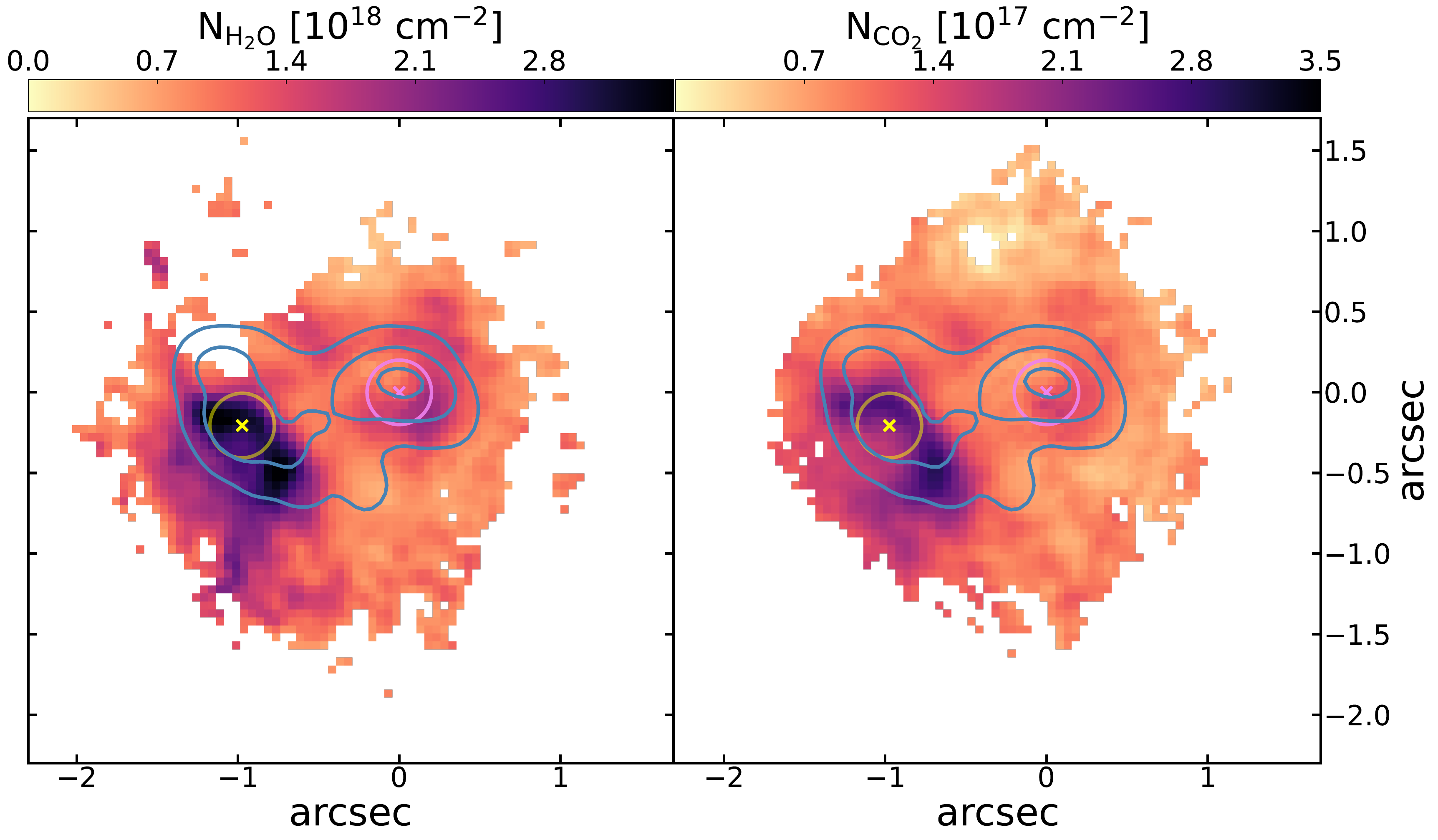}
\caption {Showcase of Method~3 applied to Arp~220 to derive columns densities from H$_2$O  ($\sim 3.02\,\mu$m) and CO$_2$  ($\sim 4.26\,\mu$m) ice features. Left panels:  H$_2$O and CO$_2$ absorptions  from the eastern nucleus of Arp~220. The observed spectrum is shown in black, and the best fit model is shown with a red dashed line  for H$_2$O and orange for CO$_2$. For H$_2$O the wavelength range excluded from the fitting procedure are marked by shaded gray areas (i.e., wavelength < 2.6 $\mu$m and > 3.2 $\mu$m). The position of the H$_2$O { peak} absorption, the PAH emission band at 3.3 $\mu$m, and the CO$_2$ absorption at $\sim$4.26 $\mu$m are indicated with dash-dotted, dashed, and dotted vertical lines, respectively. The excess absorption at wavelengths redward of the H$_2$O feature is highlighted in  dark gray. Middle and right panels: spatial distribution of the H$_2$O (N$_{\rm H_2O}$; middle) and CO$_2$ (N$_{\rm CO_2}$; right) column densities  measured in the rebinned cube of Arp~220. { In both the central and right panels, north is up and east to the left and the blue lines represent the contours of the 5$\%$, 15$\%$ and 68$\%$ of the 4.8 $\mu$m continuum image (Fig.~\ref{fig:sample}).}}

\label{fig:ice}
\end{figure*}

The absorption features of H$_2$O at 3.02 $\mu$m  (OH stretch mode) and CO$_2$ at 4.26 $\mu$m  (CO stretch mode) ices are diagnostics of the extinction { toward} the continuum, given that
\begin{equation}\label{eq:ice}
    \rm \frac{F_{obs}}{F_{int}} = 10^{-0.4\,A_{\lambda}}
\end{equation}
Therefore, we normalized the
observed absorption features (F$\rm _{obs}$) by the local continuum level (F$\rm _{int}$), estimated through
a spline-interpolated baseline defined by anchor points on both sides of the absorption features, obtaining the absorption strengths.
The normalized CO$_2$ absorption is fitted using Equation~\ref{eq:ice}, modeling A$_\lambda$ with a Gaussian profile whose amplitude, centroid, and width are free parameters (see left panel of Fig.~\ref{fig:ice}). We attempted to fit the observed CO$_2$ absorption using the pure--CO$_2$ absorbance profile provided by the Leiden ice database for astrochemistry (LIDA; \citealt{Oberg07}), however, the laboratory profiles are redshifted relative to the observed feature  and generally exhibit broader wings. 
We therefore adopted a Gaussian model, which provides an adequate approximation to the observed CO$_2$ ice absorption profile. In cases where a fraction of the CO$_2$ is present in the gas phase, as indicated by blueshifted wings, the Gaussian profile does not reproduce the asymmetric gaseous component and represents only the icy counterpart. We included the covering factor as a free parameter to the nuclear fit of NGC~3256 S, which exhibits a saturated profile. For the remaining targets, we verified that including the covering factor does not affect the best-fit parameters, as their absorption profiles are not saturated.

The normalized H$_2$O absorption is fitted using the absorbance spectrum of a pure-H$_2$O laboratory profile from LIDA (similar to \citealt{Bernete24, Bernete26}), shown in the left panel of Fig.~\ref{fig:ice} (red dashed profile). We considered the five available profiles of pure-H$_2$O at 15, 45, 75, 105 and 135 K  with an ice thickness of 3000L\footnote{1 L (Langmuir) = 1 $\times$ 10$^{-6}$ Torr s $\approx$ 1 monolayer of molecules (\citealt{Oberg07}).} and, for the nuclear spectra, we selected the  profile yielding the lowest $\chi^2$. In those nuclei showing saturated H$_2$O absorptions (Arp~220 E, Arp~220 W, and NGC~3256 S), we included the covering factor as an additional free parameter in the fitting procedure, so that the sum of the absorbed and unabsorbed components yields a normalized flux of unity. The saturated profiles are identified by their flatness rather than by a peaked shape around 3.02 $\mu$m. We obtained covering factors of 0.96, 0.77, and 0.50 for NGC~3256 S, Arp~220 E and Arp~220 W, respectively. For the rebinned cubes, we did not include the covering factor and adopted the 15~K profile, which provides the best fit for all targets.
This choice is justified because, outside of a few nuclear spaxels, the H$_2$O absorption profiles are not saturated. For both nuclear spectra and rebinned cubes, we restricted the fitting range of H$_2$O to 2.6 -- 3.2 $\mu$m to avoid contamination from the PAH emission band at 3.4 $\mu$m and from the red wing of absorption at ~3.4 -- 4 $\mu$m (see the dark gray shaded area in the left panels of Fig.~\ref{fig:ice}). Indeed, the pure-H$_2$O laboratory profiles fail to reproduce the absorption wing observed at redder wavelengths, which is detected in all targets except NGC~3256 N. This feature is not reproduced by the mixed-ice laboratory profiles either.
We attribute the origin of this wing to a combination of effects, 
mainly scattering effects related to the grain size distribution (\citealt{Dartois22, Dartois24}), and the presence of hydrogenated amorphous carbon (e.g., \citealt{Geballe06,Dartois07}){, and mixed ice compositions (\citealt{Boogert15})}.  A detailed study of this feature will be presented in a forthcoming paper. However, the bulk of the absorption over the 2.6 -- 3.2 $\mu$m range is attributed to water ice. Therefore, a pure-H$_2$O profile is suitable for reproducing the observed absorption near 3.02 $\mu$m, where contamination from other species and scattering is negligible.

As described in Section~\ref{sec:rec}, we converted A$_\lambda$ into $\tau_\lambda$, and, following Equation~4 of \citet{Rocha22}, derived the CO$_2$ and H$_2$O column densities (N$_{\rm CO_2}$ and N$_{\rm H_2O}$). The nuclear { and extended} values  are reported in  Table~\ref{tab:nucext}, while their spatial distribution is shown in the middle and right panel of Fig.~\ref{fig:ice}, for the rebinned cube of Arp~220  (see Appendix~\ref{appensixC} for the rest of targets). To compute N$_{\rm CO_2}$, we adopted a band strength of A = 1.1 $\times$ 10$^{-16}$ cm~mol$^{-1}$ (\citealt{Gerakines95}).

 For all sources, we obtain higher N$_{\rm H_2O}$ than N$_{\rm CO_2}$, but with similar morphologies in the maps. As found in Section~\ref{sec:rec} and ~\ref{sec:H2}, the nuclear regions are characterized by higher H$_2$O and CO$_2$ column densities { than more extended regions}. Both column densities are about a factor of $\sim$2 higher in the nuclei, while their  abundance  ratio remains consistent between the two environments, with median N$_{\rm CO_2}$/N$_{\rm H_2O}$ ratios of 0.10 in the nuclei and 0.11 in the extended regions. The most prominent ice features are detected in Arp~220 E and NGC~3256 S, with nuclear N$_{\rm H_2O}$ = (30.5 -- 65.3) $\times$ 10$^{17}$ cm$^{-2}$ and N$_{\rm CO_2}$ = (2.18 -- 9.74) $\times$ 10$^{17}$ cm$^{-2}$. The nuclei of these sources are also affected by the highest  extinctions. Conversely, the lowest ice column densities are measured in NGC~3256 N, with N$_{\rm H_2O}$ = 4.87 $\times$ 10$^{17}$ cm$^{-2}$ and N$_{\rm CO_2}$ = 0.30 $\times$ 10$^{17}$ cm$^{-2}$. The nuclear models and the N$_{\rm H_2O}$ and N$_{\rm CO_2}$ maps for all targets are shown in Appendix~\ref{appensixC}.

\section{Discussion}
\label{sec:discussion}

\subsection{Comparison with classical extinction diagnostics}
\label{sec:comparison}
\begin{figure*} 
\centering
\includegraphics[width=0.99\textwidth]{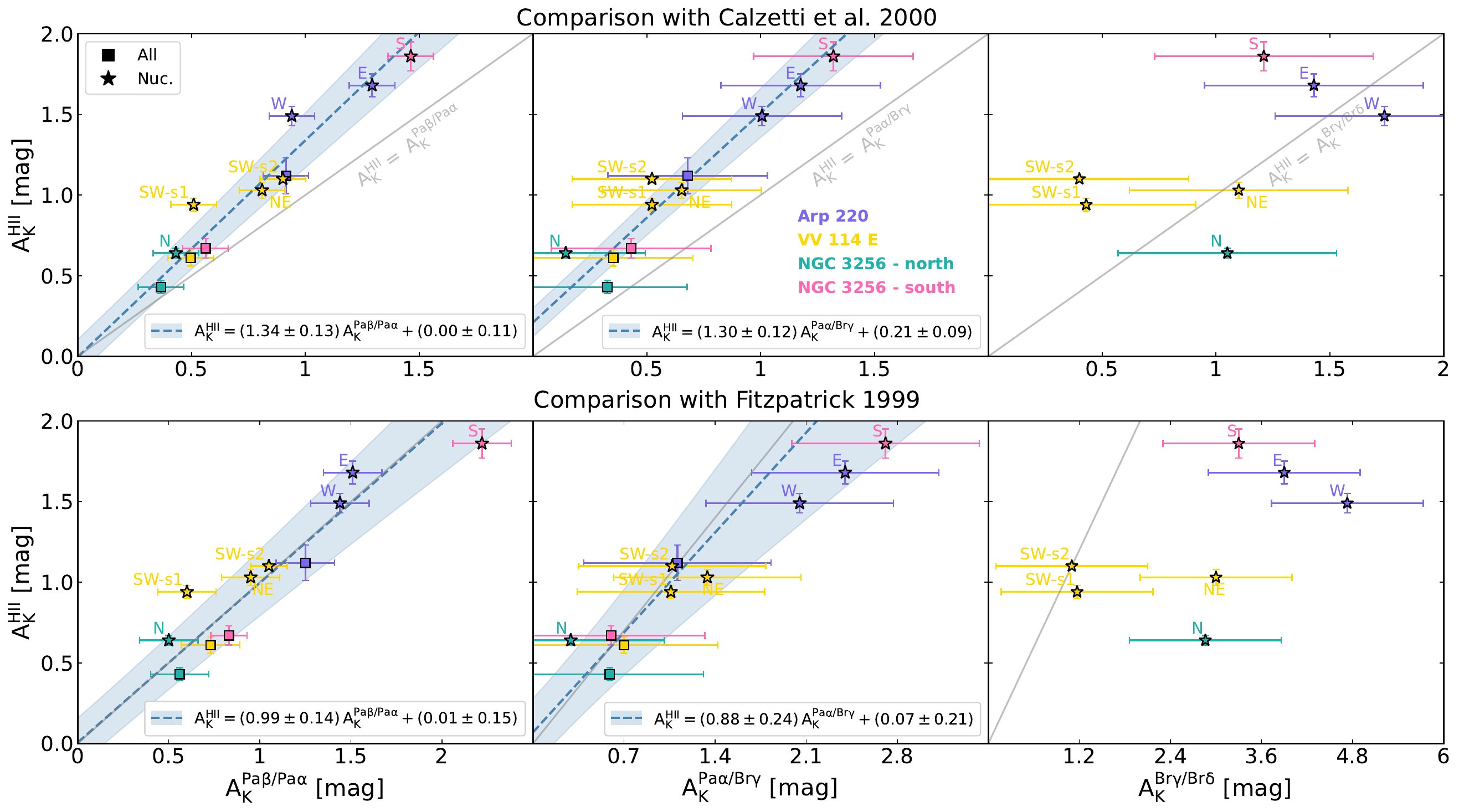}
\caption {Comparison between the A$_{\rm K}^{\rm HII}$ extinctions derived from numerous H emission lines (at least 10) following the method presented in Section~\ref{sec:rec}, and those obtained using the classical extinction curves of \citet{Calzetti00} and \cite{Fitzpatrick99} (top and bottom rows, respectively) with only a pair of recombination lines: Pa$\beta$/Pa$\alpha$, Pa$\alpha$/Br$\gamma$ and Br$\gamma$/Br$\delta$ (from left to right). In each panel, nuclear and extended measurements are represented by star and square symbols, respectively, while each region is labeled and color-coded according to the source name. The gray solid line corresponds to the one-to-one relation, whereas in the left and central panels the blue area and dashed line represent the linear fit and its corresponding uncertainty. }
\label{fig:Ak_comparison}
\end{figure*}

Extinction from H recombination lines is typically estimated using the flux ratio of two emission lines, comparing the observed ratio to the theoretical one, and evaluating the attenuation at the corresponding wavelength through an assumed extinction curve (e.g. \citealt{Cardelli89, Calzetti00, Fitzpatrick99}). In the infrared, the Paschen decrement (Pa$\beta$/Pa$\alpha$) is commonly used for this purpose  (e.g., \citealt{Ramos09, Reddy23}), alternatively, the Pa$\alpha$/Br$\gamma$ (e.g., \citealt{Alonso06, Marin09}) and/or the Br$\gamma$/Br$\delta$ (e.g., \citealt{Piqueras13, Sanchez22}) ratios are also employed.

These diagnostics rely on a limited wavelength range, which depends on the selected line pair. 
 In contrast, the method  presented in Section~\ref{sec:rec} derives extinction curves using multiple H recombination lines (a minimum of 10), detected across the three NIRSpec gratings over the range $\sim$0.97–5.14 $\mu$m. In this approach, the attenuation at a given wavelength is described by a power law (see Eq.~\ref{eq:func1}).

To quantify potential biases associated with the choice of method,
we compared the 
extinctions derived with our approach (A$_{\rm K}^{\rm HII}$; see Section~\ref{sec:rec}) to those obtained from classical diagnostics based on the line pairs listed above (i.e., A$_{\rm K}^{\rm Pa\beta/Pa\alpha}$, A$_{\rm K}^{\rm Pa\alpha/Br\gamma}$, and A$_{\rm K}^{\rm Br\gamma/Br\delta}$).   
For each line pair, the color excess is computed as:
\begin{equation}\label{eq:color}
    E(B-V) =  \frac{2.5}{k(\lambda_{\rm L_1}) - k(\lambda_{\rm L_2})} \rm \, log_{10}\left(\frac{\rm (F_{L_1}/F_{L_2})_{obs}}{(F_{L_1}/F_{L_2})_{int}}\right)
\end{equation}
where L$_1$ and L$_2$ represent the lines used for the diagnostic, with F$\rm _{L_1}$ and F$\rm _{L_2}$ being the corresponding fluxes, and $k(\lambda_{\rm L_1})$ and $k(\lambda_{\rm L_2})$ the values of the reddening curve evaluated at L$_1$ and L$_2$ wavelengths. Intrinsic line ratios are assumed from \citealt{Storey95}, using an electron temperature of 10$^4$ K and a density of n$_e$ = 100 cm$^{-3}$. 
The $k(\lambda)$ values are computed using two distinct reddening curves, of \citet{Calzetti00} and \citet{Fitzpatrick99} (the latter calibrated for the Milky Way), adopting $R_{\rm V} = 4.05$, both valid over the wavelength range spanned by the selected line pairs.
The visual extinction is obtained as A$_{\rm V}$ = R$_{\rm V} \, \times\, E(B-V)$ and converted to K-band extinction as A$_{\rm K}$ = A$_{\rm V}$/8.9 (\citealt{Cardelli89}). 

Fig.~\ref{fig:Ak_comparison} shows the comparison between A$_{\rm K}^{\rm HII}$ and the classical  diagnostics A$_{\rm K}^{\rm Pa\beta/Pa\alpha}$,  A$_{\rm K}^{\rm Pa\alpha/Br\gamma}$, and A$_{\rm K}^{\rm Br\gamma/Br\delta}$ (from left to right), derived using the \citet{Calzetti00} (top row) and the \citet{Fitzpatrick99} (bottom row) extinction curves.

A$_{\rm K}^{\rm HII}$ shows a clear linear correlation with both  A$_{\rm K}^{\rm Pa\beta/Pa\alpha}$ and  A$_{\rm K}^{\rm Pa\alpha/Br\gamma}$ obtained from the \citet{Calzetti00} law, but provides higher extinctions than the classical diagnostics. This  behavior is due to the change of slope in the  \citet{Calzetti00} extinction curve, that become steeper beyond $\sim 2\, \mu$m leading to underestimated extinctions in the near-IR. 
The best-fit linear relations are:
\begin{equation}
\rm A_{K}^{HII} = 
\begin{cases} 
    (1.34\, \pm\, 0.13) \rm A_{K}^{\rm Pa\beta/Pa\alpha} + (0.00 \, \pm 0.11) \\
    (1.30\, \pm\, 0.12) \rm A_{K}^{\rm Pa\alpha/Br\gamma} + (0.21 \, \pm 0.09)
\end{cases}
\label{eq:fit2}
\end{equation}

Although (U)LIRGs are environments that differ significantly from the Milky Way, when the classical diagnostics are computed using the \citet{Fitzpatrick99} curve, which has a slope in the near-IR similar to the beta=1.7 assumed here, A$_{\rm K}^{\rm HII}$ is consistent with the A$_{\rm K}^{\rm Pa\beta/Pa\alpha}$ values, while A$_{\rm K}^{\rm Pa\alpha/Br\gamma}$ is consistent with A$_{\rm K}^{\rm HII}$ at relatively low extinction, up to A$_{\rm K}^{\rm Pa\alpha/Br\gamma}$ $\approx$ 1.4. For higher values, measurements deviate from the one-to-one relation.
The corresponding linear relations are:
\begin{equation}
\rm A_{K}^{HII} = 
\begin{cases} 
    (0.99\, \pm\, 0.14) \rm A_{K}^{\rm Pa\beta/Pa\alpha} + (0.01 \, \pm 0.15) \\
    (0.88\, \pm\, 0.24) \rm A_{K}^{\rm Pa\alpha/Br\gamma} + (0.07 \, \pm 0.21), \,   \,A_{K}^{\rm Pa\alpha/Br\gamma} < 1.4
\end{cases}
\label{eq:fit2}
\end{equation}

For both the \citet{Calzetti00} and \citet{Fitzpatrick99} methods, the uncertainties in the extinction  increase as the differential extinction decreases. 
The Paschen decrement provides the largest differential extinction, which is reduced by a factor of $\sim$1.5 and $\sim$1.6 for the Pa$\alpha$/Br$\gamma$ and Br$\gamma$/Br$\delta$ ratios, respectively.
The accuracy of the derived extinction  therefore strongly depends on the pair of lines used.
Measurements based on multiple H lines (A$_{\rm K}^{\rm HII}$) yield significantly smaller uncertainties, whereas the largest dispersion is observed for  $\rm A_{K}^{\rm Br\gamma/Br\delta}$. Indeed, just a weak correlation is found between $\rm A_{K}^{\rm HII}$ and $\rm A_{K}^{\rm Br\gamma/Br\delta}$. A similar behavior is observed when  we directly apply the extinction law assumed in this paper (Eq.~\ref{eq:func1}) to a pair of lines (see Fig.~\ref{fig:Ak_comparison_our}).

\subsection{Ionized gas is more extincted than hot molecular gas}
\label{sec:rec_vs_h2}

\begin{figure} 
\centering
\includegraphics[width=.49\textwidth]{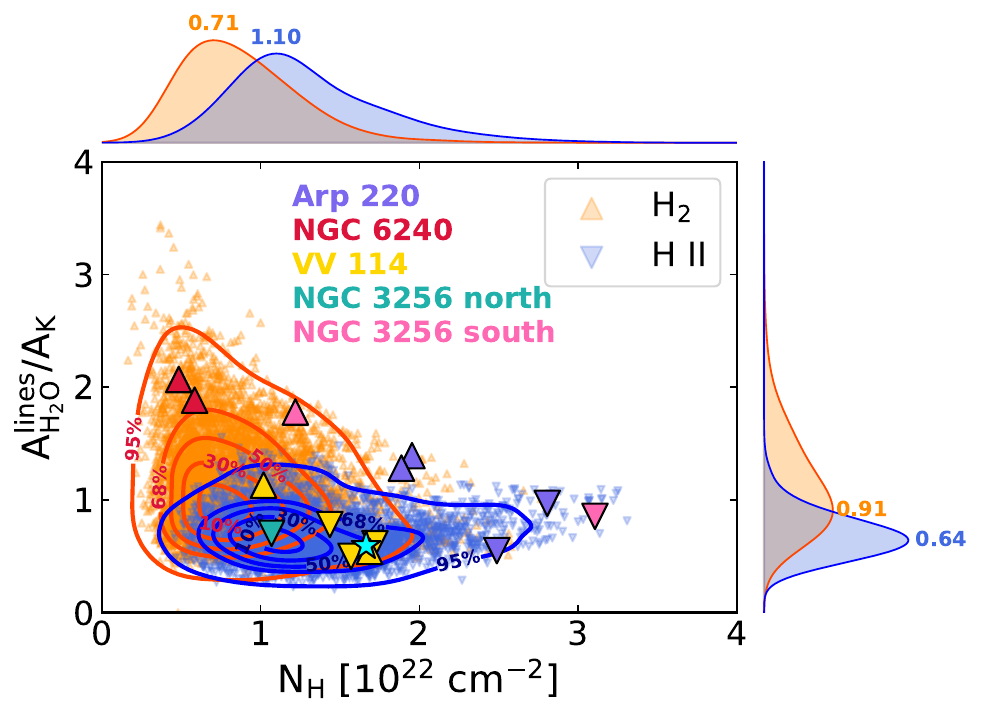}
\caption { A$_{\rm H_2O}^{\rm lines}$/A$_{\rm K}$ ratio  as a function of the hydrogen column density (N$_{\rm H}$ =  1.66 $\times$ 10 $^{22}$ A$_{\rm K}$), { combining extended values from all targets}. 
Orange triangles correspond to measurements derived from H$_2$ emission lines (i.e., A$_{\rm H_2O}^{\rm H_2}$/A$_{\rm K}^{\rm H_2}$ vs N$_{\rm H}^{\rm H_2}$), while blue inverted triangles correspond to values derived from H recombination lines (i.e., A$_{\rm H_2O}^{\rm HII}$/A$_{\rm K}^{\rm HII}$ vs N$_{\rm H}^{\rm HII}$). 
{ Nuclear values are color-coded by source.}
The density distributions along both axes are shown with their peak values. The orange and blue 2D contours enclose the indicated fractions of spaxels (10$\%$, 30$\%$, 50$\%$, 68$\%$, and 95$\%$).
The cyan star corresponds to the measurement reported by \citet{Rosenthal00} for the Orion molecular cloud, based on H$_2$ ro-vibrational emission lines.}
\label{fig:recvsh2}
\end{figure}

From Method 1 (Section~\ref{sec:rec}) and Method 2 (Section~\ref{sec:H2}), we derive  the K-band extinction from H recombination (A$_{\rm K}^{\rm HII}$) and H$_2$ emission lines (A$_{\rm K}^{\rm H_2}$), respectively. These provide direct estimates of the hydrogen column density (N$_{\rm H}$) resulting from N$_{\rm H}$ = 1.87 $\times$ 10$^{21} $A$_{\rm V}$ (\citealt{Bohlin78}), adopting A$_{\rm V}$ = 8.9 A$_{\rm K}$ (\citealt{Rieke85, Cardelli89}), and using the corresponding A$_{\rm K}^{\rm HII}$ or A$_{\rm K}^{\rm H_2}$ for each gas phase.
In addition, we measure
 the extinction affecting emission lines { associated with} the water ice absorption feature (A$_{\rm H_2O}^{\rm HII}$ and A$_{\rm H_2O}^{\rm H_2}$). 
 When the gas phase is not specified, we generally refer to these quantities as A$\rm _{H_2O}^{lines}$. 
  Using these parameters, we investigate which gas phase is more strongly extincted and how the water ice { absorption affects} the ionized and  hot molecular emission phases.

Fig.~\ref{fig:recvsh2}  shows { the A$\rm _{H_2O}^{\rm lines}$/A$_{\rm K}$ ratios } 
as a function of N$_{\rm H}$, combining spaxels from all targets and showing nuclear measurement color-coded by source. 
The H recombination lines are more strongly extincted than the H$_2$ emission lines, yielding higher hydrogen column densities. { For each gas phase, The highest N$_{\rm H}$ are measured in the most obscured nuclei, e.g., Arp~220 E and W, and NGC~3256 S, but consistently as for the extended regions, the nuclear N${\rm_ H^{HII}}$ are higher than the N${\rm_ H^{H_2}}$. } The distribution of N${\rm_ H^{HII}}$ peaks at 1.10$\times$10$^{22}$ cm$^{-2}$, while N${\rm_ H^{H_2}}$ at 0.71$\times$10$^{22}$ cm$^{-2}$, with the 95$\%$ contours of the two distributions reaching values of $\sim$3$\times$10$^{22}$ cm$^{-2}$ and $\sim$2$\times$10$^{22}$ cm$^{-2}$, respectively.
 A two-sample Kolmogorov-Smirnov test revealed a significant difference between the two distributions, with a p-value $\approx$ 0 indicating that the hypothesis of identical distributions can be rejected. 

 Consequently, the A$\rm _{H_2O}^{lines}$/A$_{\rm K}$ ratio,
is higher when derived from H$_2$ lines than from recombination lines, up to a factor of { three} for a given hydrogen column density, { as the values of A$\rm _{H_2O}^{HII}$ and A$\rm _{H_2O}^{H_2}$ are comparable between them in both the nuclear and extended structure (see Table~\ref{tab:nucext}). }
 These differences indicate that the ionized and  hot molecular gas occupy distinct regions within these galaxies, and are therefore subject to different extinctions, { but that the dust grains in which water ice accumulates might be common to both gas-phases}.

{ \citet{Donnan24} reported extinctions derived from pure rotational H$_2$ lines, based on the attenuation of the H$_2$ 0--0 S(3) line by the 9.8 $\mu$m silicate absorption feature, together with H extinctions derived from the classical Brackett decrement using Br$\beta$  (2.62 $\mu$m) and Br$\gamma$  (2.17 $\mu$m) for NGC~3256 S and N, VV~114 NE and  SW-1. They found comparable extinctions for the warm molecular and ionized gas phases in all these nuclei 
except NGC~3256 N, where the extinction toward the warm molecular gas  is a factor of 1.8 higher than that toward the ionized component.
Their analysis is complementary to ours, as they probe extinction in the warm molecular phase through pure rotational H$_2$ lines, while we investigate the hot molecular gas using ro-vibrational H$_2$ transitions.  A direct comparison with their H extinctions is not straightforward because their estimates rely on only two H lines and may therefore be subject to larger uncertainties. Indeed, adopting the extinction law provided by \citet{Donnan23}, the differential extinction decreases by a factor of $\sim$1.4 when using Br$\beta$/Br$\gamma$ instead of Pa$\beta$/Pa$\alpha$.

Both methods presented in Section~\ref{sec:rec} and ~\ref{sec:H2} enable characterizing how water ice absorption contributes to the extinction of the hot molecular and ionized gas (A$_{\rm H_2O}^{\rm H_2}$ and A$_{\rm H_2O}^{\rm HII}$, respectively), not estimated by classical methods.}
 To the best of our knowledge, the { A$_{\rm H_2O}^{\rm H_2}$ value } 
 has been reported only in the case study by \citet{Rosenthal00} for the Orion Molecular Cloud, { using the same method presented in Section~\ref{sec:H2}}. These authors report A$_{\rm H_2O}^{\rm H_2}$ = 0.58 for A$_{\rm K}^{\rm H_2}$ = 1, { which results in a lower ratio compared to our results from H$_2$ lines (see Fig.~\ref{fig:recvsh2}). } For the same extinction, we generally measure higher values of A$_{\rm H_2O}^{\rm H_2}$, { up to 1.5}.
Consequently, the { A$\rm _{H_2O}^{H_2}$/A$_{\rm K}$ ratio } derived by \citet{Rosenthal00}  
 is lower than our average ratio of A$\rm _{H_2O}^{H_2}$/A$_{\rm K}$ = 0.91.
 { We provide the first extragalactic measurements of both A$\rm _{H_2O}^{H_2}$ and A$\rm _{H_2O}^{HII}$.}

\subsubsection{A common H$_2$O icy envelope}

\begin{figure} 
\centering
\includegraphics[width=0.49\textwidth]{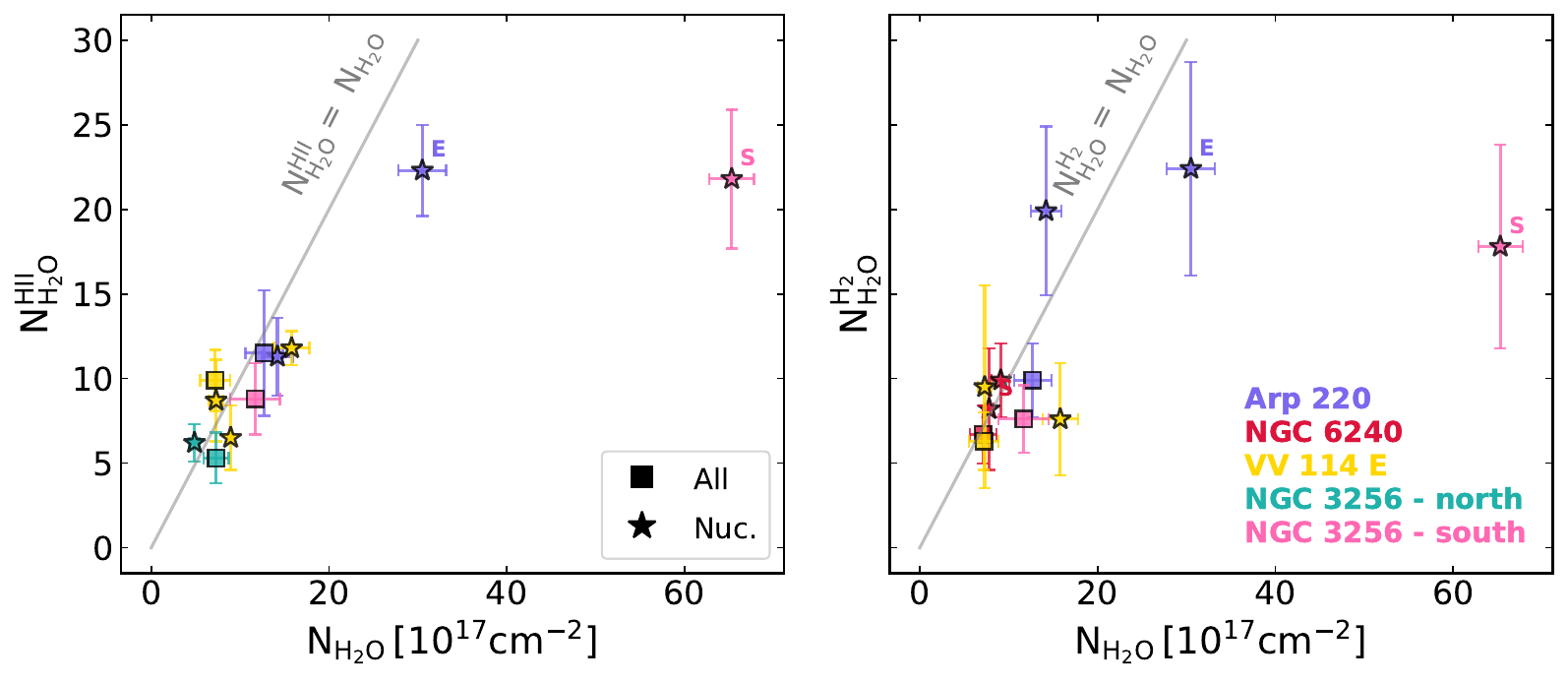}
\caption {Comparison between water ice columns densities derived from emission lines (N$_{\rm H_2O}^{\rm HII}$ and N$_{\rm H_2O}^{\rm H_2}$) and from the continuum absorption features at $\sim$3.02 $\mu$m (N$_{\rm H_2O}$). 
Right panel: N$_{\rm H_2O}^{\rm HII}$ vs. N$_{\rm H_2O}$. Left panel: N$_{\rm H_2O}^{\rm H_2}$ vs. N$_{\rm H_2O}$. In both panels nuclear and extended regions are indicated by star and square symbols, respectively, and color-coded by source.} 
\label{fig:Nh2o_comp}
\end{figure}

In the previous section, we discussed that the hot molecular and ionized gas are characterized by different levels of extinction, while being similarly affected  by water ice absorption. For completeness, we compare the water ice column densities N$\rm _{H_2O}^{H_2}$ and N$\rm _{H_2O}^{HII}$, derived from A$\rm _{H_2O}^{H_2}$ and A$\rm _{H_2O}^{HII}$, respectively (see Section~\ref{sec:H2} and ~\ref{sec:rec}), with  the water ice column density derived from the continuum absorption feature, i.e., N$\rm _{H_2O}$ (see Section~\ref{sec:ices}).

\begin{figure*} 
\centering
\includegraphics[width=0.84\textwidth]{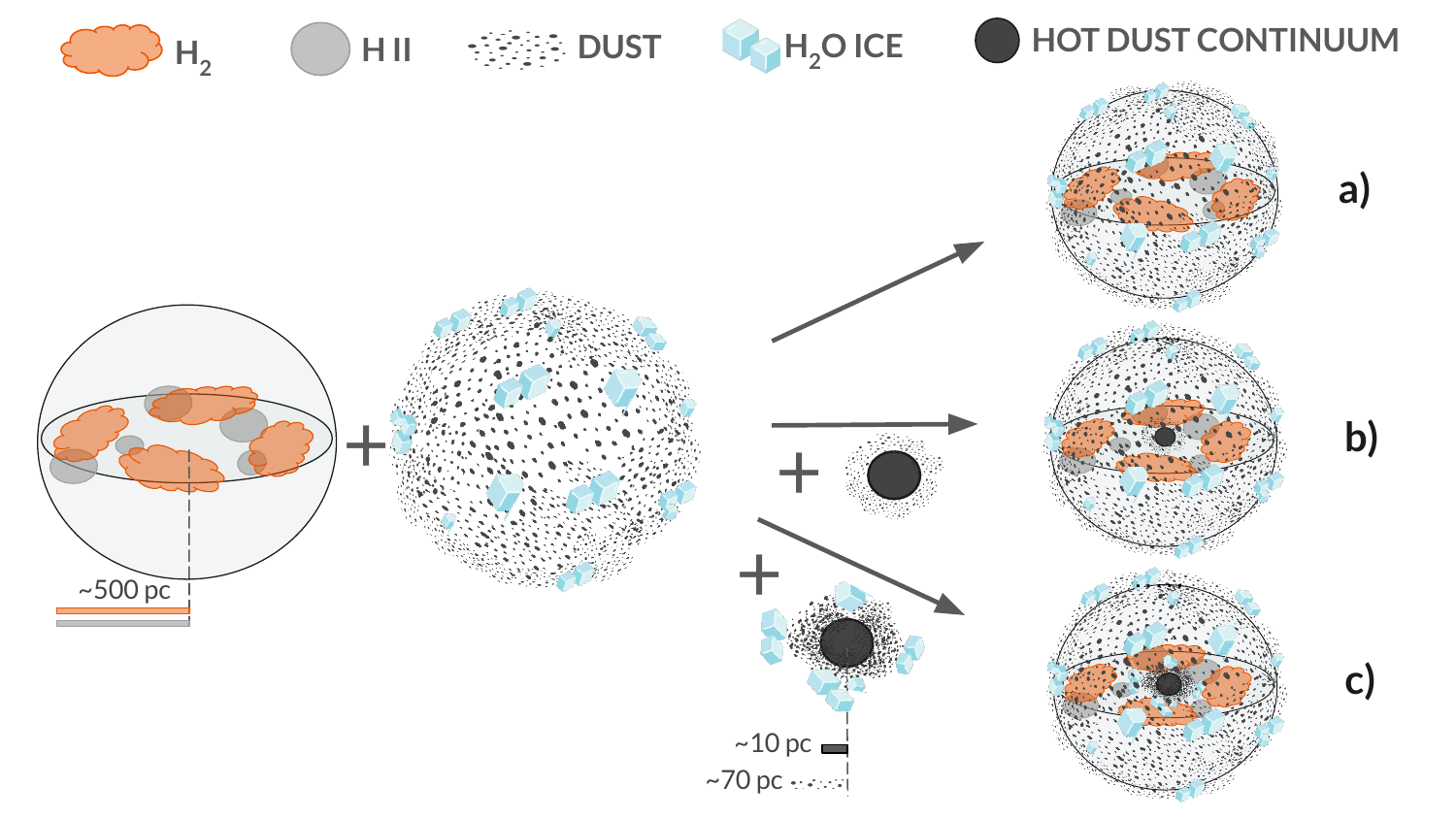}
\caption{ Schematic representation (not to scale) of the proposed scenarios, showing the common features on the left and the three cases on the right. Top right: scenario a), in which the H$_2$ molecular gas is more diffuse than the compact H~II emission, with a common shell of dust and H$_2$O ice surrounding both gas phases. Central right: scenario b), which is equivalent to scenario a) but includes an inner hot dust continuum source. Bottom right: scenario c), which is equivalent to scenario b) but with the hot dust continuum source embedded within higher column densities, enabling an additional layer of water ice to accumulate on its outer layer (i.e., the scenario proposed for Arp~220 E and NGC~3256 S).
}

\label{fig:dibujo}
\end{figure*}

This comparison is presented in Fig.~\ref{fig:Nh2o_comp}, where the left and right panels show N$_{\rm H_2O}^{\rm HII}$ and N$_{\rm H_2O}^{\rm H_2}$, respectively, as a function of N$_{\rm H_2O}$. In both panels,  nuclear measurements are denoted by star symbols, while square symbols represent the median values from extended regions.
There is a good one-to-one correlations between the water ice column densities derived from the lines and those obtained from the continuum. This suggests that the H$_2$O ice originates from dust grains distributed in an outer shell that surrounds the diffuse H$_2$, the continuum, and the H I emissions, the latter being concentrated in more dust extincted regions (as A$_{\rm K}^{\rm HII} >$ A$_{\rm K}^{\rm H2}$; see Fig.~\ref{fig:dibujo}). 
The nuclear regions of Arp~220 E and NGC~3256 S reach the highest values of N$_{\rm H_2O}$ = (30.5 $\pm$ 2.7) $\times 10^{17}$ cm$^{-2}$ and N$_{\rm H_2O}$ = (65.3 $\pm$ 2.5) $\times 10^{17}$ cm$^{-2}$, respectively, representing extreme environments that deviate from the 1:1 correlation. Arp~220 E deviates by a factor of 1.4, and NGC~3256 S by factors ranging from 3 up to 3.7. 
Such differences may be driven by the presence of a compact, obscured hot dust continuum source in the nucleus, emitting in the near-IR, with an additional layer of H$_2$O ice accumulating on the external layer of the dusty torus.
This might be the case of an AGN or SF core for  Arp~220 E  (e.g., \citealt{Perna24}), while a deeply embedded AGN is hosted in NGC~3256 S  (e.g. \citealt{Pereira24}; \textcolor{blue}{Pereira-Santaella et al. in prep.}). This proposed scenario is illustrated in panel c) of Fig.~\ref{fig:dibujo}. The extremely high column densities would produce the so-called greenhouse effect (see \citealt{gonzalez19}), according to which continuum radiation is trapped by the high column densities, enhancing the dust temperatures in the innermost regions. Conversely,  low dust temperatures are reached in the outer layer thanks to the high thermal cooling efficiency of dust, such steep gradient of temperature ensures suitable conditions for the accumulation of water ice on the outer layer of dust. 
Panel b) of Fig.~\ref{fig:dibujo} represents cases in which the column densities surrounding the hot dust continuum are not sufficiently high to produce this effect. According to our results, this applies to Arp~220 W, NGC~6240 N and S, and to the nuclei of VV~114. We exclude the possibility that orientation could prevent us from detecting such an additional water ice layer, as in the dusty environment of (U)LIRGs the compact hot-dust continuum sources have small opening angles, 
so we would not observe different behavior depending on our line-of-sight. Panel a) represents cases in which a compact continuum source is not observed, as in NGC~3256 N, where the background continuum emission is likely mostly caused by star formation activity.

\subsection{H$_2$O  to CO$_2$ ice abundances}\label{ices_comp}


From the  continuum method presented in Section~\ref{sec:ices}, we derive the H$_2$O and CO$_2$ ice column densities using the  absorption features in the continuum at 3.02 $\mu$m and $\sim$4.26 $\mu$m, respectively. H$_2$O ice column densities are higher than those of CO$_2$ in all targets. 
In Fig.~\ref{fig:h2oco2} we show their  abundance  ratio (N$_{\rm CO_2}$/N$_{\rm H_2O}$) 
through histograms corresponding to each source, where ice column densities have been measured spaxel-by-spaxel in the rebinned cubes, with median relative abundances of N$_{\rm CO_2}$/N$_{\rm H_2O}$ between 0.09 and 0.13 (see dashed vertical lines in all panels).

Lower CO$_2$ ice column densities than those of H$_2$O are expected, as  water ice  constitutes the primary  component of ISM {  ices} (\citealt{Garrod11}), forming through more efficient reactions than CO$_2$ (e.g., \citealt{Molpeceres23}). 
Indeed, higher H$_2$O than CO$_2$ ice columns densities are measured in the Milky Way (e.g., \citealt{Boogert15}). 
However, the N$_{\rm CO_2}$/N$_{\rm H_2O}$ ratios we find are lower than typical Galactic values. Combining measurements from { the four (U)LIRGs}, we obtain a median relative abundance of N$_{\rm CO_2}$/N$_{\rm H_2O}$ = 0.10 (dashed vertical line in the last panel of Fig.~\ref{fig:h2oco2}). This is more than a factor of two lower than the median relative abundances { of N$_{\rm CO_2}$/N$_{\rm H_2O}$ $\approx$ 0.22} reported for Galactic interstellar ices { (e.g., \citealt{Gibb04, Bergin05, Smith25}; see also \citet{Gerakines99} who reported a ratio of 0.17$\pm$ 0.03}).

Most extragalactic studies measuring N$_{\rm CO_2}$/N$_{\rm H_2O}$ focus on the Large and Small Magellanic Clouds and generally find values consistent with those observed in the Milky Way  (e.g., \citealt{Oliveira09, Seale11, demarchi25}).
Using AKARI NIR spectroscopy, \citet{Yamagishi15} report N$_{\rm CO_2}$/N$_{\rm H_2O}$ ratios of 0.05–0.20 for seven nearby star-forming galaxies. Our results are consistent with this range.

\section{Summary and conclusions}  
\label{sec:conclusions}

We investigated the extinction and  the H$_2$O ($\sim$3.02 $\mu$m) and CO$_2$ ($\sim$4.26 $\mu$m) ice content in four nearby interacting (U)LIRGs using JWST/NIRSpec IFU observations covering the 0.97 -- 5.14 $\mu$m  wavelength range. 
The broad spectral coverage allows us to simultaneously detect numerous H recombination  ($>$ 10) and H$_2$  ($\geq$ 15) emission lines,  including those near $\sim$3 $\mu$m that are sensitive to water ice, as well as the H$_2$O (3.02 $\mu$m) and CO$_2$ ($\sim$4.26 $\mu$m) ice absorption features in the continuum.  This combination of emission and absorption features enable us to constrain the obscuration affecting different gas phases and the continuum emission (see Section~\ref{sec:rec}, ~\ref{sec:H2}, and ~\ref{sec:ices}), in both nuclear and extended regions.
 The main conclusions from our analysis  
 are as follows:

\begin{itemize}

\item \textit{Comparison with classical H extinction diagnostics}
We found that extinctions derived from multiple (at least ten) H lines (i.e., A$_{\rm K}^{\rm HII}$) are consistent with those obtained from the classical Paschen decrement (e.g., Pa$\beta$/Pa$\alpha$ ratio) when adopting the \citet{Fitzpatrick99} extinction law. 
However, this agreement is not observed when using other line ratios (e.g., Pa$\alpha$/Br$\gamma$ or Br$\gamma$/Br$\delta$) and/or adopting the \citet{Calzetti00} extinction law. In Section~\ref{sec:comparison}, we provide correction relations for cases in which a linear correlation is found between the classical diagnostics listed above (i.e., A$_{\rm K}^{\rm Pa\beta/Pa\alpha}$, A$_{\rm K}^{\rm Pa\alpha/Br\gamma}$, and A$_{\rm K}^{\rm Br\gamma/Br\delta}$) and A$_{\rm K}^{\rm HII}$.

\item \textit{Different extinction levels for ionized and hot H$_2$ gas phases}. Extinction derived from H recombination lines are systematically higher than those obtained from H$_2$ emission lines.  In the nuclei we obtain median values of A$_{\rm K}^{\rm HII} = (1.25 \pm 0.02)$ and A$_{\rm K}^{\rm H_2} = (0.75 \pm 0.10)$, and for the extended regions A$_{\rm K}^{\rm HII} = (0.61 \pm 0.04)$ and A$_{\rm K}^{\rm H_2} = (0.50 \pm 0.07)$.   
 In contrast, the  H$_2$O ice  column densities are comparable between the two gas phases, with N$_{\rm H_2O}^{\rm H_2}$ = (13.5$\pm$1.8) cm$^{-2}$ and N$_{\rm H_2O}^{\rm HII}$ = (12.6$\pm$0.9) cm$^{-2}$ in the nuclei, and N$_{\rm H_2O}^{\rm H_2}$ = (7.8$\pm$2.0) cm$^{-2}$ and N$_{\rm H_2O}^{\rm HII}$ = (7.6$\pm$2.0) cm$^{-2}$ in the extended regions.
 This suggests that the icy shell surrouds the diffuse H$_2$, continuum, and H~II emission, with the latter distributed in more deeply dust-embedded regions (A$_{\rm K}^{\rm HII}$ >  A$_{\rm K}^{\rm H2}$).

\item \textit{Extragalactic versus Galactic H$_2$ extinction and H$_2$O ice abundance}. 
  We present the first extragalactic measurements of  A$_{\rm K}^{\rm H_2}$ and A$_{\rm H_2O}^{\rm H_2}$, finding average ratios of A$_{\rm H_2O}^{\rm H_2}$/A$_{\rm K}^{\rm H_2}$ = 1.31 and 1.10 in  the nuclear and extended regions, respectively. 
These ratios are more than a factor of two  higher than the Galactic value measured in the Orion Molecular Cloud by \citet{Rosenthal00} (A$_{\rm H_2O}^{\rm H_2}$/A$_{\rm K}^{\rm H_2}$ = 0.58).  Consequently, the water ice abundance reported by these authors, N$_{\rm H_2O}^{\rm H_2}$ = 5.9 $\times$ 10$^{-5}$, is lower than our average value, N$_{\rm H_2O}^{\rm H_2}$ = (9.0 $\pm$ 3.0) $\times$ 10$^{-5}$. 
This difference is possibly due to distinct physical conditions that, in (U)LIRGs, favor more effective shielding of water ice by dust grains compared to the Orion photodis-sociation region.

\item \textit{A common icy mantle and obscured cores}
 In addition to providing phase-specific extinctions (A$_{\rm K}^{\rm HII}$ and A$_{\rm K}^{\rm H_2}$), this study accounts for the additional attenuation due to water ice affecting emission lines around  $\sim 3\,\mu$m (N$_{\rm H_2O}^{\rm HII}$ and N$_{\rm H_2O}^{\rm H_2}$).
 We combine these measurements with the water ice column density derived from the continuum (N$_{\rm H_2O}$). The values of N$_{\rm H_2O}^{\rm HII}$, N$_{\rm H_2O}^{\rm H_2}$ and N$_{\rm H_2O}$ are comparable, revealing a common icy mantle that surrounds the H~II and the more diffuse H$_2$ emitting regions. Simultaneously, we identified two heavily obscured dust continuum sources, Arp~220 E and NGC~3256 S, for which the N$_{\rm H_2O}$ values are, respectively, a factor 1.4 and more than 3 times higher than the values derived from  N$_{\rm H_2O}^{\rm HII}$ and N$_{\rm H_2O}^{\rm H_2}$. This suggests that, in addition to the common outer layer, an additional layer of water ice is deposited at the edge of the dusty region surrounding the hot dust continuum sources. 

\item \textit{H$_2$O excess absorption}. 
 The red wing of the water ice feature exhibits excess absorption in most targets (all except NGC~3256 N). Its depth varies among sources and affect both the continuum and the emission lines.
It is not reproduced by either pure or mixed-ice laboratory profiles. We suggest that this excess may arise from scattering effects related to the dust grain size distribution, and/or from the presence of hydrogenated amorphous carbon. A detailed analysis of this feature will be presented in a forthcoming paper.

\item \textit{Underabundant CO$_2$ ice in extragalactic environments}. 
The median abundance ratio of CO$_2$ to H$_2$O ice (N$_{\rm CO_2}$/N$_{\rm H_2O} \approx 0.10$) is lower than typical Galactic values where (N$_{\rm CO_2}$/N$_{\rm H_2O} \approx 0.22$), { but in agreement with extragalactic measurements reported by \citet{Yamagishi15} of N$_{\rm CO_2}$/N$_{\rm H_2O}$ = 0.05--0.20}.

\end{itemize}

Overall, this work provides the  first spatially resolved extragalactic study of phase-dependent extinction and ice column densities in nearby galaxies. These results demonstrate the unique capability of JWST/NIRSpec to probe the structure and physical conditions of heavily obscured galaxies, revealing differences in obscuration between nuclear and extended regions as well as among distinct gas phases and the continuum-emitting dust.

\begin{acknowledgements}

GS and MPS acknowledge support under grant
CNS2023-145506 funded by MCIN/AEI/10.13039/501100011033
and the European Union NextGenerationEU/PRTR. MPS acknowledges support under grant RYC2021-033094-I funded by
MCIN/AEI/10.13039/501100011033 the European Union NextGenerationEU/PRTR and thanks the Spanish
MCINN for funding support under grant PID2023-146667NB-I00
funded by MCIN/AEI/10.13039/501100011033. AAH and LHM acknowledge financial support by the grant PID2021-124665NB-I00 funded by the Spanish Ministry of Science and Innovation and the State Agency of Research MCIN/AEI/10.13039/501100011033 PID2021-124665NB-I00 and ERDF. MP acknowledges support under grants PID2021-127718NB-I00, PID2024-159902NA-I00, and RYC2023-044853-I, funded by the Spain Ministry of Science and Innovation/State Agency of Research MCIN/AEI/10.13039/501100011033 and el FSE+. IL acknowledges support from PRIN-MUR project “PROMETEUS”  financed by the European Union -  Next Generation EU, Mission 4 Component 1 CUP B53D23004750006. SA acknowledges grants PID2025-169195NB-I00 and  PID2021-127718NB-I00 funded by the Spanish Ministry of Science and Innovation/State Agency of Research (MICIN/AEI/ 10.13039/501100011033). IGB is supported by the Programa Atracci\'on de Talento Investigador ``C\'esar Nombela'' via grant 2023-T1/TEC-29030 funded by the Community of Madrid, and by the research project PID2024-159902NA-I00 funded by the Spanish Ministry of Science and Innovation/State Agency of Research (MCIN/AEI/10.13039/501100011033) and FSE+. EG-A thanks the Spanish MICINN for support under projects PID2025-168521OB-C41 and PID2023-146667NB-I00.

\end{acknowledgements}

\medskip
\bibliography{./sample}

\begin{appendix}

\section{Continuum images at $\sim$2$\mu$m} \label{appensix0}
\begin{figure*}
\centering
\includegraphics[width=1.\textwidth]{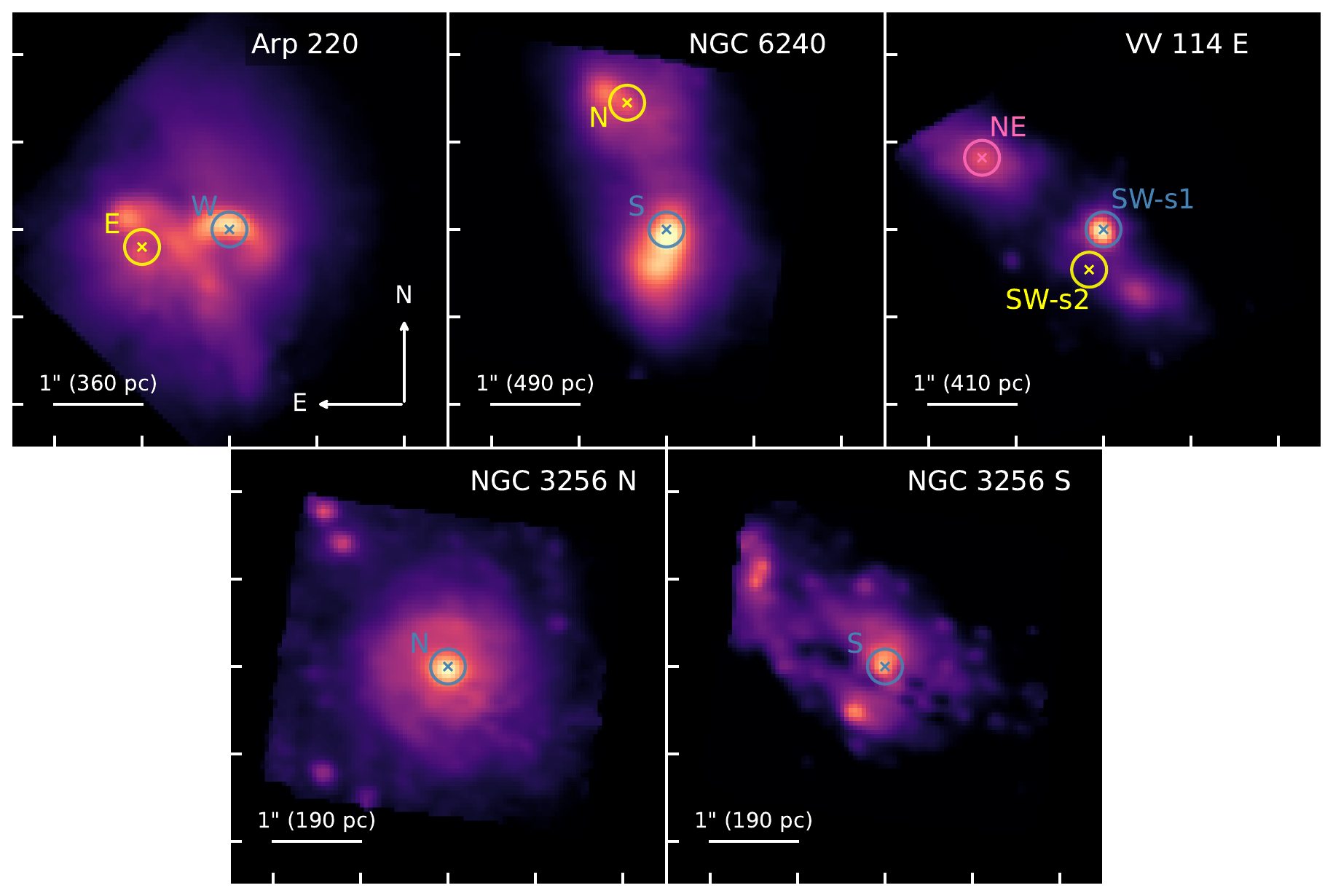}
\caption { Same as Fig.~\ref{fig:sample} but with the NIRSpec continuum selected at an observed wavelength of $\sim$2 $\mu$m  (dominated by the stellar emission).}
\label{fig:sample_4.8}
\end{figure*}

The continuum images at observed $\sim$2 $\mu$m in Fig.~\ref{fig:sample_4.8} are complementary to those  at observed 4.8 $\mu$m in Fig.~\ref{fig:sample}. The galaxy morphology at 2 $\mu$m is dominated by stellar emission, while  some structures, such as NGC~6240 N or VV~114 SW-s1, are only clearly observed  at $\sim$ 4.8 $\mu$m, where the contribution from hot dust emission dominates.

\section{H extinction in Arp~220, VV~114, and NGC ~3256} \label{appensixA}

\begin{figure}[H]
\centering
\includegraphics[width=0.405\textwidth]{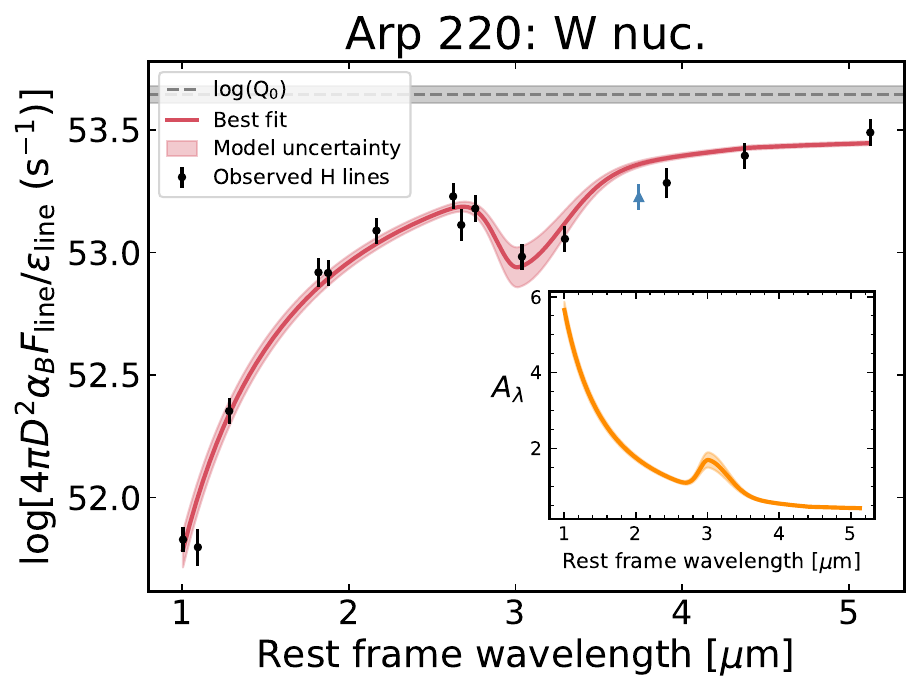}
\caption {Same as Fig.~\ref{fig:rec} but for Arp~220 W.}
\label{fig:rec_arpw}
\end{figure}

\begin{figure*}
\centering
\includegraphics[width=0.405\textwidth]{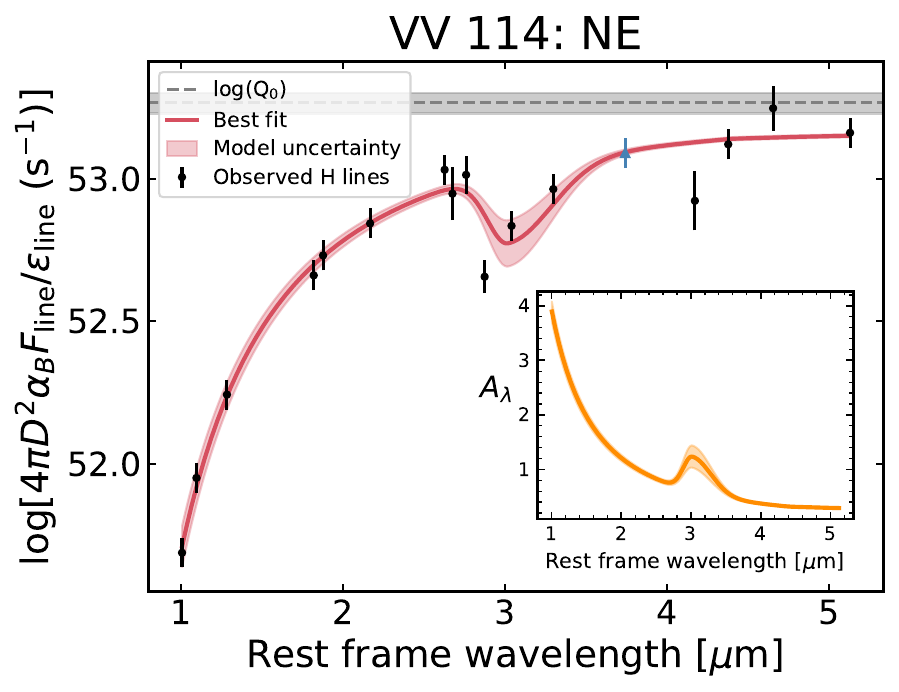}
\includegraphics[width=0.425\textwidth]{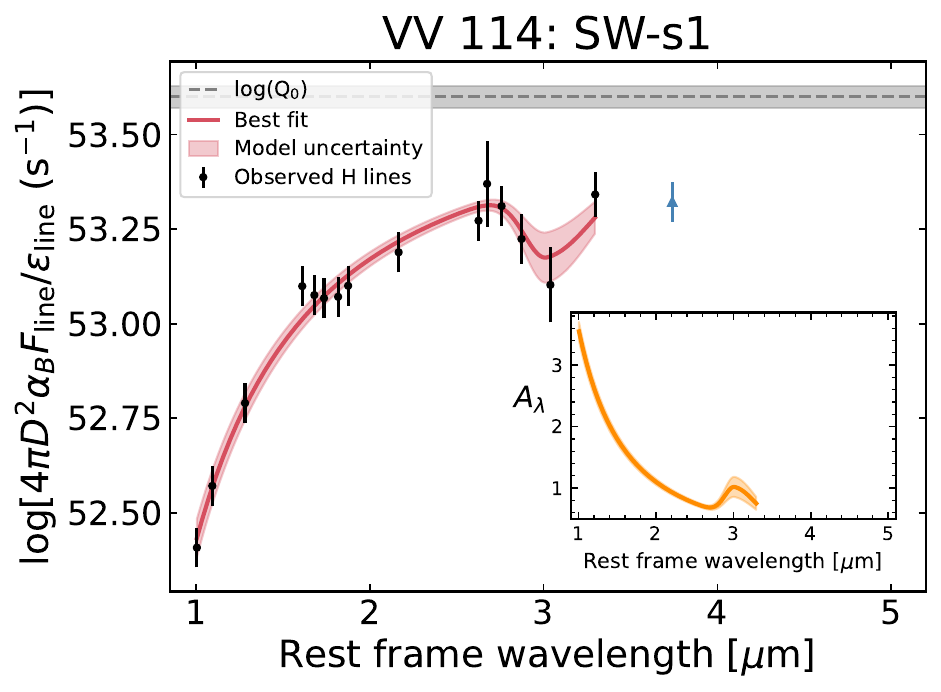}
\includegraphics[width=0.405\textwidth]{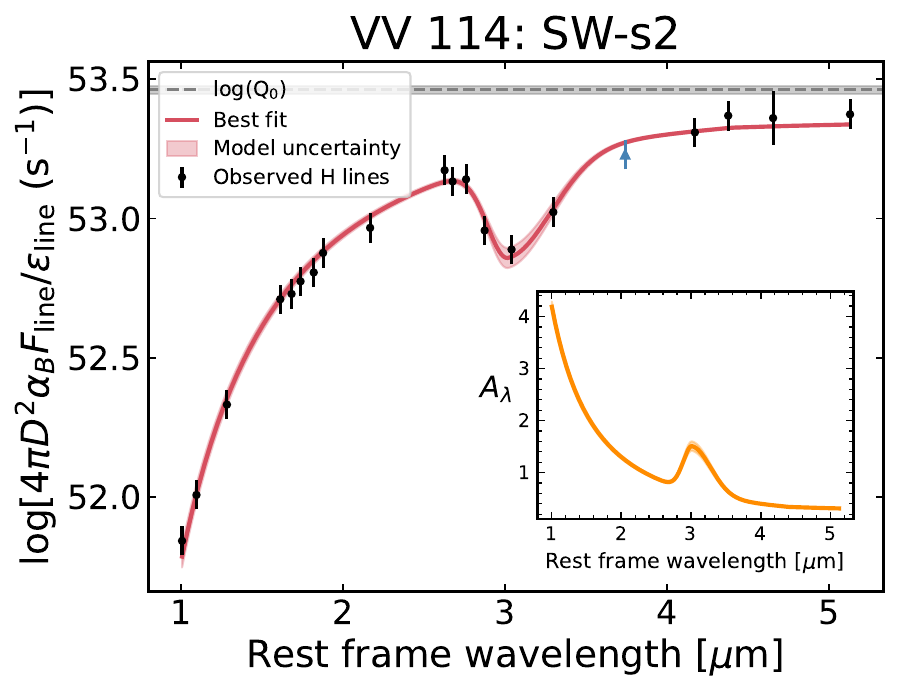}
\includegraphics[width=0.59\textwidth]{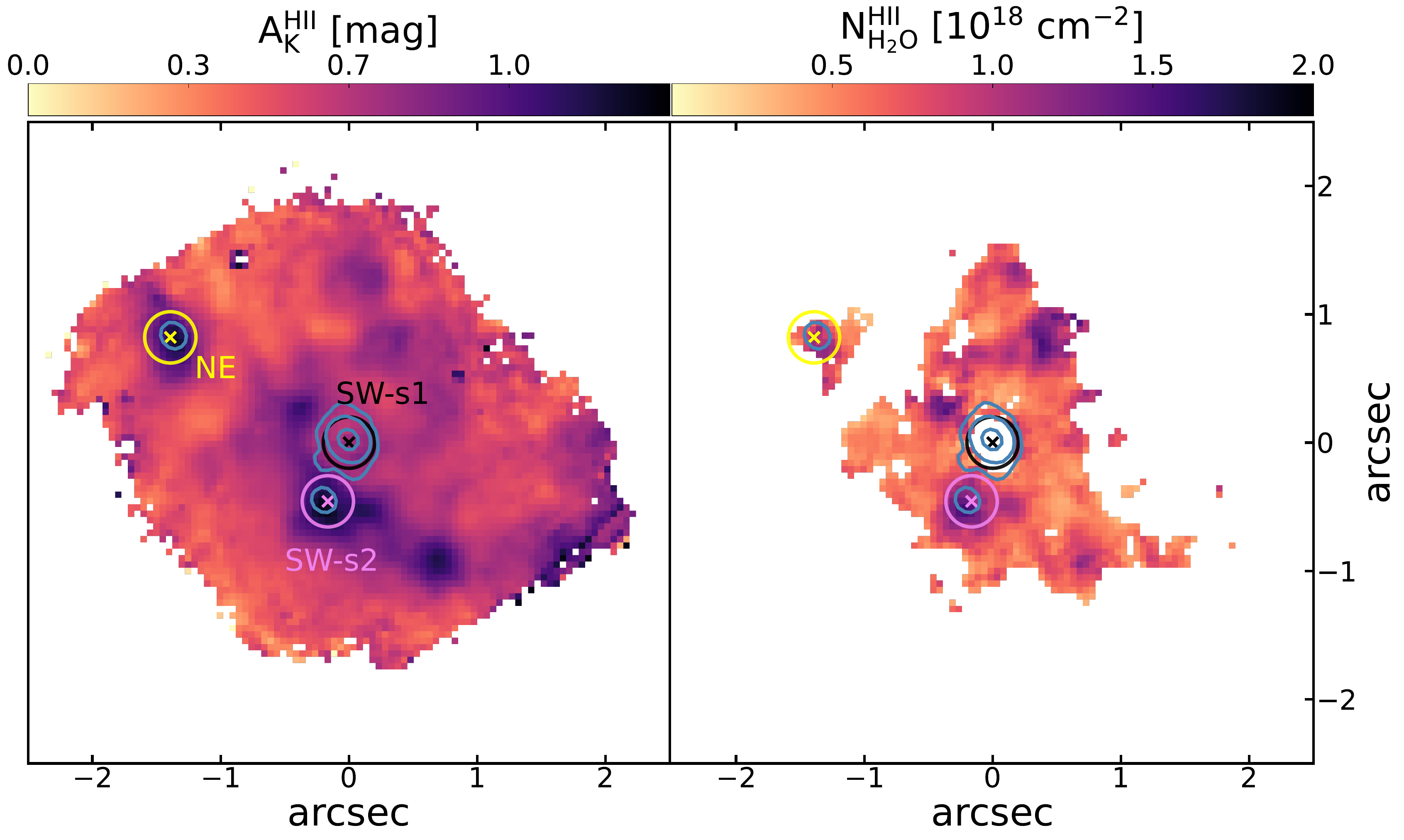}
\caption {Same as Fig.~\ref{fig:rec} but for VV~114 E.}
\label{fig:rec_vv114}
\end{figure*}

\begin{figure*}
\centering
\includegraphics[width=0.425\textwidth]{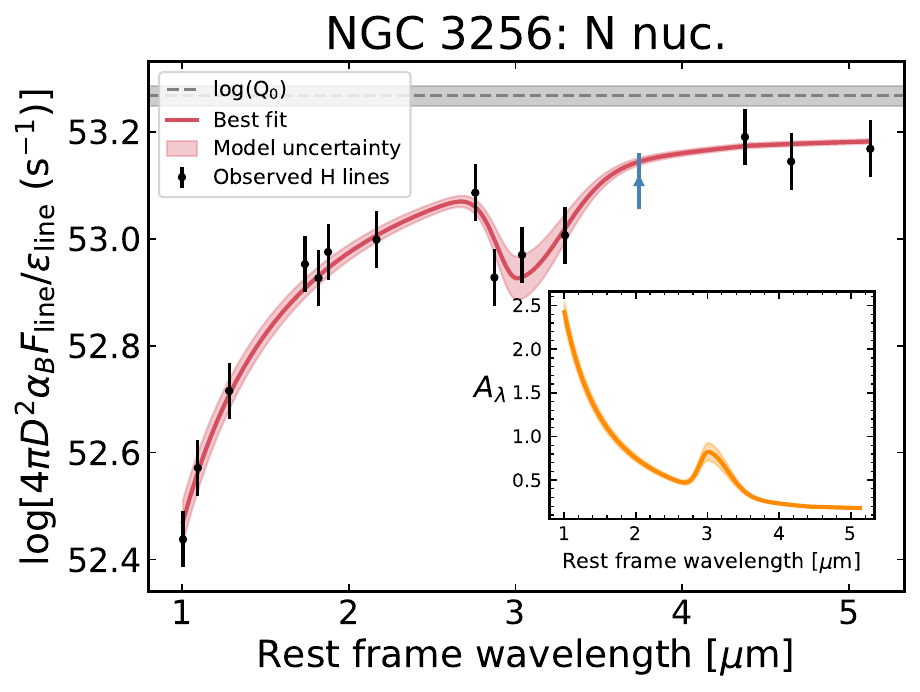}
\includegraphics[width=0.57\textwidth]{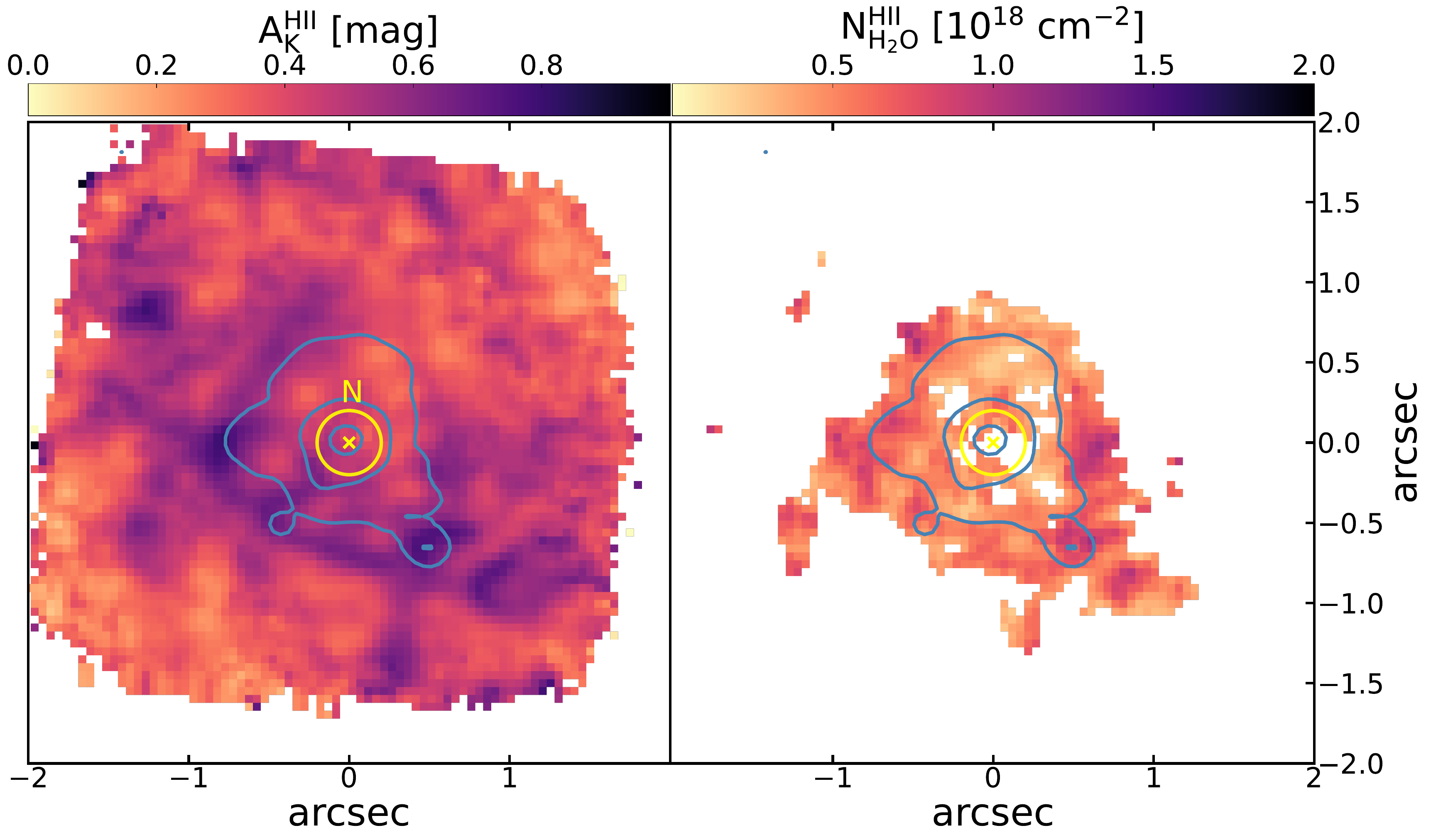}
\caption {Same as Fig.~\ref{fig:rec} but for NGC~3256--north.}
\label{fig:rec_ngc3256n}
\end{figure*}

\begin{figure*}
\centering
\includegraphics[width=0.425\textwidth]{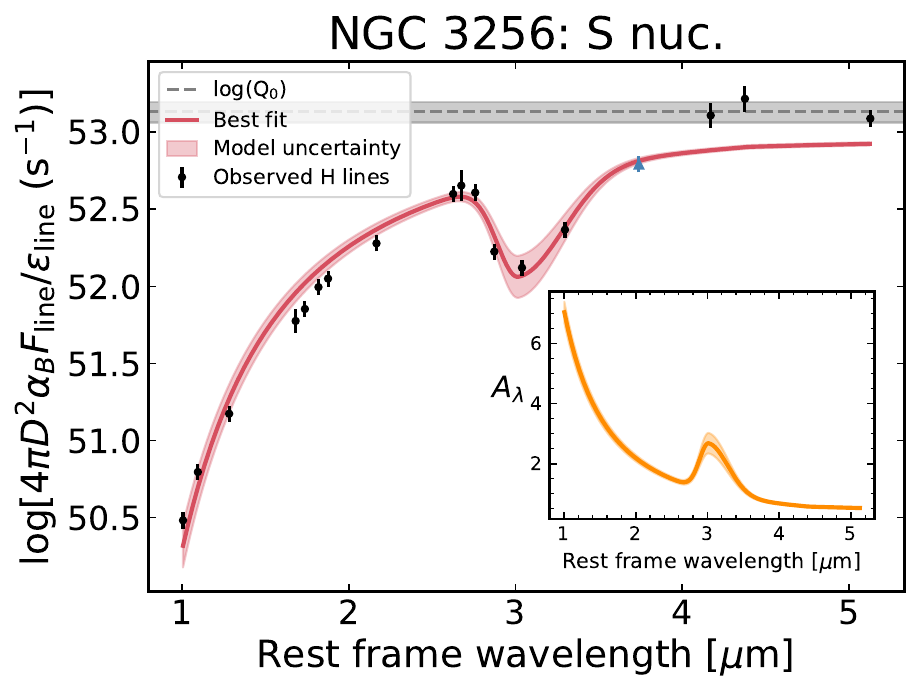}
\includegraphics[width=0.57\textwidth]{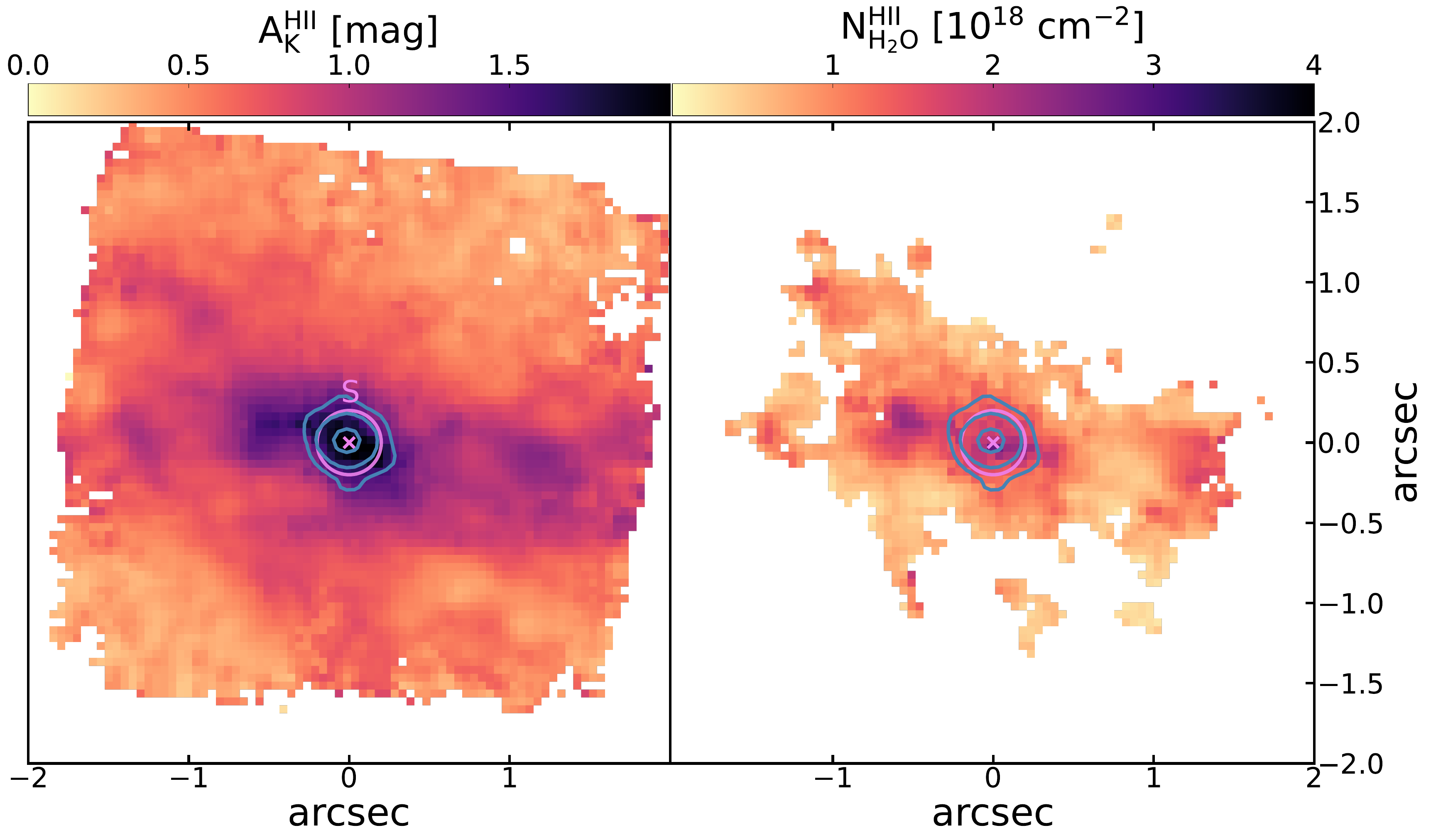}
\caption {Same as Fig.~\ref{fig:rec} but for NGC~3256--south.}
\label{fig:rec_ngc3256s}
\end{figure*}

In this Section, we present the plots and maps of the H extinctions derived using the method described in Section~\ref{sec:rec}. We show the counterparts of the  plots presented in Fig.~\ref{fig:rec} for Arp~220 E and its extended emission, but for Arp~220 W, NGC 3256 N, NGC 3256 S, VV~114 E NE, SW-s1, and SW-s2, together with the  maps of A$_{\rm K}^{\rm HII}$ and N$_{\rm H_2O}^{\rm HII}$ (see Fig.~\ref{fig:rec_arpw}, \ref{fig:rec_vv114}, \ref{fig:rec_ngc3256n}, \ref{fig:rec_ngc3256s}).  In the case of VV~114 SW-s1, we do not detect H recombination lines with S/N > 5 above $\sim$3.3 $\mu$m, except for the H 8-5 transition, which is excluded from the fitting procedure (see Section~\ref{sec:rec}). For NGC~6240, we do not detect the minimum number of ten H emission lines required to apply this method.

\section{H$_2$ extinction in Arp~220, NGC 6240, VV~114, and NGC~3256} \label{appensixB}

\begin{figure}[H]
\centering
\includegraphics[width=0.405\textwidth]{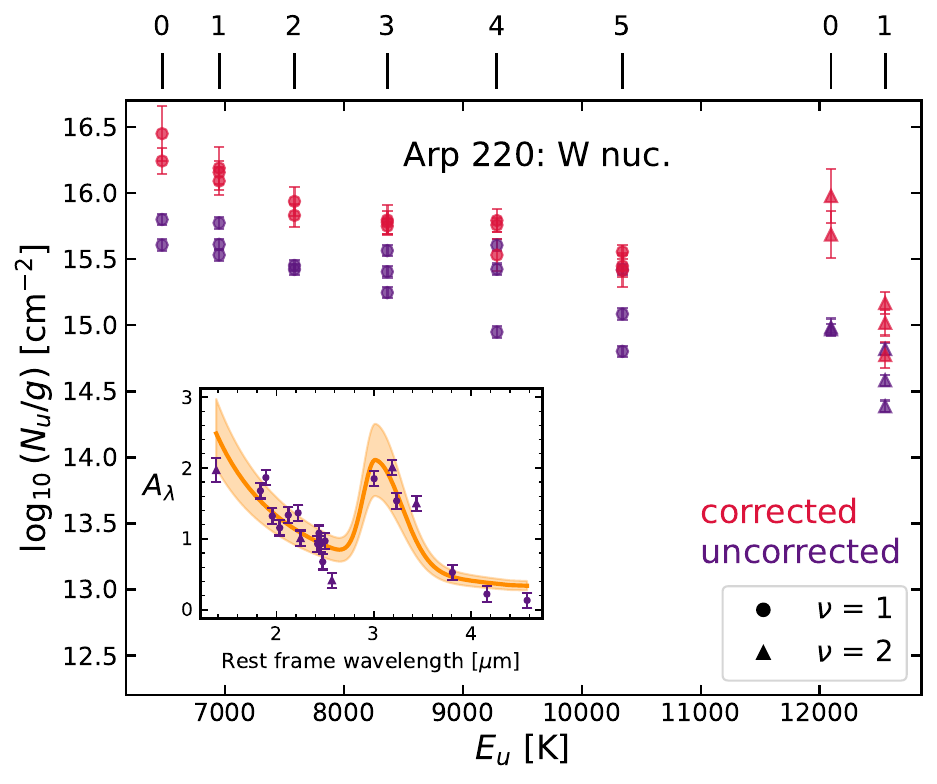}
\caption {Same as Fig.~\ref{fig:h2} but for Arp~220--west. }
\label{fig:h2_arpw}
\end{figure}

\begin{figure*} 
\centering
\includegraphics[width=0.405\textwidth]{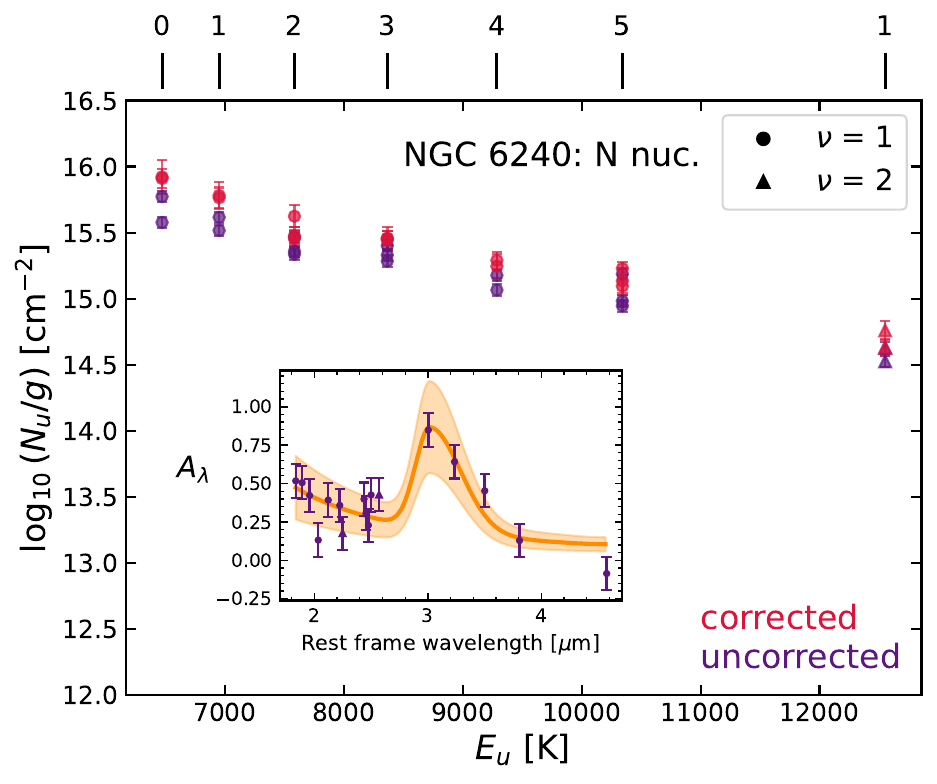}
\includegraphics[width=0.405\textwidth]{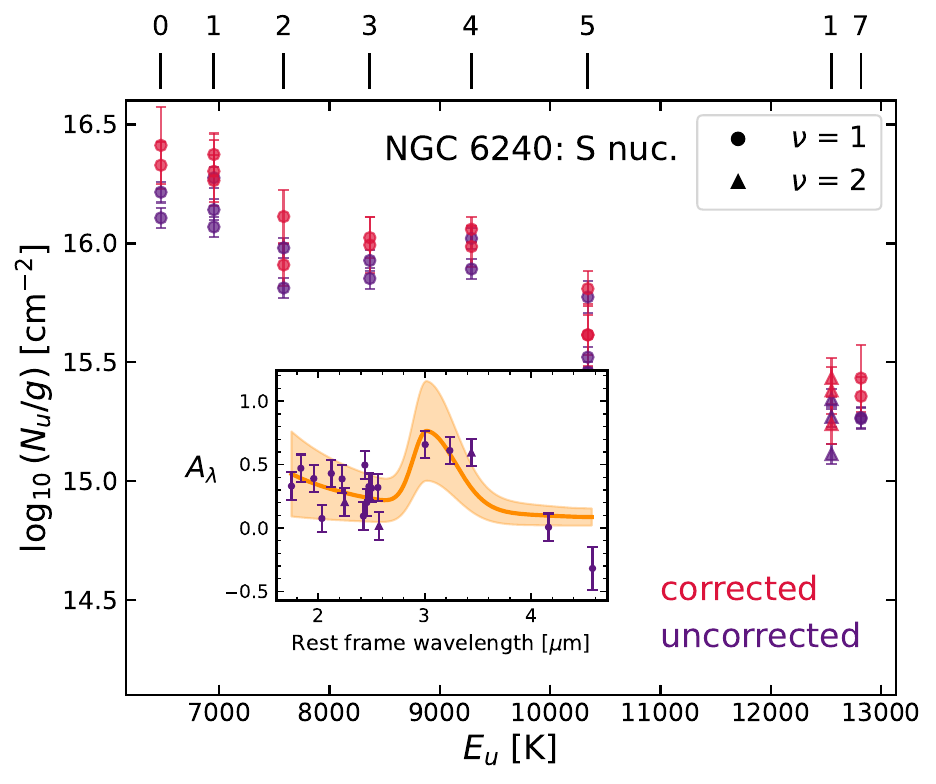}
\includegraphics[width=0.59\textwidth]{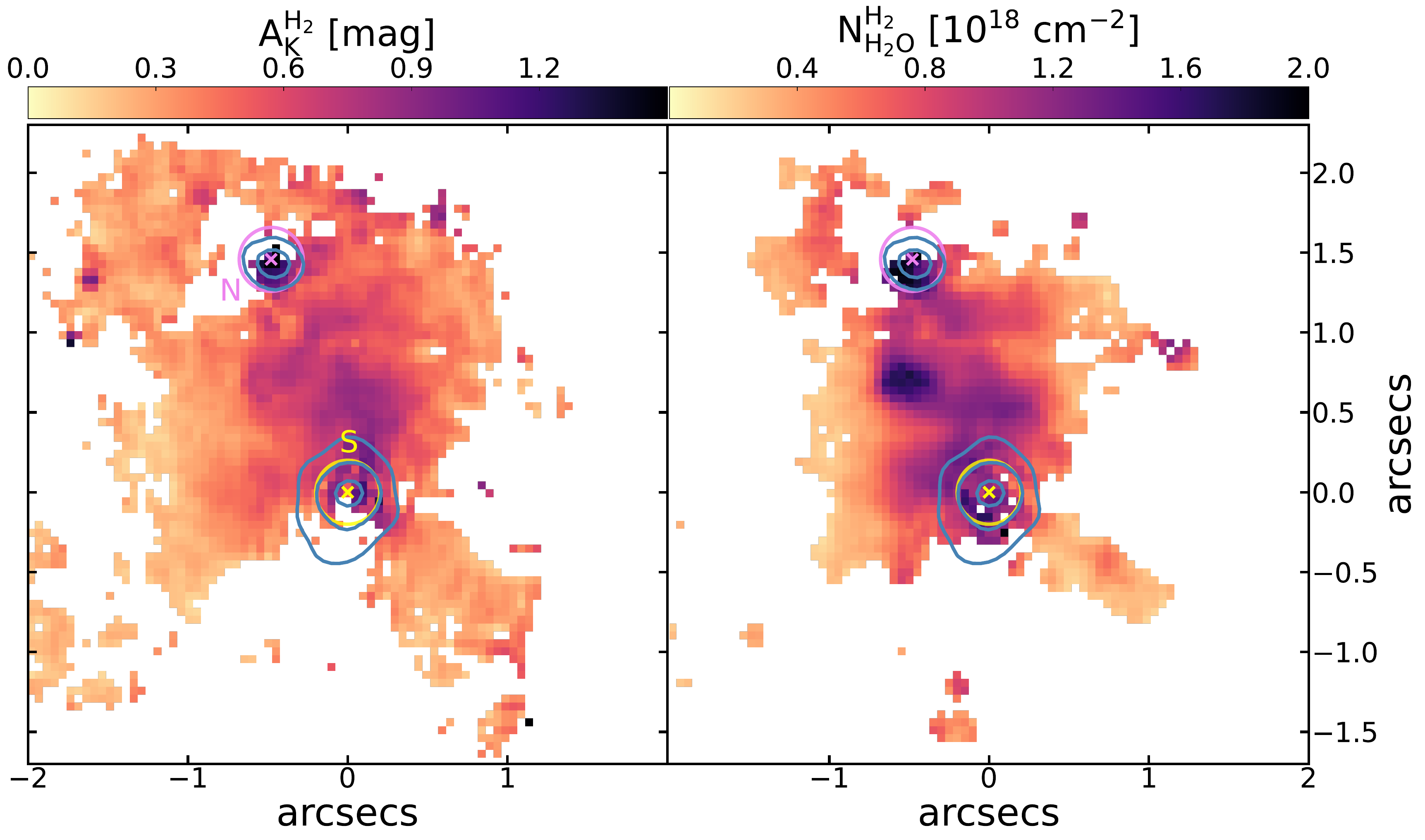}
\caption {Same as Fig.~\ref{fig:h2} but for NGC~6240. }
\label{fig:h2_6240}
\end{figure*}

\begin{figure*} 
\centering
\includegraphics[width=0.405\textwidth]{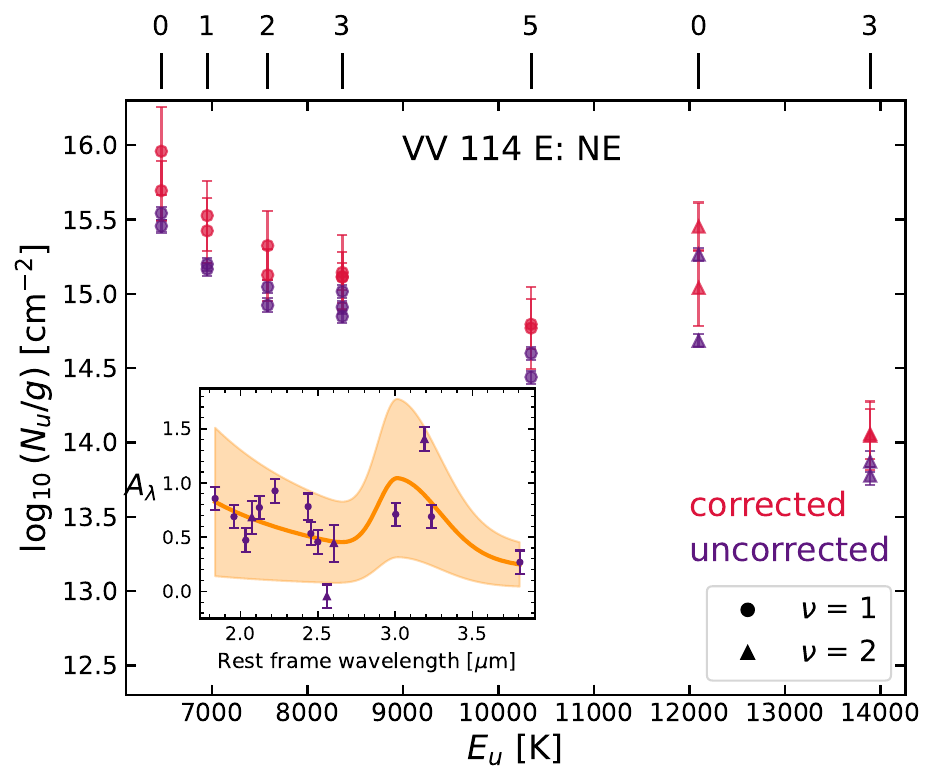}
\includegraphics[width=0.405\textwidth]{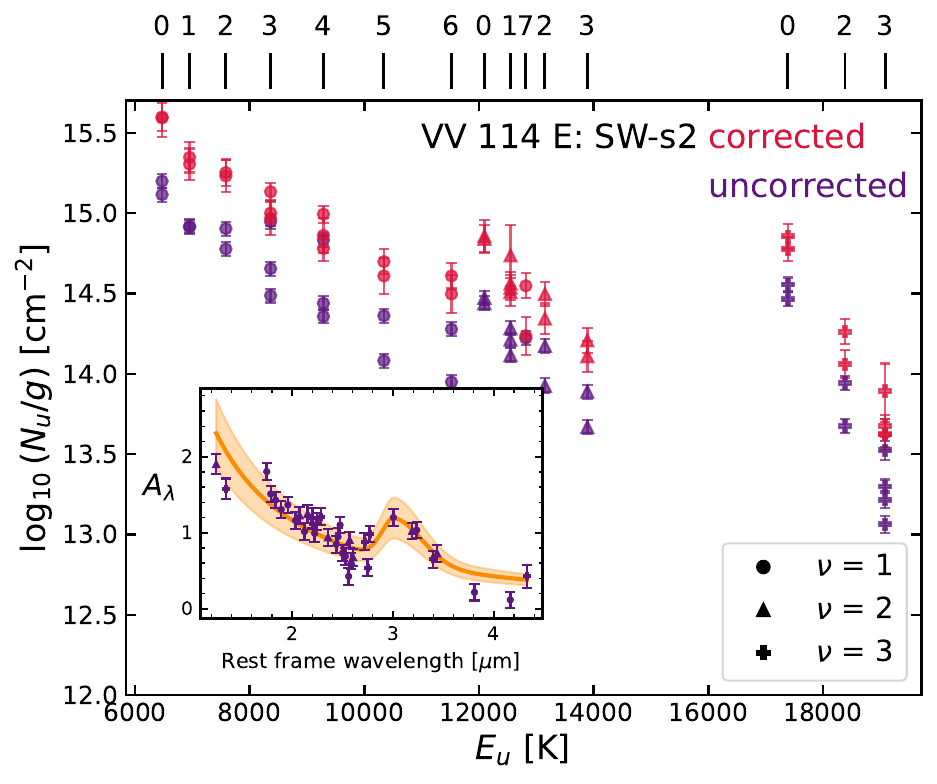}
\includegraphics[width=0.59\textwidth]{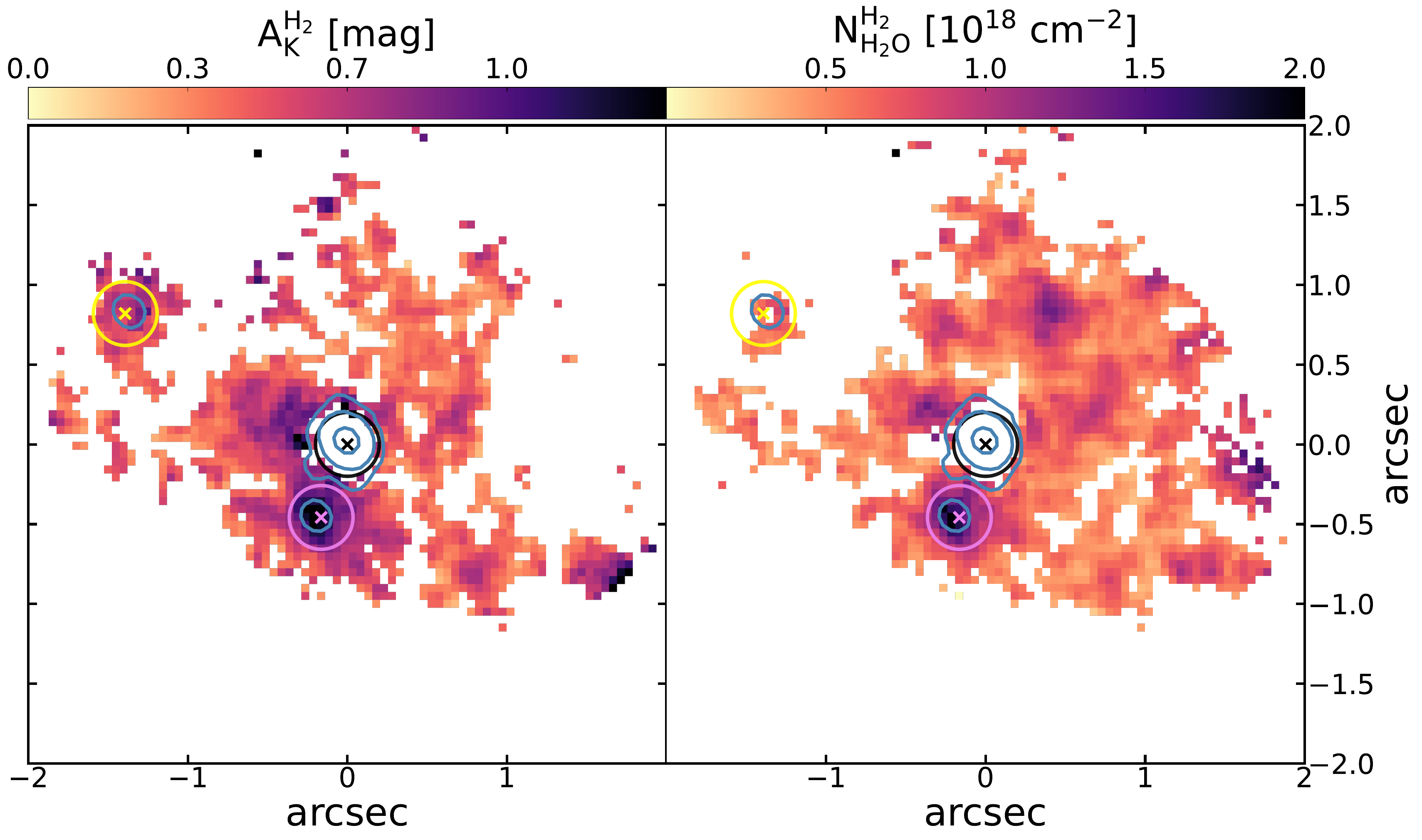}
\caption {Same as Fig.~\ref{fig:h2} but for VV~114 E. }
\label{fig:h2_vv114}
\end{figure*}

\begin{figure*} 
\centering
\includegraphics[width=0.405\textwidth]{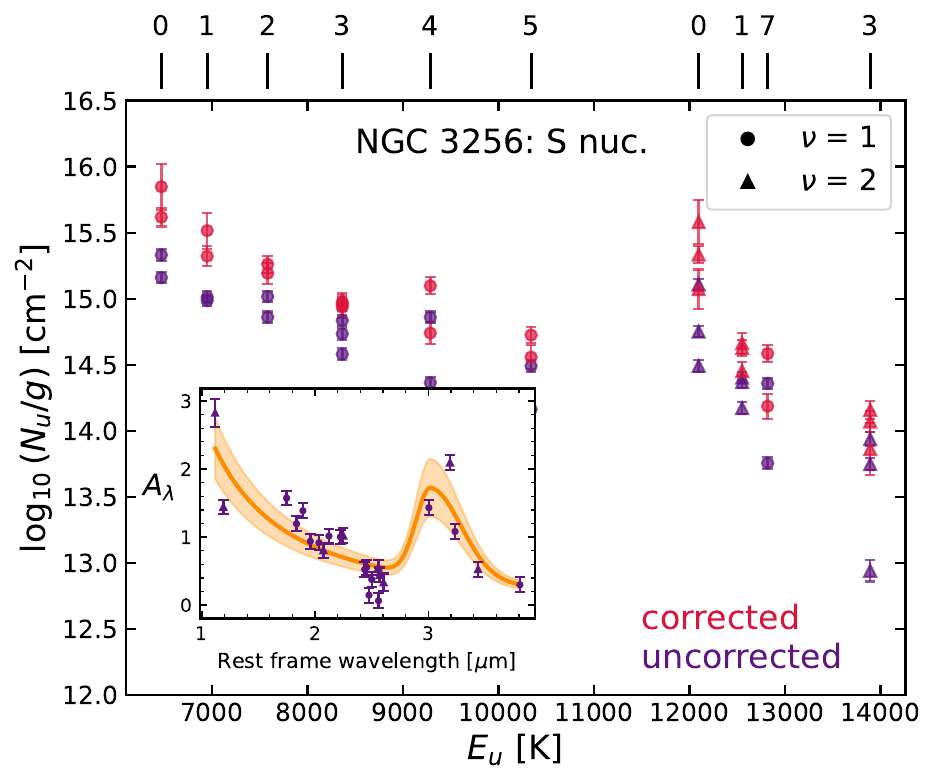}
\includegraphics[width=0.59\textwidth]{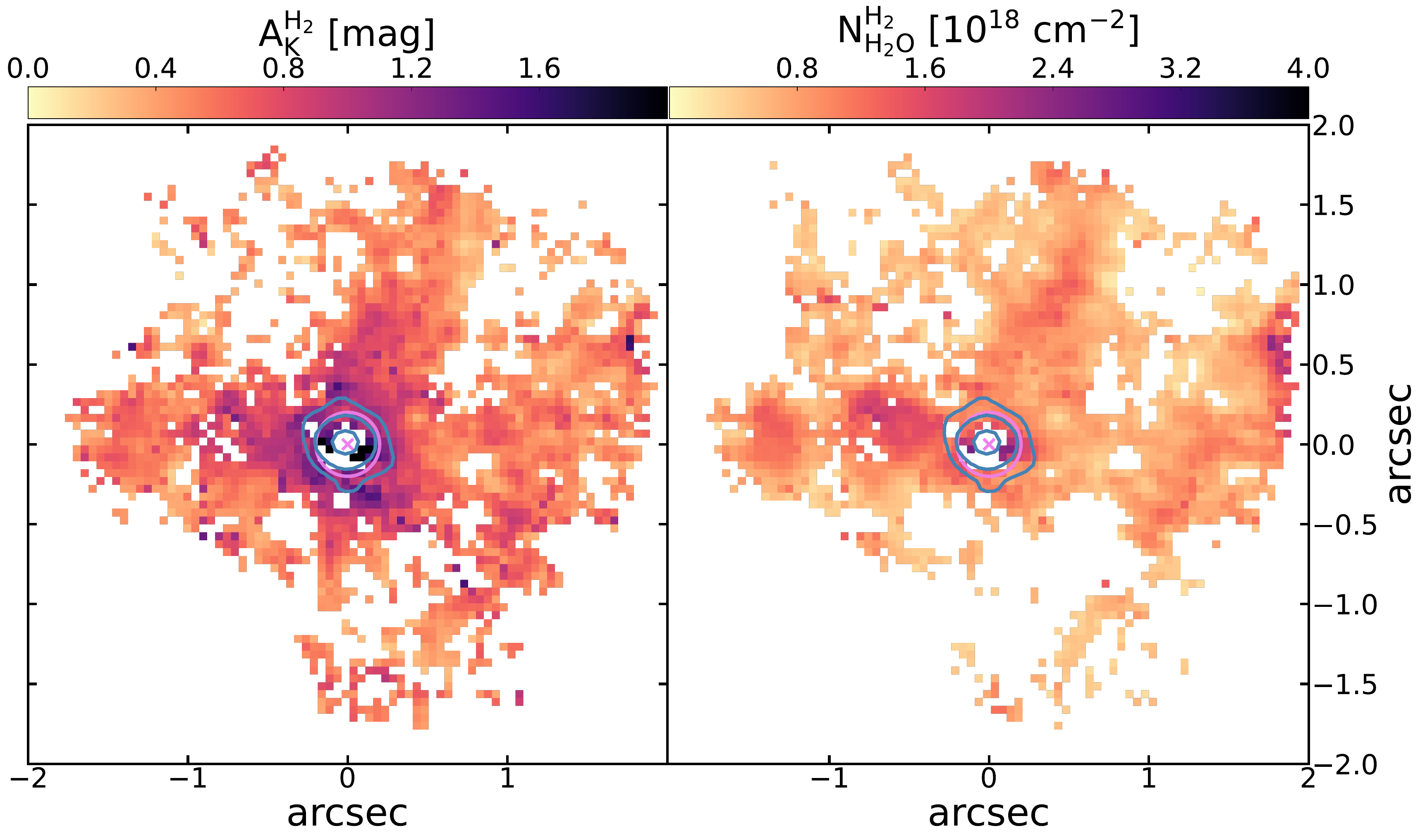}
\caption {Same as Fig.~\ref{fig:h2} but for NGC~3256--south. }
\label{fig:h2_ngc3256s}
\end{figure*}

In this Section, we present the plots and maps of the hot H$_2$ extinctions derived using the method described in Section~\ref{sec:H2}. We show the counterparts of the  plots presented in Fig.~\ref{fig:h2} for Arp~220 E and its extended emission, but for Arp~220 W, VV~114 E NE, SW-s1, and SW-s2, NGC~6240 N and S, and NGC 3256 S, together with the  maps of A$_{\rm K}^{\rm H_2}$ and N$_{\rm H_2O}^{\rm H_2}$ (see Fig.~\ref{fig:h2_arpw}, ~\ref{fig:h2_6240}, ~\ref{fig:h2_vv114}, ~\ref{fig:h2_ngc3256s}).

It is worth noticing that the A$_{\rm K}^{\rm H_2}$ and N$_{\rm H_2O}^{\rm H_2}$ maps of NGC~3256 south show an enhancement along the northern axis. This region corresponds to the location where the redshifted molecular outflow is detected (see \citealt{Pereira24}). The central and right panel of Fig.~\ref{fig:h2_ngc3256s} show how the outflow is illuminated in the near-IR through a 
 dust screen lying between our line of sight and the outflowing material. In  NGC~6240, the ongoing interaction plays a key role in shaping the extinction maps. Due to the interaction, most of the molecular gas is concentrated between the northern and southern nuclei (\citealt{Cicone18}), resulting in an enhancement  of A$_{\rm K}^{\rm H_2}$ and A$_{\rm K}^{\rm H_2}$ in this region.  In the case of NGC~3256 N, we do not detect a sufficient number of H$_2$ ro-vibrational lines to measure the extinction of the hot molecular phase, neither in the nuclear and extended regions.

\section{H$_2$O and CO$_2$ ice content in Arp~220, NGC 6240, VV~114, and NGC~3256} \label{appensixC}

\begin{figure}[H]
\centering
\includegraphics[width=0.415\textwidth]{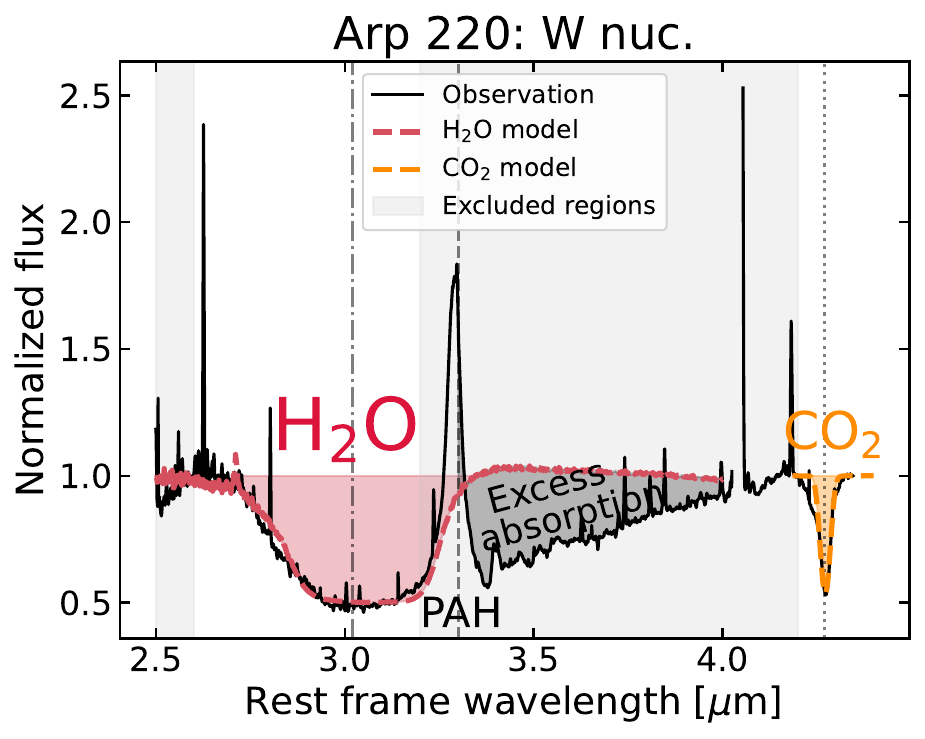}
\caption {Same as Fig.~\ref{fig:ice} but for Arp~220--west.} 
\label{fig:arp220wices}
\end{figure}

\begin{figure*} 
\centering
\includegraphics[width=0.27\textwidth]{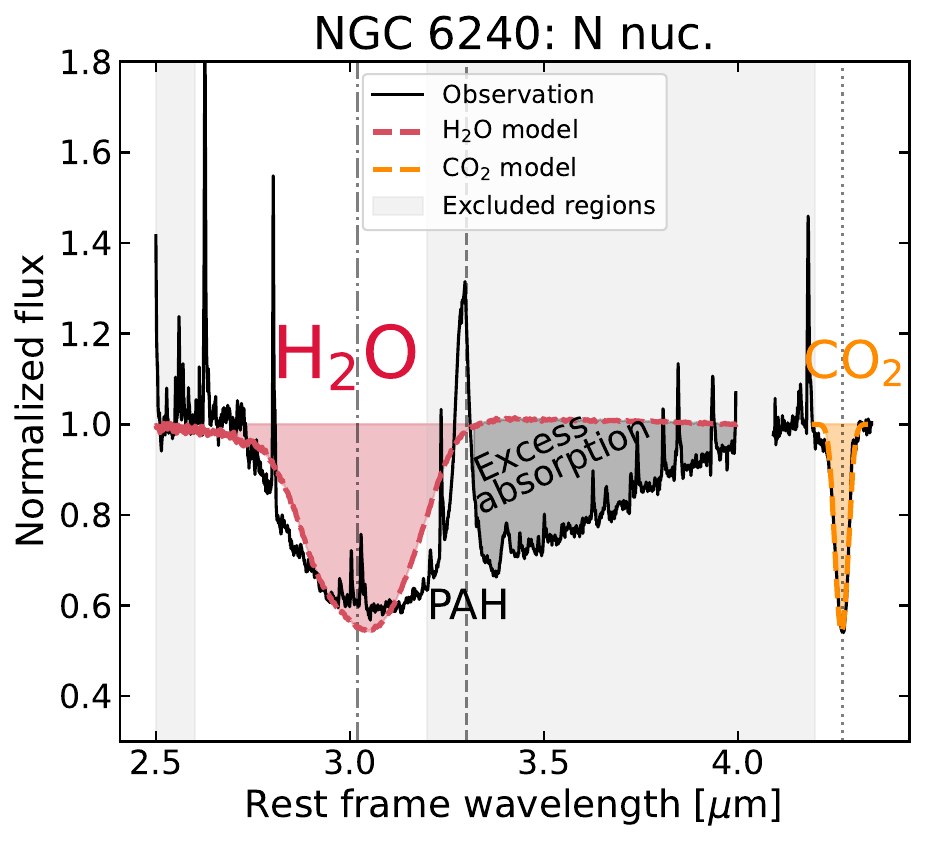}
\includegraphics[width=0.25\textwidth]{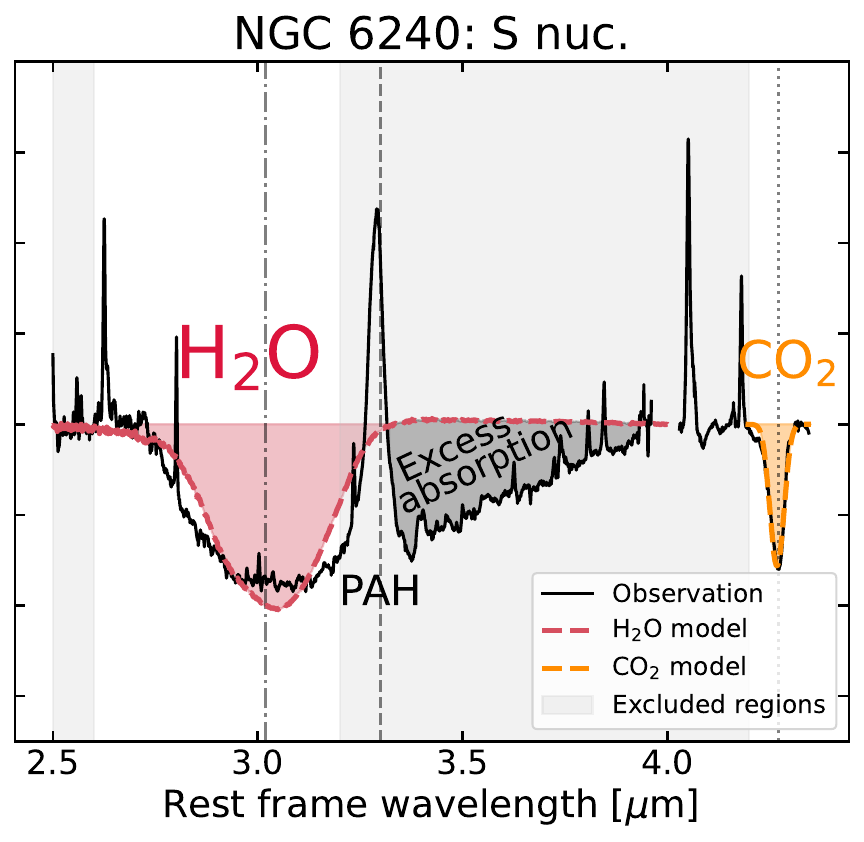}
\includegraphics[width=0.46\textwidth]{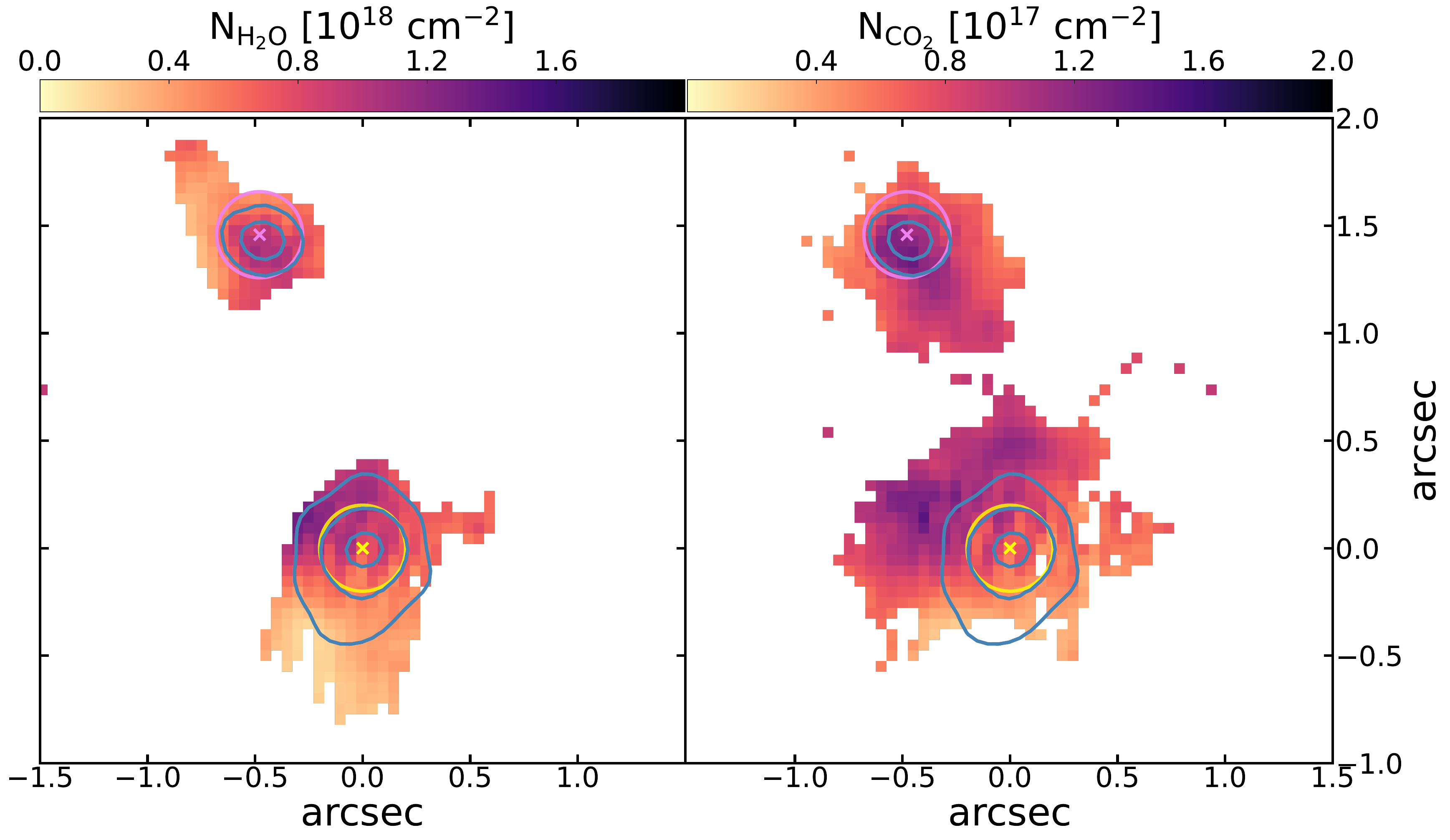}
\caption {Same as Fig.~\ref{fig:ice} but for NGC~6240--north.} 
\label{fig:ngc6240nices}
\end{figure*}

\begin{figure*} 
\centering
\includegraphics[width=0.26\textwidth]{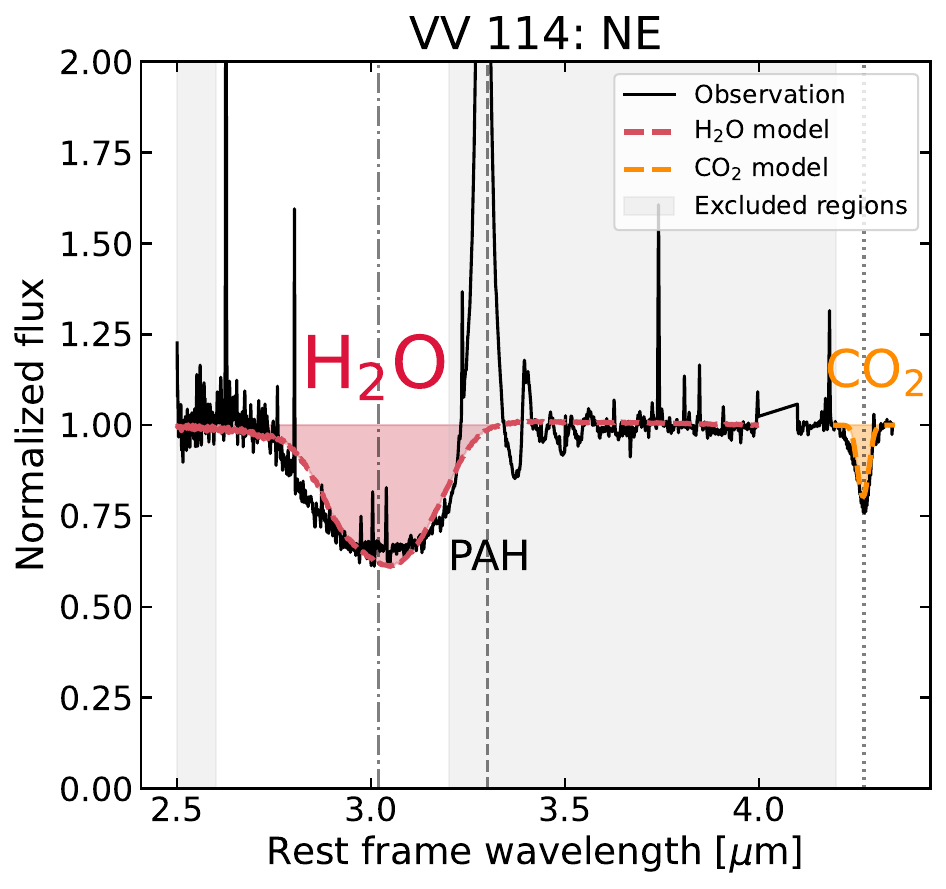}
\includegraphics[width=0.24\textwidth]{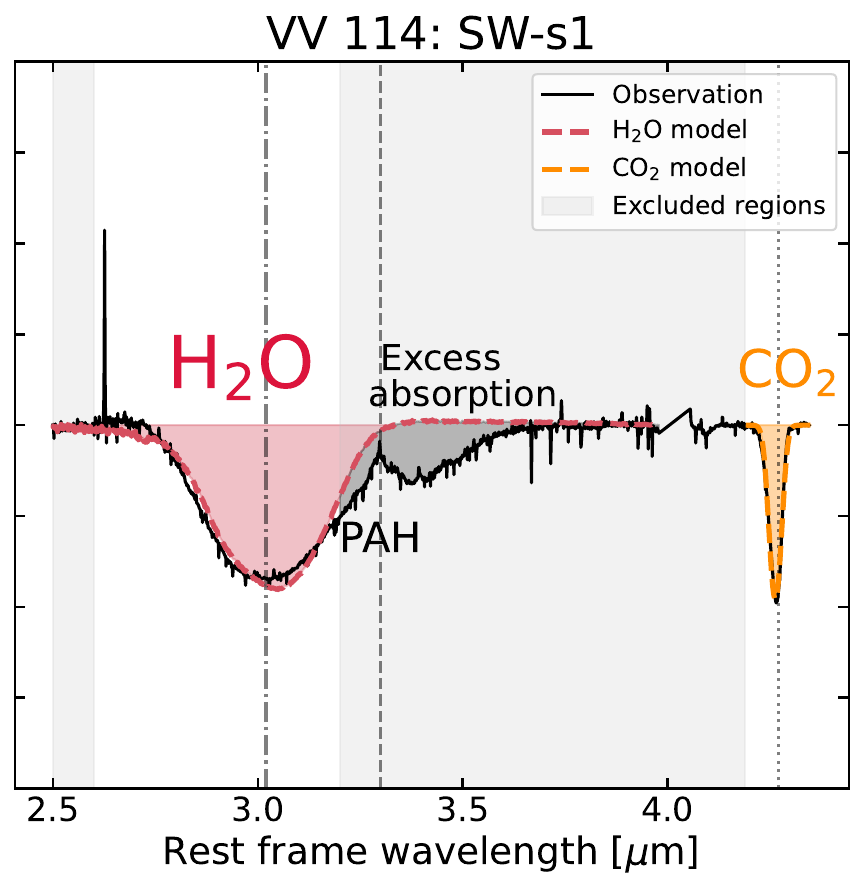}
\includegraphics[width=0.24\textwidth]{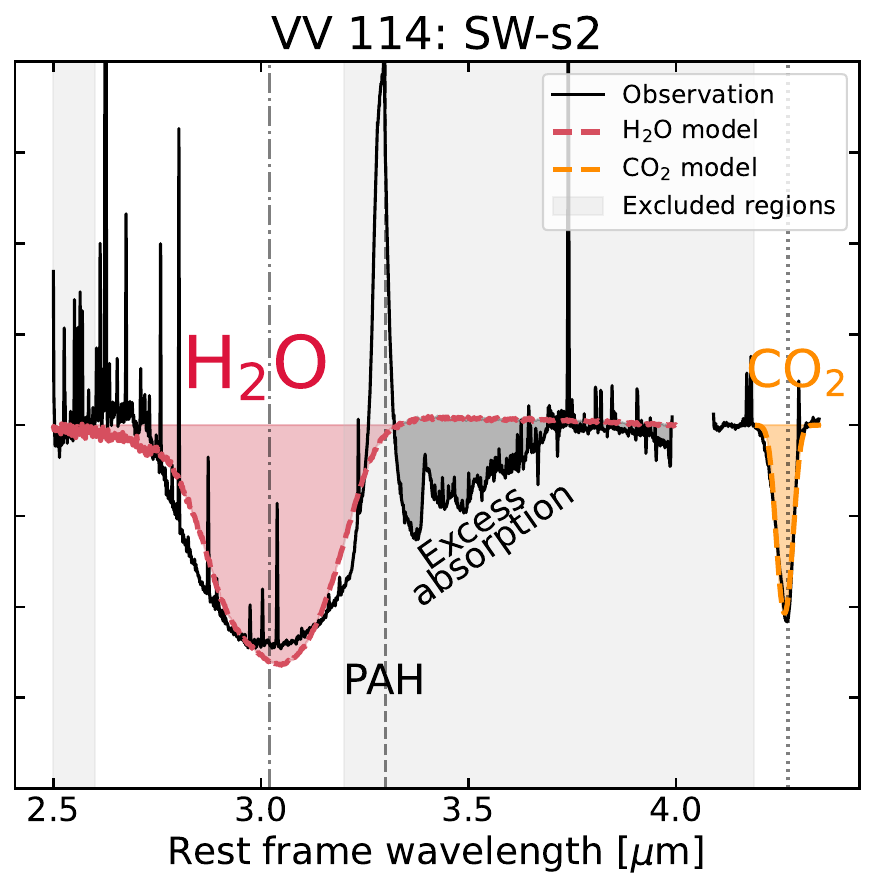}
\includegraphics[width=0.58\textwidth]{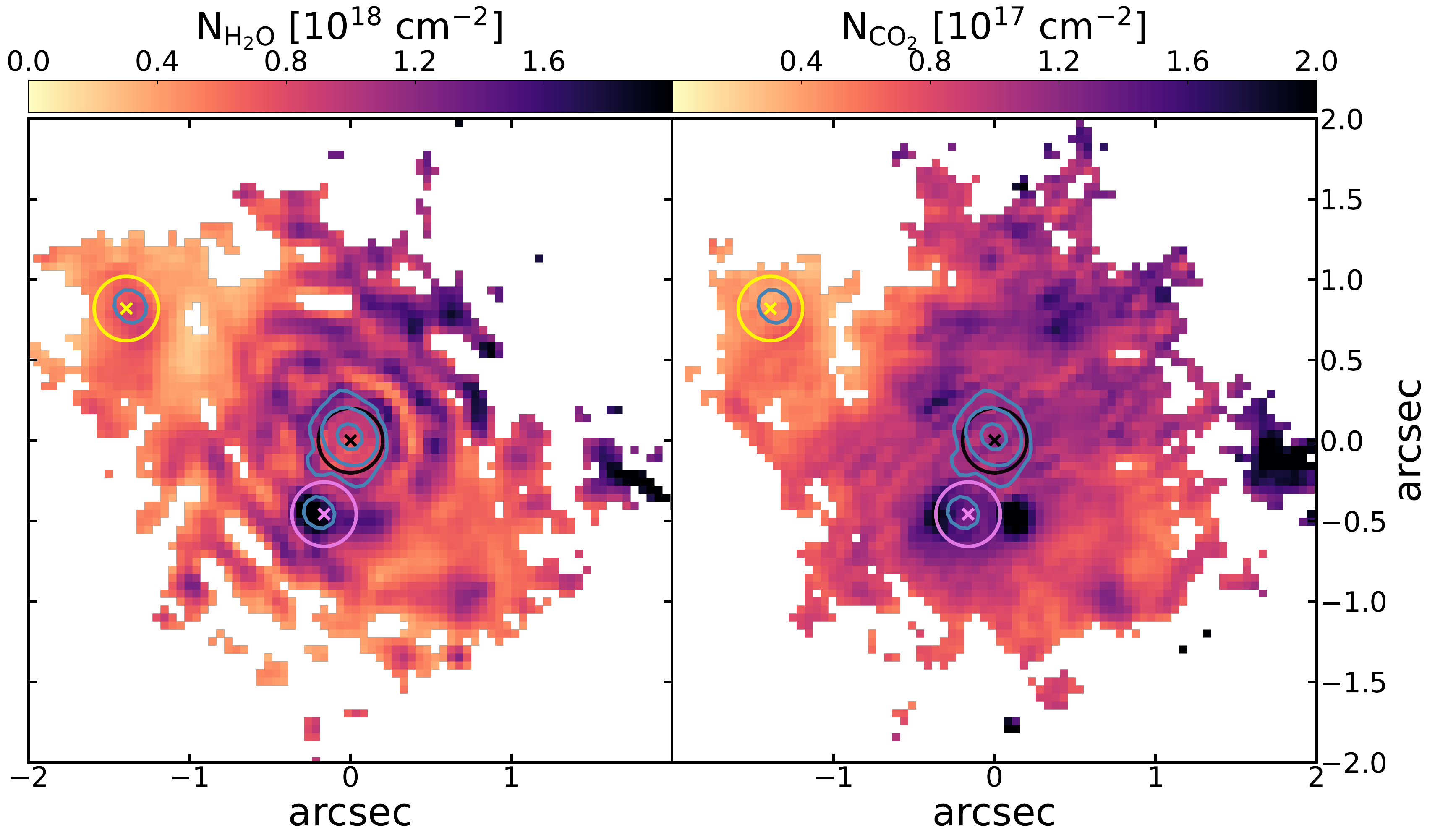}
\caption {Same as Fig.~\ref{fig:ice} but for  VV~114 E.} 
\label{fig:vv114ices}
\end{figure*}

\begin{figure*} 
\centering
\includegraphics[width=0.415\textwidth]{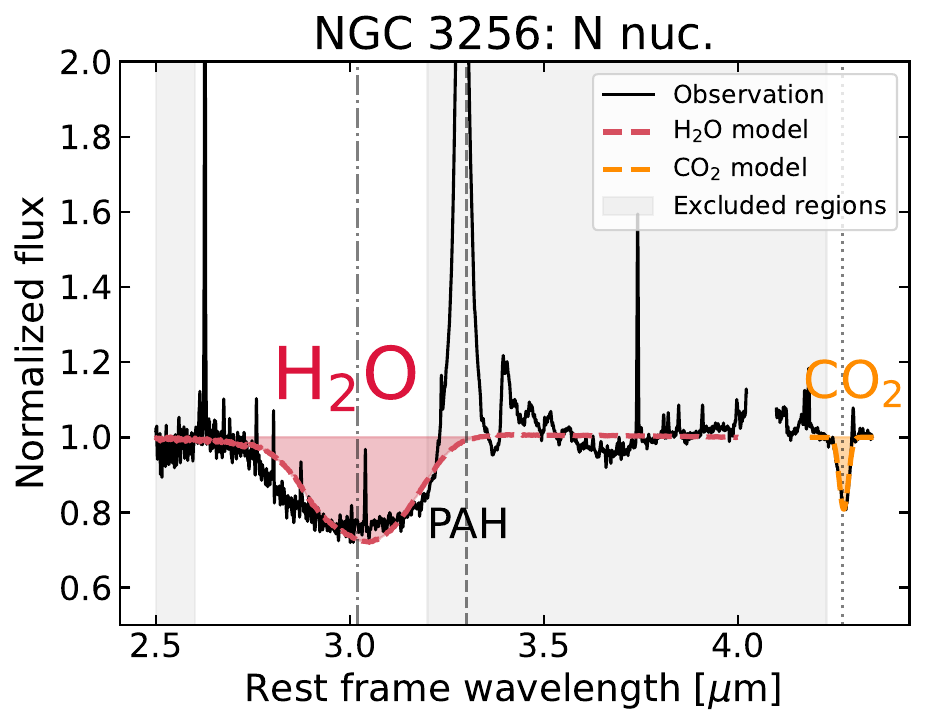}
\includegraphics[width=0.58\textwidth]{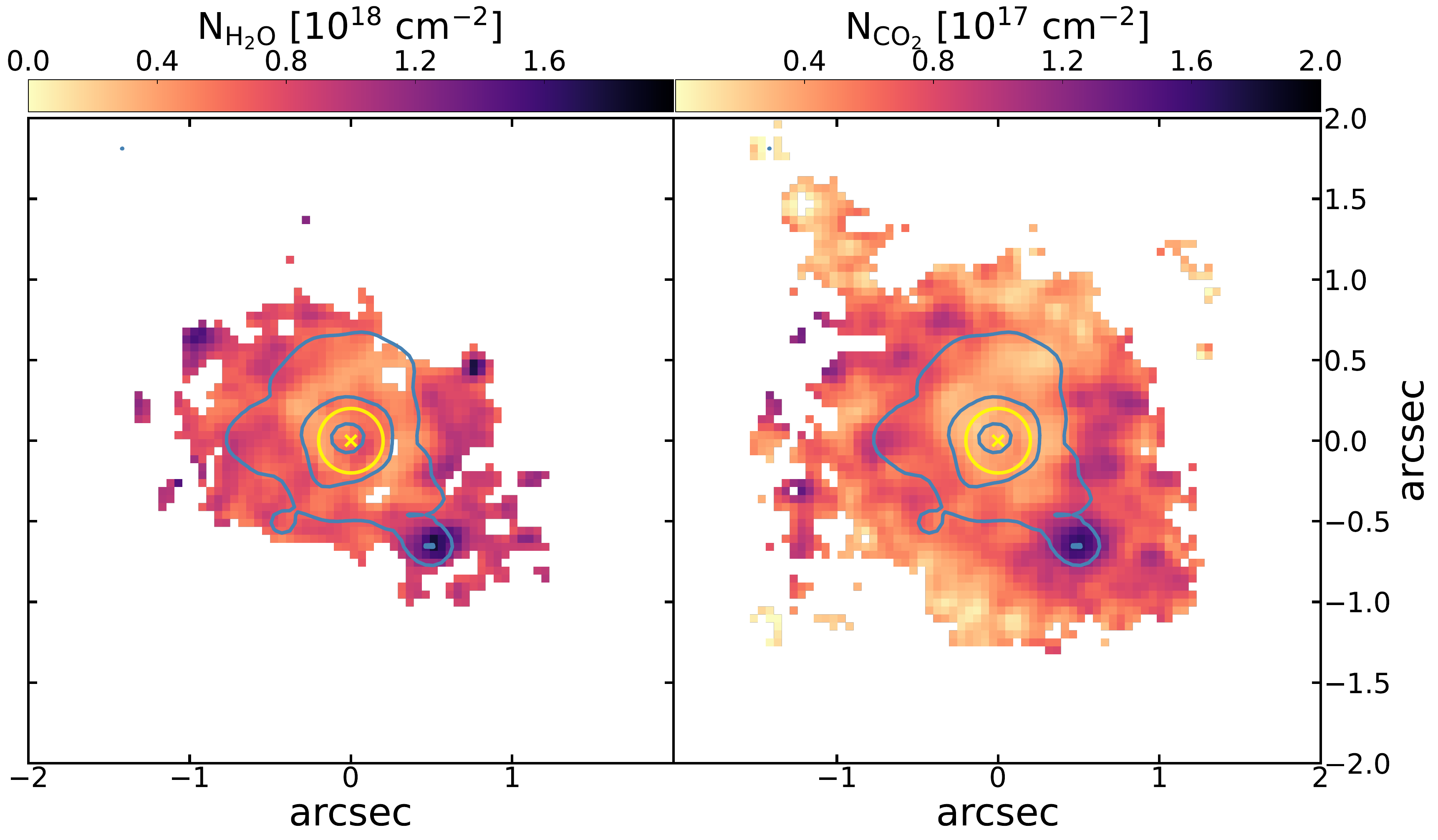}
\caption {Same as Fig.~\ref{fig:ice} but for NGC~3256--north.} 
\label{fig:ngc3256nices}
\end{figure*}

\begin{figure*} 
\centering
\includegraphics[width=0.415\textwidth]{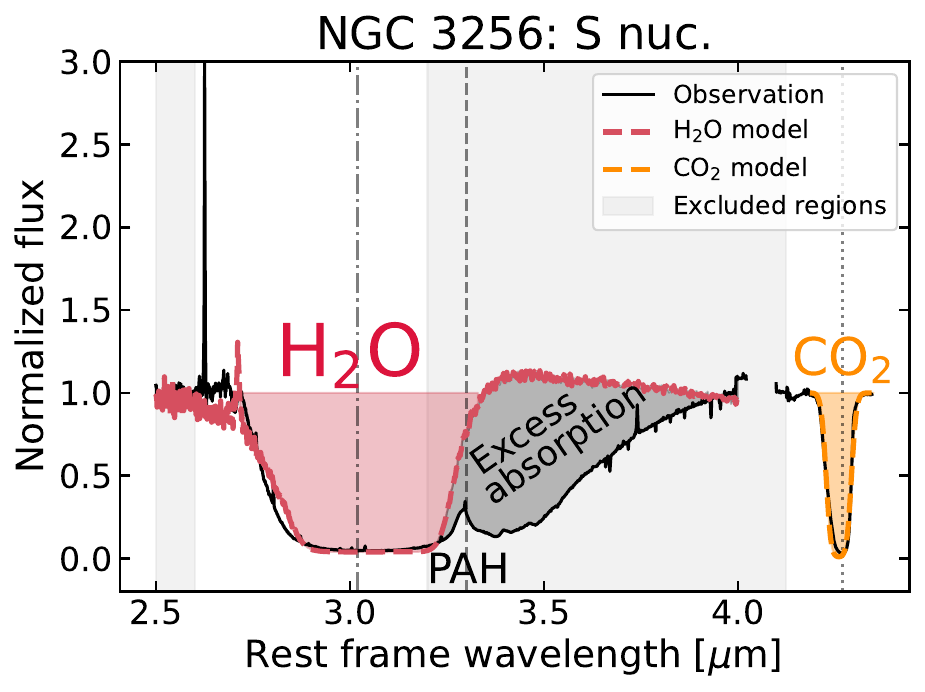}
\includegraphics[width=0.58\textwidth]{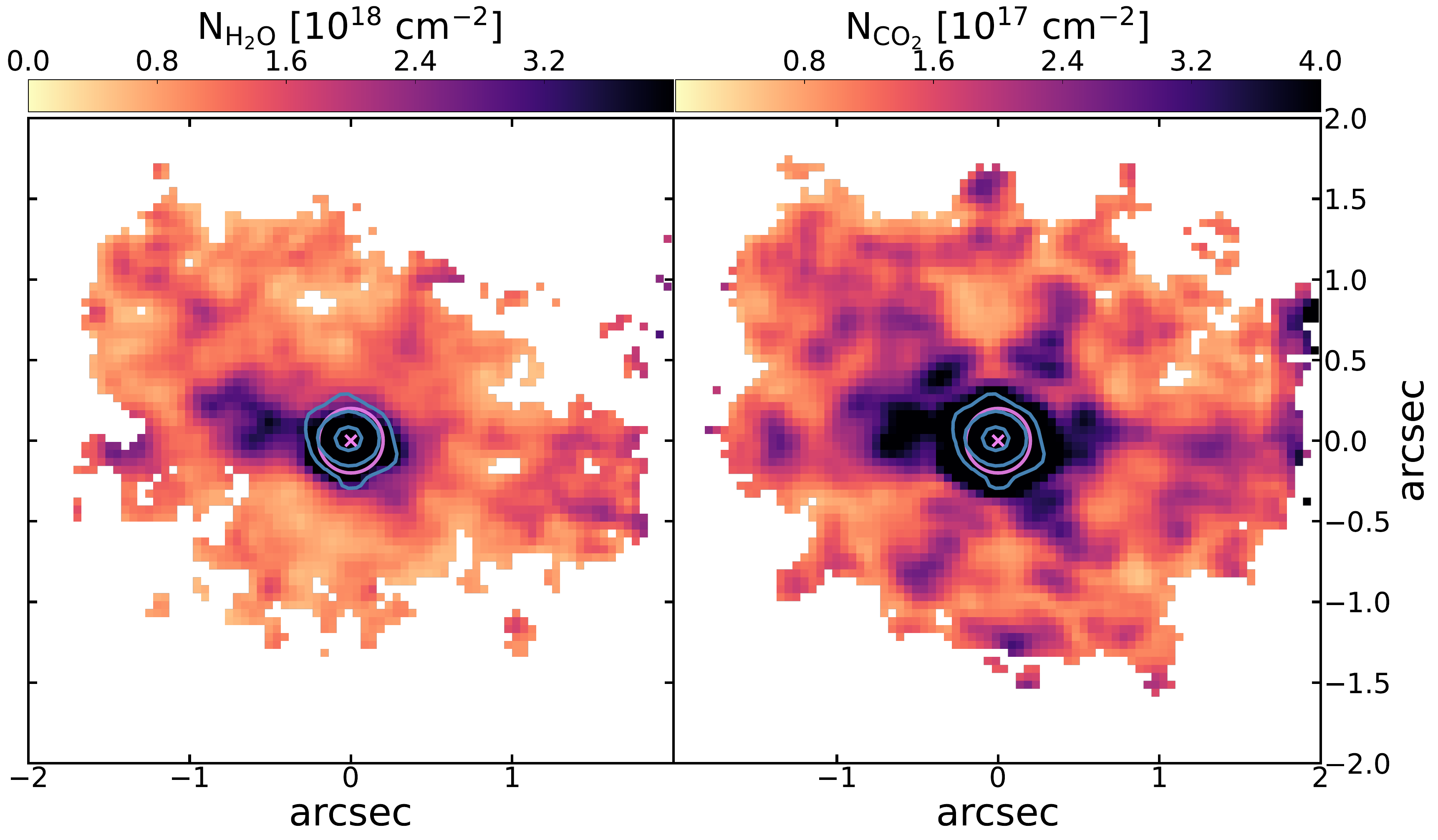}
\caption {Same as Fig.~\ref{fig:ice} but for NGC~3256--south.} 
\label{fig:ngc3256sices}
\end{figure*}

In this Section, we present the H$_2$O ($\sim$3.02 $\mu$m) and CO$_2$  ($\sim$4.26 $\mu$m) ice absorption features detected in the continuum, together with the spatial distribution of their column densities, following the method described in Section~\ref{sec:ices}. We show the counterparts of the  plots presented in Fig.~\ref{fig:ice} for Arp~220 E and its extended emission, but for Arp~220 W, VV~114 E NE, SW-s1, and SW-s2, NGC~6240 N and S, NGC 3256 S, and NGC~3256 N together with the  maps of N$_{\rm H_2O}$ and N$_{\rm CO_2}$ (see Fig.~\ref{fig:arp220wices}, ~\ref{fig:ngc6240nices}, ~\ref{fig:vv114ices}, ~\ref{fig:ngc3256nices}, and ~\ref{fig:ngc3256sices}).  Some maps are characterize by the PSF contamination, especially the N$_{\rm H_2O}$ map of VV~114, and the N$_{\rm CO_2}$ map of NGC~3256 S. For these cases, we masked the contaminated spaxels when measuring the column densities and relative N$_{\rm CO_2}$/N$_{\rm H_2O}$ abundances reported in Table~\ref{tab:nucext} and Fig.~\ref{fig:h2oco2}.

\section{Comparison of A$_{\rm K}^{\rm HII}$ derived from pair and numerous ($\geq$ 10) H lines } \label{appensixD}

\begin{figure*} 
\centering
\includegraphics[width=1\textwidth]{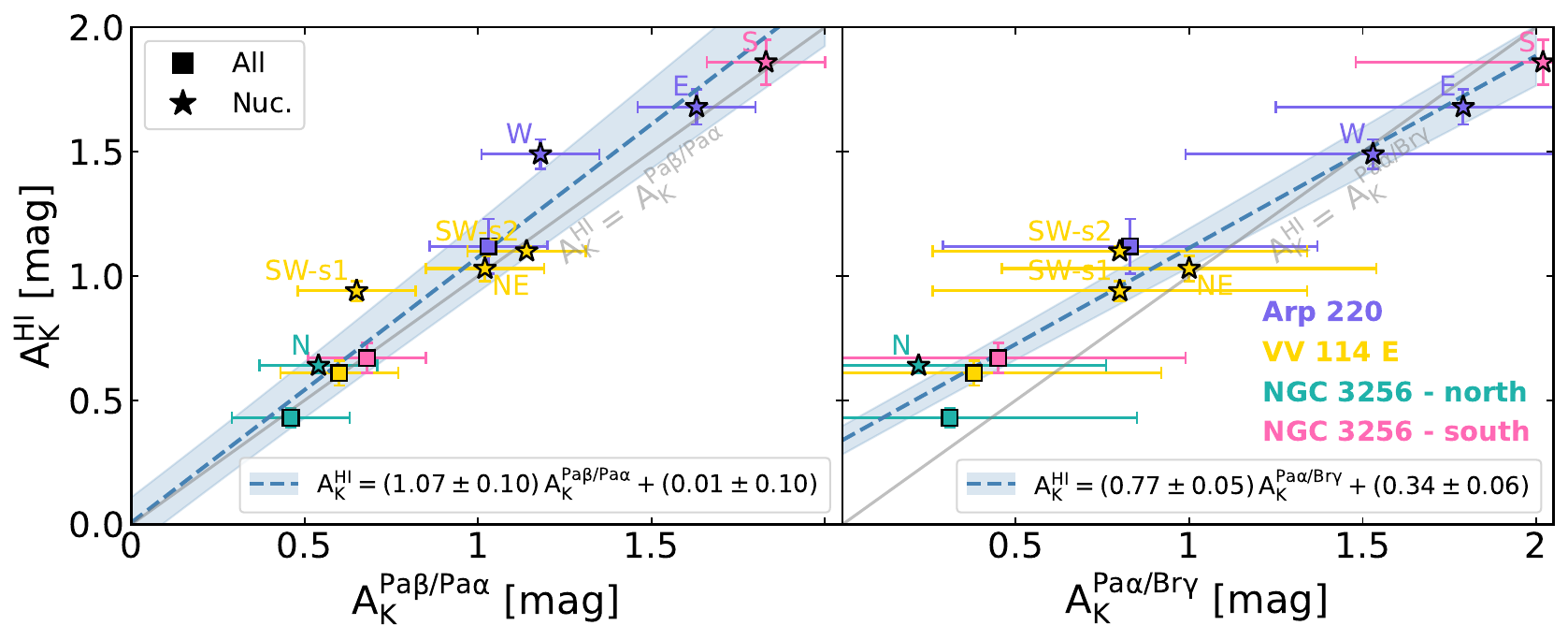}
\caption {Same as Fig.~\ref{fig:Ak_comparison} but applying the the extinction law assumed in
this paper (Eq.~\ref{eq:func1}) to  pair of lines, Pa$\beta$/Pa$\alpha$ (left panel) and Pa$\alpha$/Br$\gamma$ (right panel).} 
\label{fig:Ak_comparison_our}
\end{figure*}

We compare the ionized-gas extinctions estimated using a minimum of ten H emission lines with those derived from only a pair of recombination lines, applying the same Eq.~\ref{eq:func1} adopted in this study. We find that the dispersion increases as the differential extinction between the line pairs decreases. Accordingly, we obtain consistent extinctions between A$_{\rm K}^{\rm HII}$ and A$_{\rm K}^{\rm Pa\beta/Pa\alpha}$, while a larger dispersion is observed for A$_{\rm K}^{\rm Pa\alpha/Br\gamma}$. This behavior is similar to that found for classical extinction diagnostics (see Section~\ref{sec:comparison}).

\section{CO$_2$ to H$_2$O relative abundances} \label{appensix0}
\begin{figure*} 
\centering
\includegraphics[width=1.\textwidth]{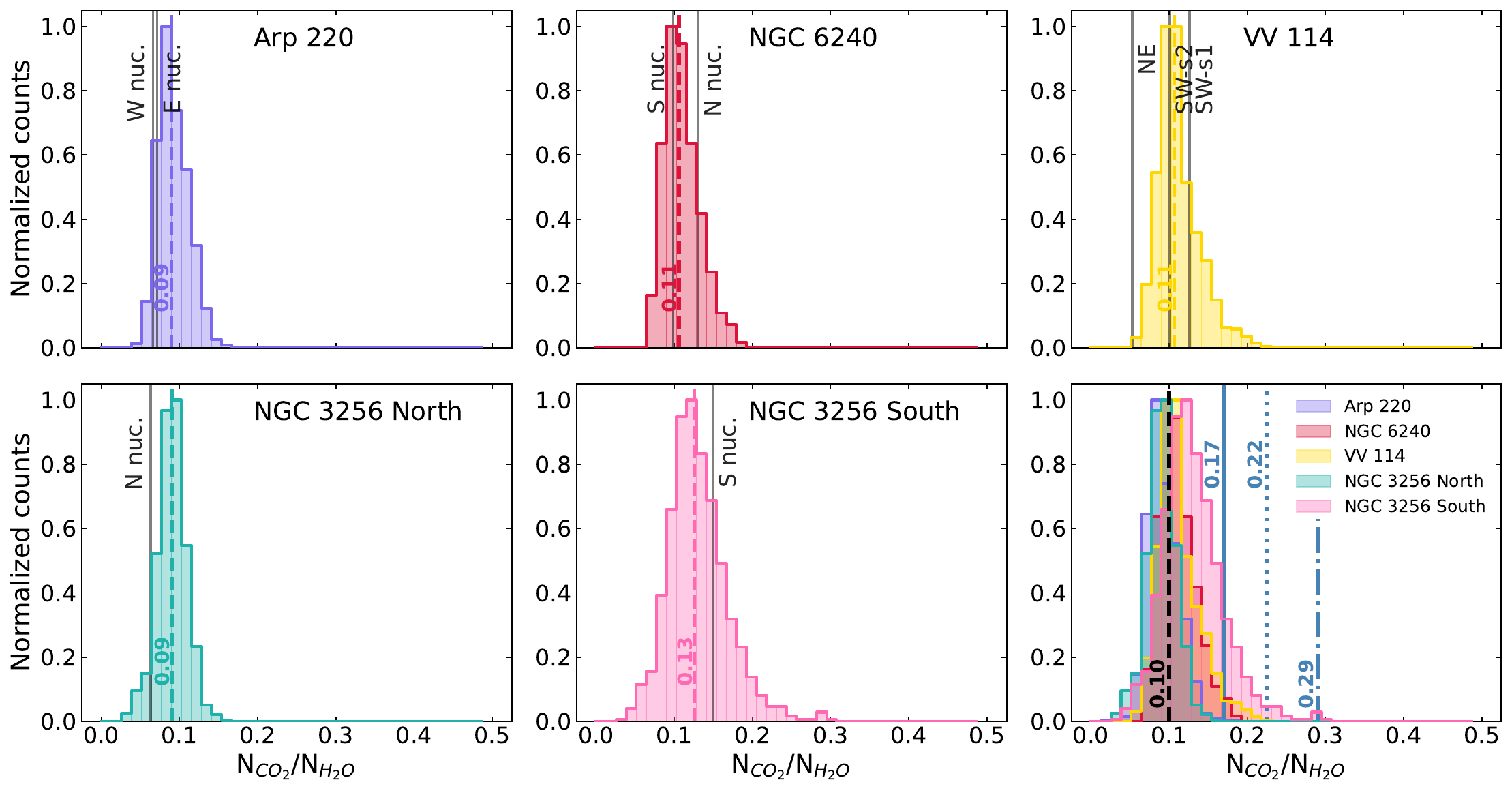}
\caption {Normalized histograms of the N$_{\rm CO_2}$/N$_{\rm H_2O}$  ice abundance ratio measured spaxel-by-spaxel in the rebinned cubes of Arp~220, NGC~6240, VV~114, NGC~3256 north,   and NGC~3256 south  (panels from top left to bottom center). The last panel combines all histograms and shows the global median abundance  ratio (N$_{\rm CO_2}$ = 0.10 N$_{\rm H_2O}$). Median values of each distribution are indicated by a vertical dashed line, with the corresponding fraction labeled. In the last panel, the blue solid, dotted, and dash-dotted vertical lines mark the median values obtained by \citet{Gerakines99} for 14 massive protostar and the Galactic center (using Infrared Space Observatory data), by \citet{Gibb04} for 18 Galactic sources  (expanding the same  data sample of \citealt{Gerakines99}), and by \citet{Smith25} for spatially resolved measurements in the Chamaeleon~I molecular cloud (using JWST data), respectively.}
\label{fig:h2oco2}
\end{figure*}

In Fig.~\ref{fig:h2oco2} we show the relative CO$_2$ to H$_2$O ice abundances and their comparison with Galactic values. A detailed discussion is provided in Section~\ref{ices_comp}.

\end{appendix}

\end{document}